\documentclass[oldversion]{aa}
\usepackage{txfonts}
\usepackage{graphicx}
\usepackage{natbib}
\bibpunct{(}{)}{;}{a}{}{,}
\begin{document}

\def \lessa {\mathrel{\vcenter
     {\offinterlineskip \hbox{$<$}\hbox{$\sim$}}}}
\def \grtra {\mathrel{\vcenter
     {\offinterlineskip \hbox{$>$}\hbox{$\sim$}}}}
\newcommand{\iau}{Int.Astron.U.}
\newcommand{\asj}{Astron.Soc.Jap.}

\title{Equation-of-state dependent features in shock-oscillation modulated
neutrino and gravitational-wave signals from supernovae}
\titlerunning{Equation-of-state dependent supernova signals}

\author{A.~Marek, H.-Th.~Janka, and E.~M\"uller}
\authorrunning{Marek et al.}
\institute{Max-Planck-Institut f\"ur Astrophysik,
Karl-Schwarzschild-Str.1, 85748 Garching,
Germany}
\offprints{H.-Th. Janka, \email{thj@mpa-garching.mpg.de}}
\date{\today}

\abstract{
We present two-dimensional (axisymmetric)
neutrino-hydrodynamic simulations of the long-time
accretion phase of a 15$\,M_\odot$ progenitor star after core
bounce and before the launch of a supernova explosion, when 
non-radial hydrodynamic instabilities like convection occur
in different regions of the collapsing stellar core and
the standing accretion shock instability (SASI) leads
to large-amplitude oscillations of the stalled shock with a
period of tens of milliseconds. Our simulations were
performed with the {\sc Prometheus-Vertex} code, which includes a
multi-flavor, energy-dependent neutrino transport scheme and employs
an effective relativistic gravitational potential. Testing
the influence of a stiff and a soft equation of state for hot
neutron star matter, we find that the non-radial mass motions 
in the supernova core impose a time variability on the neutrino 
and gravitational-wave signals with larger amplitudes, as well as 
higher frequencies in the case of a more compact nascent neutron
star. 
After the prompt shock-breakout burst of electron neutrinos, a
more compact accreting remnant produces higher neutrino luminosities
and higher mean neutrino energies. The observable neutrino emission
in the SASI sloshing direction exhibits a modulation of several 
ten percent in the luminosities and around 1$\,$MeV in the mean 
energies with most power at typical SASI frequencies 
between roughly 20 and 100$\,$Hz. The modulation
is caused by quasi-periodic variations in the mass accretion rate
of the neutron star in each hemisphere. At times later
than $\sim$50--100$\,$ms after bounce, the gravitational-wave 
amplitude is dominated by the growing low-frequency ($\la$200$\,$Hz)
signal associated with anisotropic neutrino emission.
A high-frequency wave signal results from nonradial
gas flows in the outer layers of the anisotropically accreting 
neutron star. Right after bounce such nonradial mass motions occur
due to prompt post-shock convection in both considered cases 
and contribute mostly to the early wave production around 100$\,$Hz.
Later they are instigated by the SASI and by convective overturn 
that vigorously stir the neutrino-heating and cooling layers, 
and also by 
convective activity developing below the neutrinosphere. The 
gravitational-wave power then peaks at about 300--800$\,$Hz, connected 
to changes in the mass quadrupole moment on a timescale of milliseconds.
Distinctively higher spectral frequencies originate from the
more compact and more rapidly contracting neutron star. Both the
neutrino and gravitational-wave emission therefore carry information
that is characteristic of the properties of the nuclear equation of
state in the hot remnant. The detectability of the SASI effects
in the neutrino and gravitational-wave signals is briefly discussed.
\keywords{
Supernovae: general -- Hydrodynamics -- Neutrinos -- Gravitational waves 
-- Dense matter
}}

\maketitle

\section{Introduction}
\label{sec:introduction}

Neutrinos and gravitational waves are the most direct potential
probes of the processes that occur deep inside of a dying star,
accompanying or causing the initiation of the stellar explosion. 
Neutrinos were already detected in connection with supernova
SN~1987A (Bionta et al.\ 1987, Hirata et al.\ 1987, Alexeyev et al.\ 
1988), although with poor statistics so that the extraction of
information for constraining the explosion mechanism was not possible.
The simultaneous measurement of signals of both types remains a 
very realistic hope for the next Galactic supernova.

Capturing neutrinos and gravitational waves from the same source has the
advantage of providing complementary insight into the conditions
of the stellar core. While neutrino signals reflect the density 
structure and thermodynamic
conditions in the high-density plasma of the collapsing core and 
forming neutron star, gravitational waves carry crucial
information about the dynamics and nonradial 
motions of the stellar matter, e.g.\ of its rotational
state or of hydrodynamic instabilities such as convection that deform and 
stir the condensing central compact remnant of the explosion.

The prompt burst of electron neutrinos, for example, signals the 
breakout of the supernova shock from the neutrinosphere and the
shock heating of matter in this region. 
The neutrino emission after core bounce is a sensitive
probe of the mass accretion rate on the forming neutron star, hence
of the density structure of the infalling layers of the dying star 
(e.g., Liebend\"orfer et al.\ 2003, Buras et al.\ 2006b).
The onset of the explosion is expected to show up as a 
more or less sudden drop in the mass accretion rate of the
forming neutron star. A soft supernuclear equation of state will lead to a 
more compact and hotter remnant, radiating higher neutrino luminosities 
and more energetic neutrinos (e.g., Janka et al.\ 2005, Marek 2007).
And a possible phase transition to non-nucleonic matter
in the neutron star core may impose characteristic features like a 
second neutrino burst (Sagert et al.\ 2008) or, if triggering
the collapse to a black hole, may cause an abrupt termination of the neutrino
emission (e.g., Burrows 1988; Keil \& Janka 1995;
Sumiyoshi et al.\ 2006, 2007; Fischer et al.\ 2008).

Theoretical work on the gravitational-wave signals from stellar core collapse
and explosion has a long history of successively refined numerical models.
In particular the infall and bounce phases, which are theoretically 
relatively well understood parts of the evolution, have received a
lot of interest, because a strong and characteristic signature could
make them a promising source of a detectable gravitational-wave
burst (for a review-like introduction to the subject and a nearly complete
list of publications, see Dimmelmeier et al.\ 2008). For this to be the case,
the core of the progenitor star must develop a sufficiently large deformation
during its infall, and for that it must rotate enough rapidly. This,
however, does not seem to be compatible with predictions from the latest 
generation of stellar evolution models for the vast majority of massive 
stars, which are expected to have lost most of their angular momentum before
collapse (Heger, Woosley, \& Spruit 2005). Only in rare, very special cases,
possibly accounting for the few tenths of a percent of all stellar core 
collapses that produce gamma-ray bursts, such stars seem to be able to 
retain a high angular momentum in their core and to thus produce relativistic
jets and highly asymmetric and extremely energetic, probably 
magnetohydrodynamically driven explosions (for a recent review, see e.g. 
Woosley \& Bloom 2006). If this hypothetical connection was true, 
the core bounce phase would not really offer grand perspectives for the
measurement of gravitational waves.

In contrast, only relatively little work has so far been done on 
determining the wave signals from the post-bounce evolution of a supernova,
although these signals are likely to carry important information about the 
still incompletely understood explosion mechanism and the associated core
dynamics (see the review by Ott 2008). M\"uller \& Janka (1997)
and later M\"uller et al.\ (2004) on the basis of significantly improved
numerical models, showed that in the case of delayed, neutrino-driven 
explosions convective overturn behind the stalled shock as well as 
convection inside the nascent neutron star can account for sizable
gravitational-wave signals, which should be detectable with a high
probability from a Galactic supernova when the Advanced Laser
Interferometer Gravitational-Wave Observatory (LIGO~II) is running
(see also Fryer et al.\ 2002, 2004).

While the delayed neutrino-heating mechanism relies on the support by
strong nonradial hydrodynamic instabilities in the region of 
neutrino-energy deposition, magnetohydrodynamic explosions, if
linked to rapid rotation, are expected to
develop relatively soon after core bounce and thus to occur faster
than the growth of the mentioned nonradial instabilities. 
They tap the reservoir of differential
rotation in a collapsing stellar environment and therefore
require rapidly spinning cores (see Burrows et al.\ 2007b, Thompson
et al.\ 2005). Thus they are good candidates for sizable
gravitational-wave pulses from the moment of 
core bounce (e.g., Kotake et al.\ 2006, Ott et al.\ 2004).
The newly proposed acoustic explosion mechanism, in which 
large-amplitude core gravity-mode oscillations of the neutron star 
convert accretion power to pressure and shock waves that feed the
supernova shock with acoustic energy (Burrows et al.\ 2006, 2007a),
is predicted to be associated with enormous and very characteristic 
gravitational-wave activity due to the fast periodic movement of 
roughly a solar mass of dense matter at late times ($\sim$1~second)
after core bounce (Ott et al.\ 2006).

In the present paper our focus is on an analysis of the features
in the neutrino and gravitational-wave signals that might provide
evidence for the action of the so-called 
standing accretion shock instability
(SASI), which has been shown to lead to low-$\ell$ mode (in terms of an
expansion in spherical harmonics with order $\ell$), large-amplitude
nonradial shock deformation and violent sloshing motions of the
stalled supernova shock (Blondin et al.\ 2003; Blondin \&
Mezzacappa 2007; Bruenn et al.\ 2006;
Scheck et al.\ 2004, 2006, 2008; Ohnishi et al.\ 2006,
Foglizzo et al.\ 2007; Yamasaki \& Foglizzo 2008). This
SASI activity is found to reach the nonlinear regime at 
roughly 100$\,$ms
after core bounce and to grow in amplitude over possibly hundreds
of milliseconds (Scheck et al.\ 2008; Marek \& Janka 2007; Burrows
et al.\ 2006, 2007a). It does not only act as seed of powerful 
secondary convection but can also provide crucial aid for
the neutrino-heating mechanism by pushing the accretion 
shock to larger radii and by thus stretching the time accreted
matter is exposed to neutrino heating in the gain layer 
(Buras et al.\ 2006b, Scheck et al.\ 2008, Marek \& Janka 2007,
Murphy \& Burrows 2008). Moreover, the SASI is found to
play the driving force of the g-modes pulsations of the neutron 
star core that are the essential ingredient of the acoustic 
mechanism. 

Observational signatures of the presence of the SASI would 
therefore be extremely important for our understanding of how
massive stars begin their explosion. Supernova asymmetries and pulsar
kicks are one, yet not unambiguous observational hint.
Neutrinos and gravitational waves,
which originate directly from the region where the blast is
initiated, however, may remain the only way to obtain direct
information. It is therefore a highly relevant question to ask whether
any characteristic structures are imprinted on the neutrino and 
gravitational-wave emission by the SASI activity in the supernova
core. In contrast to convection, which exhibits the fastest
growth for the higher-$\ell$ modes 
(see Foglizzo et al.\ 2006), the SASI is
expected to possess the largest growth rates for the dipolar
and quadrupolar deformations (corresponding to $\ell = 1,\,2$; Blondin
\& Mezzacappa 2006; Foglizzo et al.\ 2007; Yamasaki \& Foglizzo 2008;
Foglizzo 2001, 2002). Even in the fully nonlinear situation the
geometry and motion of the shock and post-shock layer are found 
to be governed by these lowest modes. 

We present here an analysis of the neutrino and gravitational-wave
signals that are calculated on the basis of the two-dimensional 
post-bounce and 
pre-explosion simulations of a 15$\,M_\odot$ star recently published
by Marek \& Janka (2007). We constrain ourselves to two 
nonrotating models, which allow us to discuss the differences
that can be expected from a stiff and a soft nuclear equation
of state (results of the corresponding 1D simulations can also be found
in Marek \& Janka 2007). We find that the SASI sloshing of 
the shock and the associated quasi-periodic mass motions lead to a 
time-modulation of the neutrino emission and to gravitational-wave
amplitudes whose size and characteristic frequency depend
on the compactness of the proto-neutron star during the first half of a  
second after core bounce. Though the SASI contributions to the power
spectra are superimposed by a significant high-frequency ``noise''
due to convective fluctuations, at least the combination of 
measurements should make it possible to identify the SASI activity
in the supernova core.

Our paper is organized as follows.
In Sect.~\ref{sec:numerics} we will briefly outline
the main numerical and physics ingredients of our simulations,
and the basic features of the two simulations we compare. In
Sect.~\ref{sec:results} we discuss our results with respect to
the SASI effects on the shock motion, neutrino emission, and 
gravitational-wave signal, and in Sect.~\ref{sec:conclusions} 
we will summarize our findings and draw conclusions, 
including a discussion whether the SASI modulations are
detectable by neutrino and gravitational-wave experiments.

\section{Code, input, models}
\label{sec:numerics}

The 2D simulations discussed in this paper were performed with the
{\sc Prometheus-Vertex} code, whose numerical aspects and the 
implemented microphysics were described by Rampp \& Janka (2002)
and Buras et al.\ (2006a), and the publications quoted in those papers.
The detailed list of ingredients was also provided by Marek \& Janka 
(2007), where the simulations were already introduced as Models
M15LS-2D and M15HW-2D and compared with other cases in a greater
set of calculations. We therefore repeat
only a few essential aspects of immediate relevance here and refer
the reader to Sect.~2 of the Marek \& Janka (2007) paper for more 
complete information.

The hydrodynamics module of the code is based on a conservative
and explicit Eulerian implementation of a Godunov-type scheme
with higher order spatial and temporal accuracy. It solves the
nonrelativistic equations of motions for the stellar fluid, whose
self-gravity is described by an ``effective relativistic 
potential'' for an approximative treatment of general relativistic
gravity (see Marek et al.\ 2006; the discussed simulations were 
performed with the potential of Case~A from this work). 
The neutrino transport, which is coupled to the hydrodynamics part
via lepton number, energy, and momentum source terms, is computed
with our ``ray-by-ray plus scheme'' (see Buras et al.\ 2006a).
It accounts for the full neutrino-energy dependence of the transport 
but treats its dependence on the direction of the neutrino 
momentum in an approximative way, which is numerically less 
demanding and more efficient than a full multi-dimensional 
version of the transport. Recent multi-angle simulations in
two-dimensional situations (Ott et al.\ 2008) --- though done at 
the expense of a sophisticated description of the energy 
dependence --- show that a detailed
angular treatment produces considerably less lateral smearing of
the outward directed radiation field
than flux-limited diffusion. We suspect that our ray-by-ray 
description compares much better with multi-angle results than
flux-limited diffusion does.

The progenitor star used for our simulations was 
model s15s7b2 from Woosley \& Weaver (1995), which is a standard 
nonrotating
15$\,M_\odot$ star widely used for supernova simulations. We 
employed two different nuclear equations of state (EoS) for our 
studies: (1) a soft version of the Lattimer \& Swesty (1991)
EoS (``L\&S EoS'') with
an incompressibility modulus of bulk nuclear matter of 180$\,$MeV 
and a symmetry energy parameter of 29.3$\,$MeV, and (2)
the considerably stiffer EoS of Hillebrandt \& Wolff 
(1985; ``H\&W EoS''; see also Hillebrandt et al.\ 1984), whose
parameter values are 263$\,$MeV and 32.9$\,$MeV, 
respectively\footnote{Note that below a certain density, which is
typically chosen to be $10^{11}\,$g$\,$cm$^{-3}$ after core bounce, 
we replace the high-density EoSs by an ideal-gas equation of state
with electrons, positrons, photons, and a mixture of classical,
nonrelativistic Boltzmann gases for nucleons, alpha particles, and
14 kinds of heavier nuclei (see Marek \& Janka 2007 for more
information).}$^{,}$\footnote{In order to save computer time, in particular
during the simulation phase right after core bounce when the timesteps
are constrained to very low values, the model with the H\&W EoS was 
performed with the assumption of equatorial symmetry until 125.3$\,$ms
post bounce. Only afterwards it was continued with a full 180$^\circ$ 
grid. Tests showed that this had no important influence on the results,
neither for the evolution of the shock radius, nor for the growth 
and development of hydrodynamic instabilities. In particular, we did
not find any qualitative difference in the SASI modes, also for 
odd values of $\ell$, at later times (see Marek~2007).}. 
The former leads to a radius of about 12$\,$km for cold neutron stars
with a ``typical'' (gravitational) mass of 1.4$\,M_\odot$, whereas
this radius is roughly 14$\,$km in the second case.
The Hillebrandt \& Wolff EoS is based on a largely different 
modeling approach for inhomogeneous nucleon matter than the L\&S EoS
and the more recent EoS of Shen et al.\ (1998). It employs a
Hartree-Fock calculation in contrast to the compressible liquid
drop model of the L\&S EoS and the relativistic mean field 
description of the Shen et al.\ EoS. The three EoSs yield significantly
different results in 1D core-collapse 
simulations with respect to the shock formation 
point, the luminosities and mean energies of the radiated neutrinos, 
and the evolution of the shock radius and neutron star radius
after bounce (see Janka et al.\ 2005; Marek 2007; Figs.~6 and 7 in
Janka et al.\ 2007). Since the EoS of
Shen et al.\ yields intermediate values for many of these 
quantities, we consider 
the soft Lattimer \& Swesty EoS on the one hand and the stiff 
Hillebrandt \& Wolff EoS on the other as two cases that roughly
span the range of extreme possibilities for baryonic matter around
and above nuclear saturation density near core bounce and shortly
afterwards.

\begin{figure*}[!htp]
\begin{center}
\includegraphics[width=8.5cm]{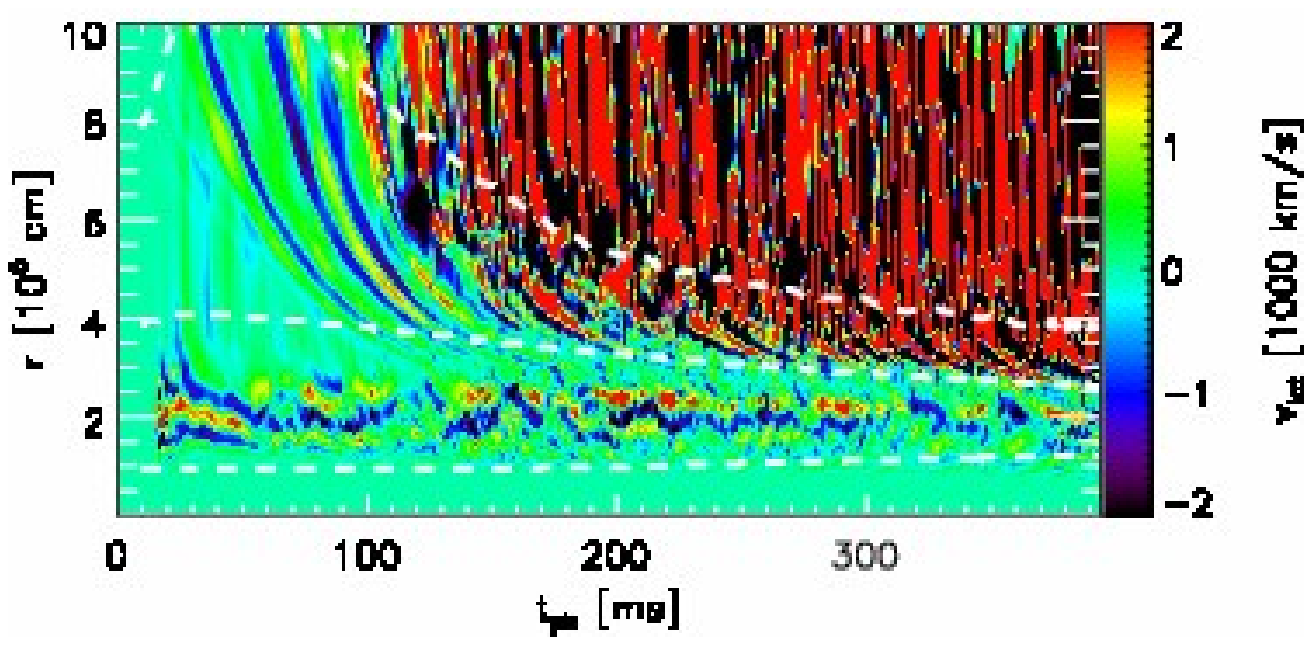} \hspace{0.3cm}
\includegraphics[width=8.5cm]{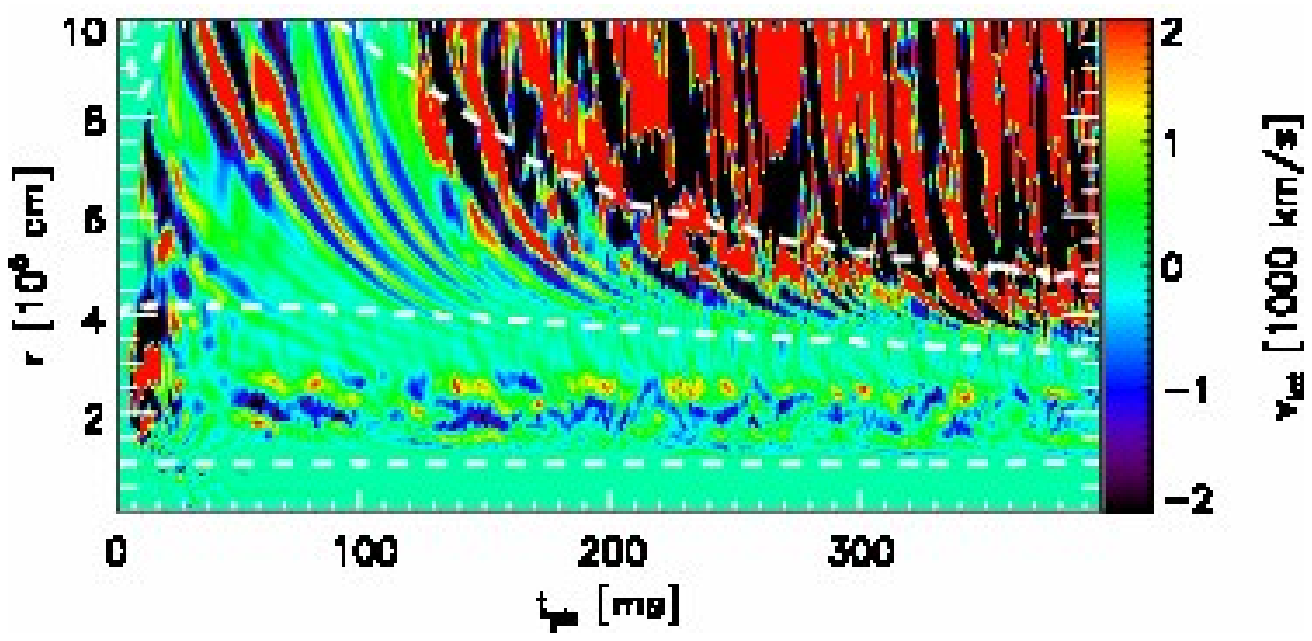} \\ \vspace{0.1cm}
\includegraphics[width=8.5cm]{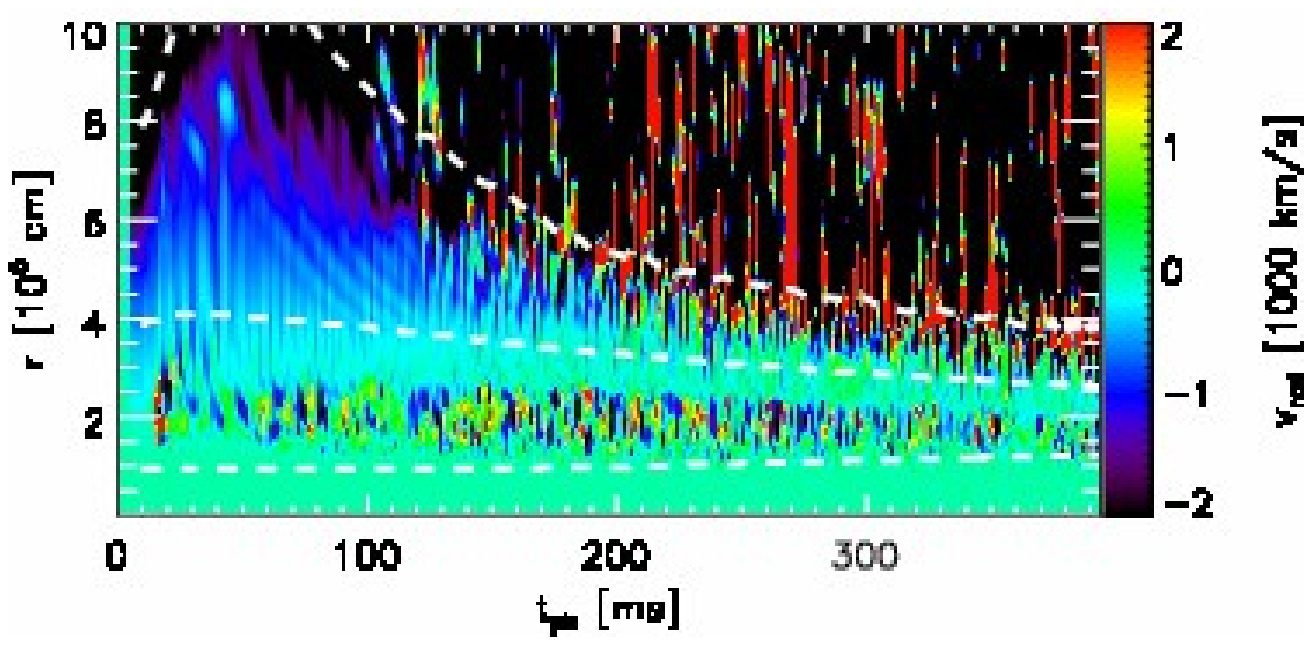} \hspace{0.3cm}
\includegraphics[width=8.5cm]{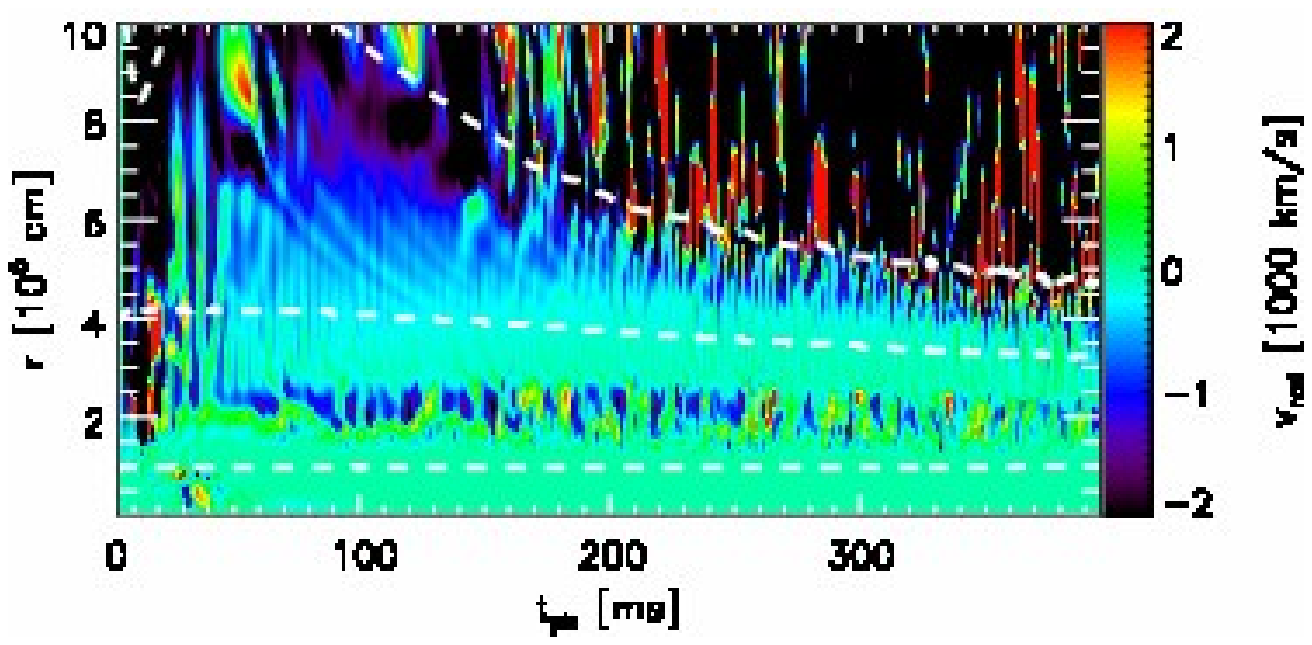} \\ 
\end{center}
\caption{Regions of mass motions due to hydrodynamic instabilities 
in the supernova core. The plots show color coded 
the lateral velocity ({\em top}) and the radial velocity ({\em bottom})
in the equatorial plane of the polar grid
as functions of post-bounce time and radius for the 2D simulation with
the L\&S EoS ({\em left}) and for the simulation with the H\&W EoS
({\em right}). The displayed range of velocity values is limited
to $\pm 2\times 10^8\,$cm$\,$s$^{-1}$. The white dashed lines mark 
(with decreasing radius) the locations of densities
$10^{10}$, $10^{12}$, and $10^{14}\,$g$\,$cm$^{-3}$. Prompt 
post-shock convection is strongest at $r \grtra 15\,$km before 
$t\sim 30\,$ms after 
core bounce, proto-neutron star convection occurs later at 
$r \lessa 30\,$km and $\rho > 10^{12}\,$g$\,$cm$^{-3}$, and SASI and
convective activity behind the shock are visible at 
$r \grtra 30$--40$\,$km.} 
\label{fig:convectionregions}
\end{figure*}

\begin{figure*}[!htp]
\begin{center}
\includegraphics[width=8.5cm]{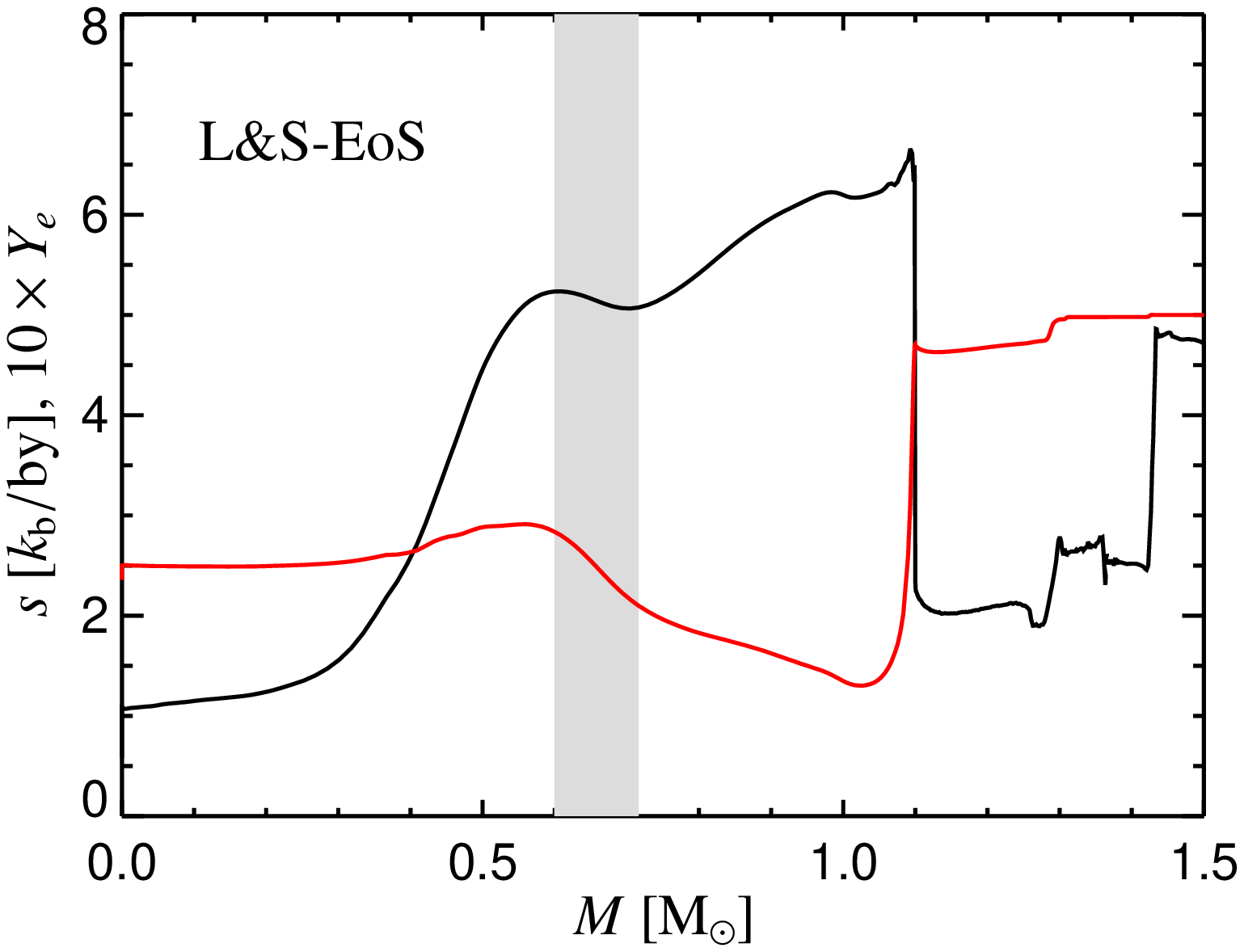}
\includegraphics[width=8.5cm]{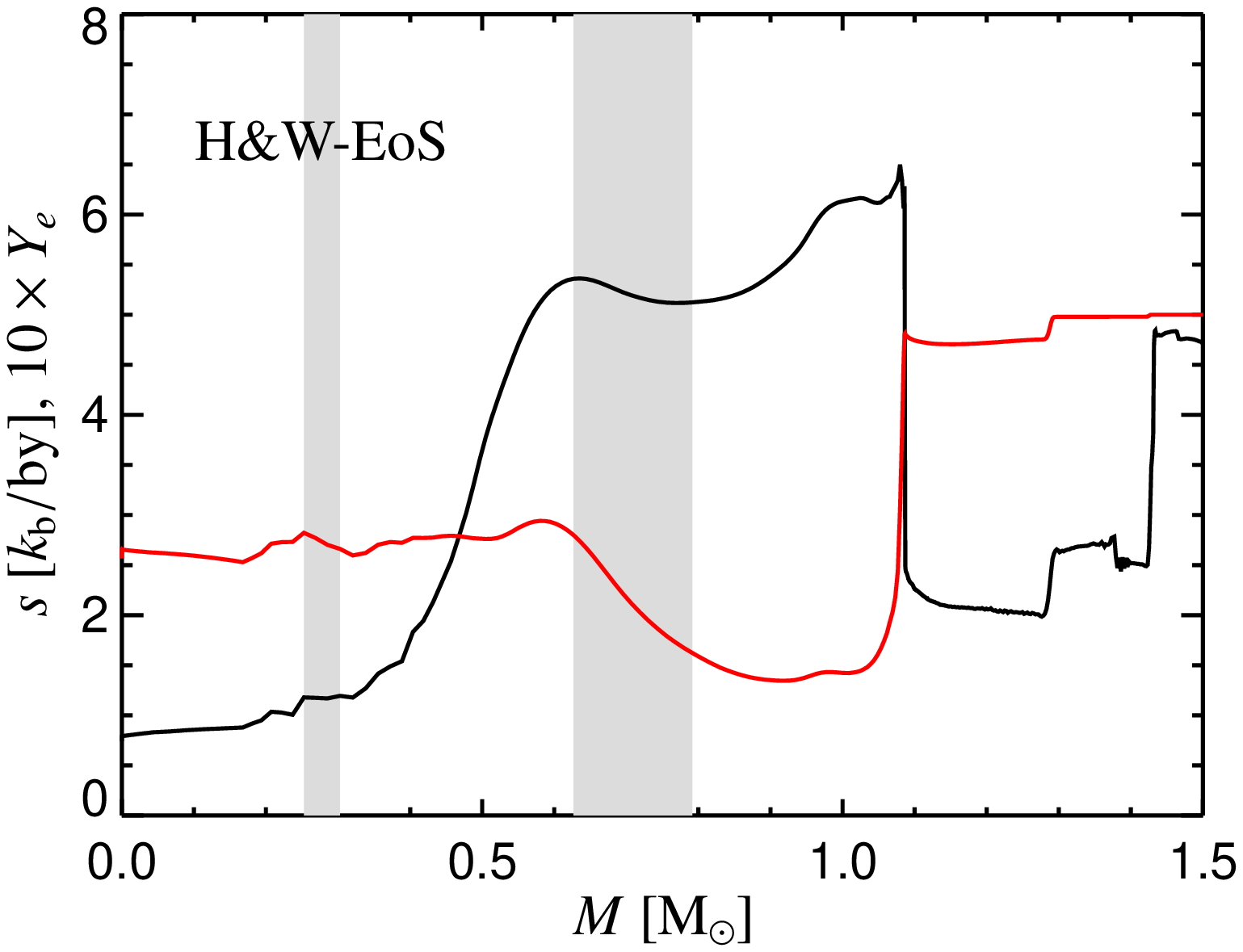}
\end{center}
\caption{Entropy (black) and electron fraction (red) profiles (the
plotted values correspond to $10\times Y_e$) as functions of
enclosed mass at 10$\,$ms after core 
bounce in spherically symmetric counterparts of the
two 2D simulations discussed in this paper. The shock is the
entropy jump near 1.1$\,M_\odot$. The vertical grey bars
indicate the mass regions that are convectively unstable according
to the Ledoux criterion.}
\label{fig:promptconvection}
\end{figure*}

\begin{figure*}[!htp]
\begin{center}
\includegraphics[width=8.5cm]{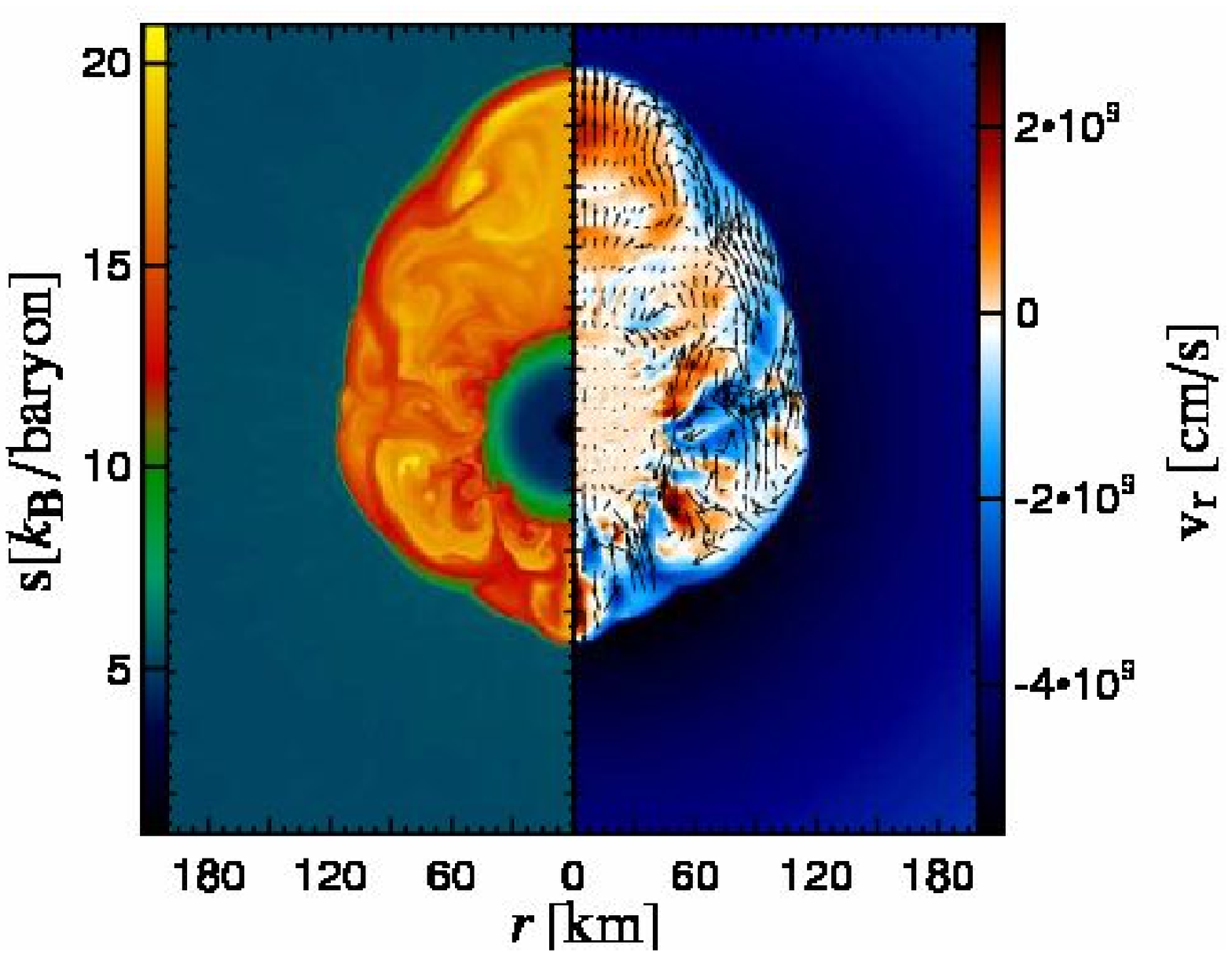} \hspace{0.2cm}
\includegraphics[width=8.5cm]{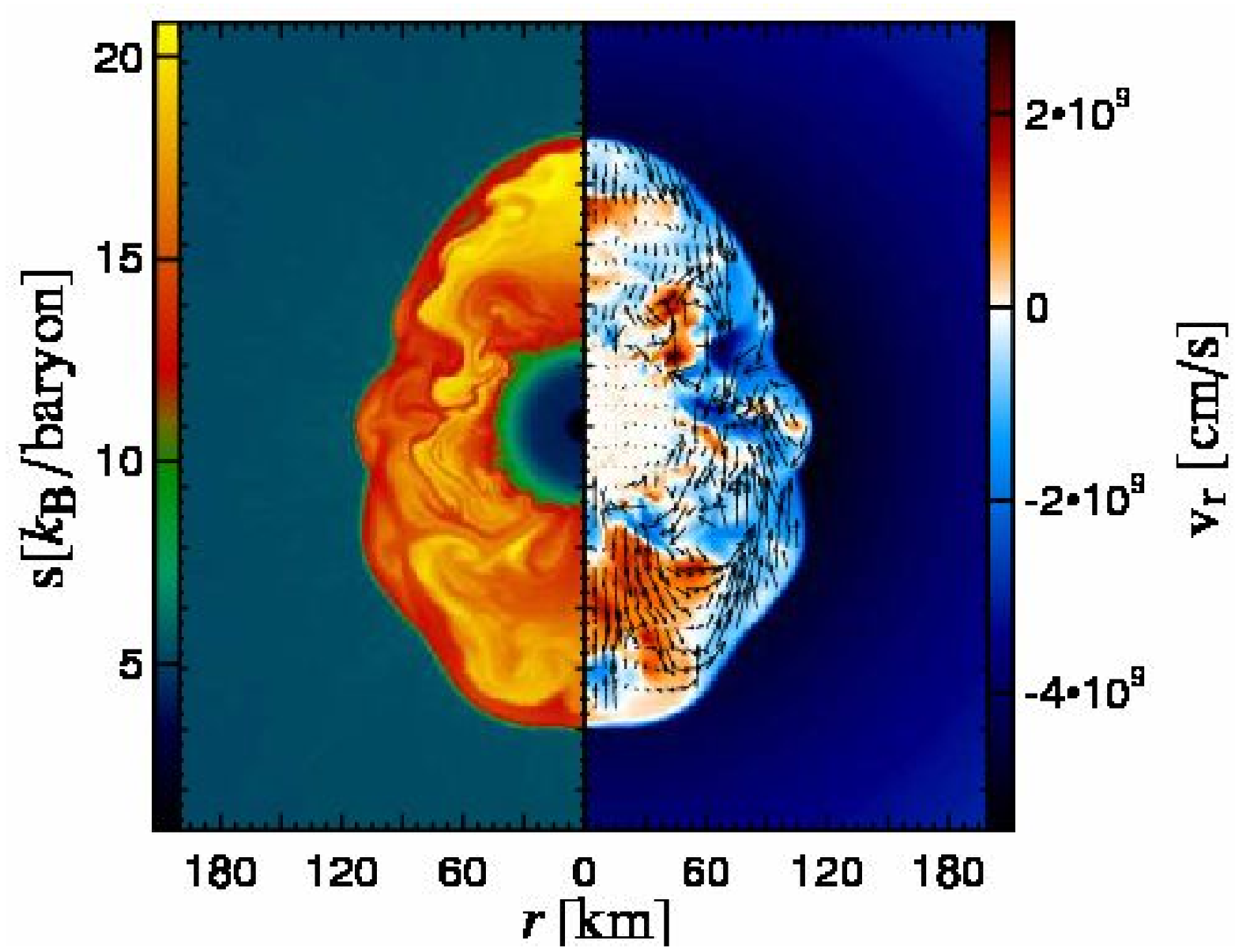}\\ \vspace{0.3cm}
\includegraphics[width=8.5cm]{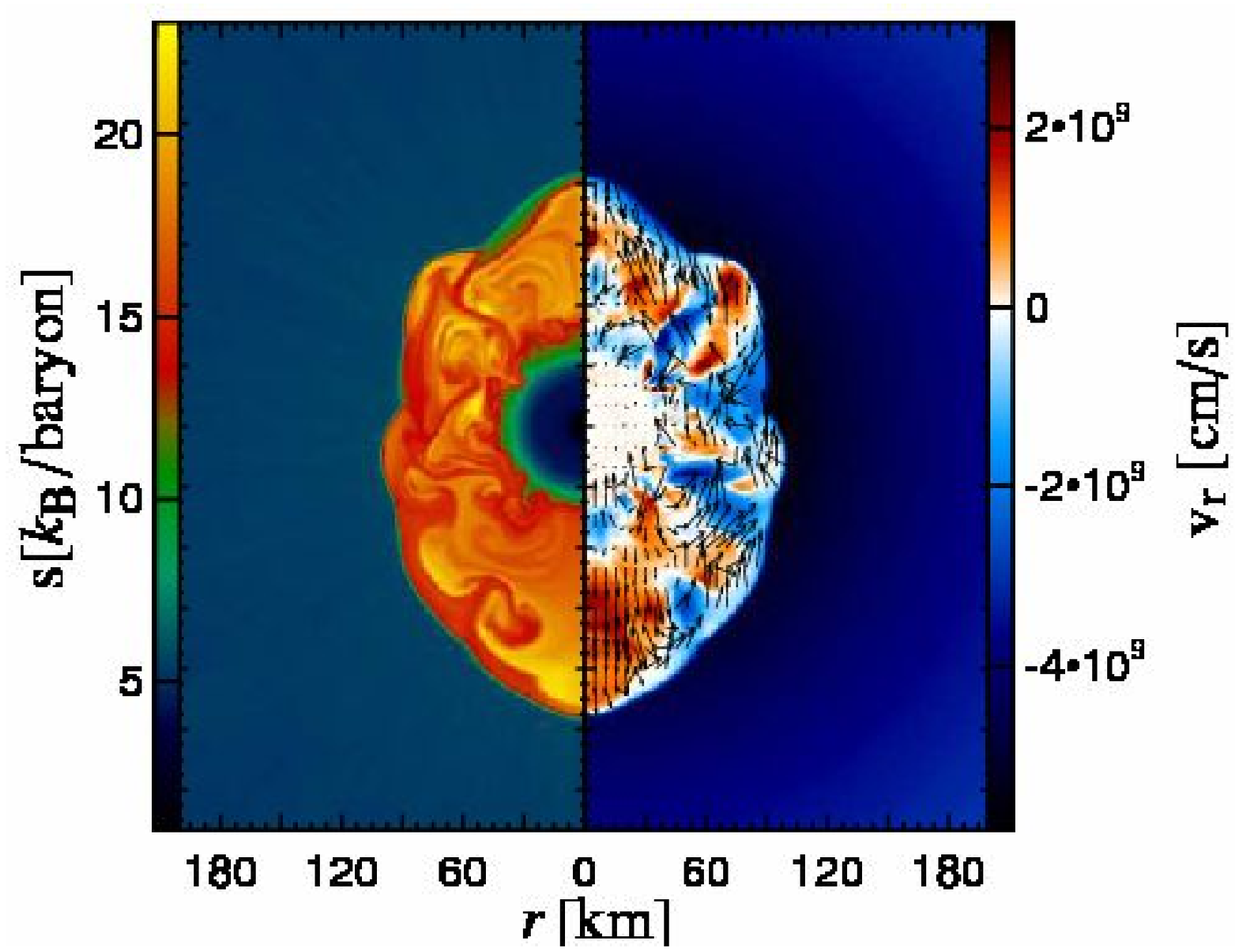} \hspace{0.2cm}
\includegraphics[width=8.5cm]{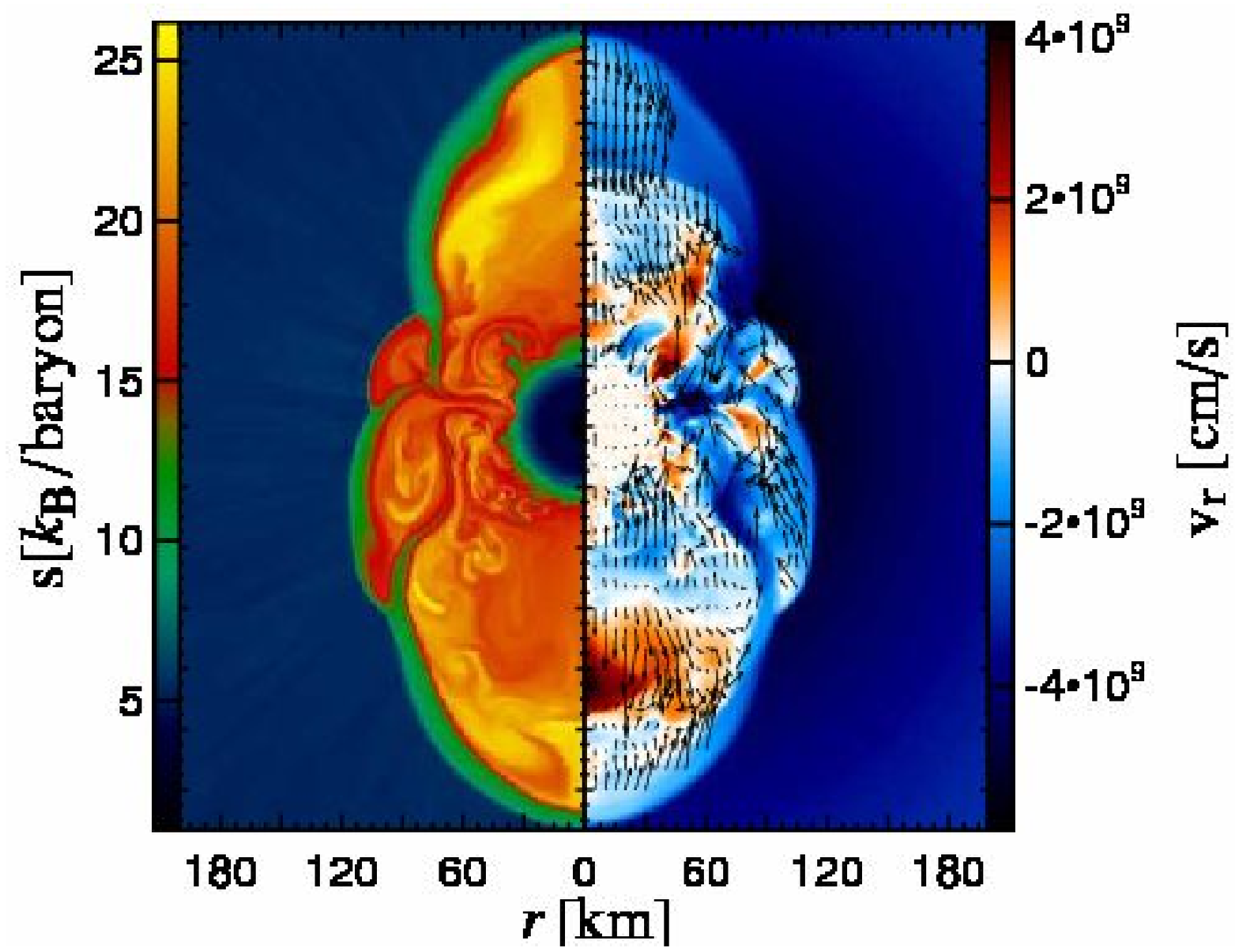}
\end{center}
\caption{Four representative snapshots from the 2D simulation with 
the L\&S EoS at post-bounce times of 247$\,$ms ({\em top left}),
255$\,$ms ({\em top right}), 322$\,$ms ({\em bottom left}), 
and 375$\,$ms ({\em bottom right}). The lefthand panel of each figure
shows color-coded the entropy distribution, the righthand panel the
radial velocity component with white and whitish hues denoting
matter at or near rest; black arrows in the righthand panel indicate
the direction of the velocity field in the post-shock region
(arrows were plotted only in regions where the absolute values of
the velocities were less than $2\times 10^9\,$cm$\,$s$^{-1}$). 
The vertical axis is the symmetry axis of the 2D simulation.
The plots visualize the accretion funnels and expansion flows in
the SASI layer, but the chosen color maps are unable to resolve
the convective shell inside the nascent neutron star.}
\label{fig:snapshots}
\end{figure*}

\begin{figure*}[!htp]
\begin{center}
\includegraphics[width=8.5cm]{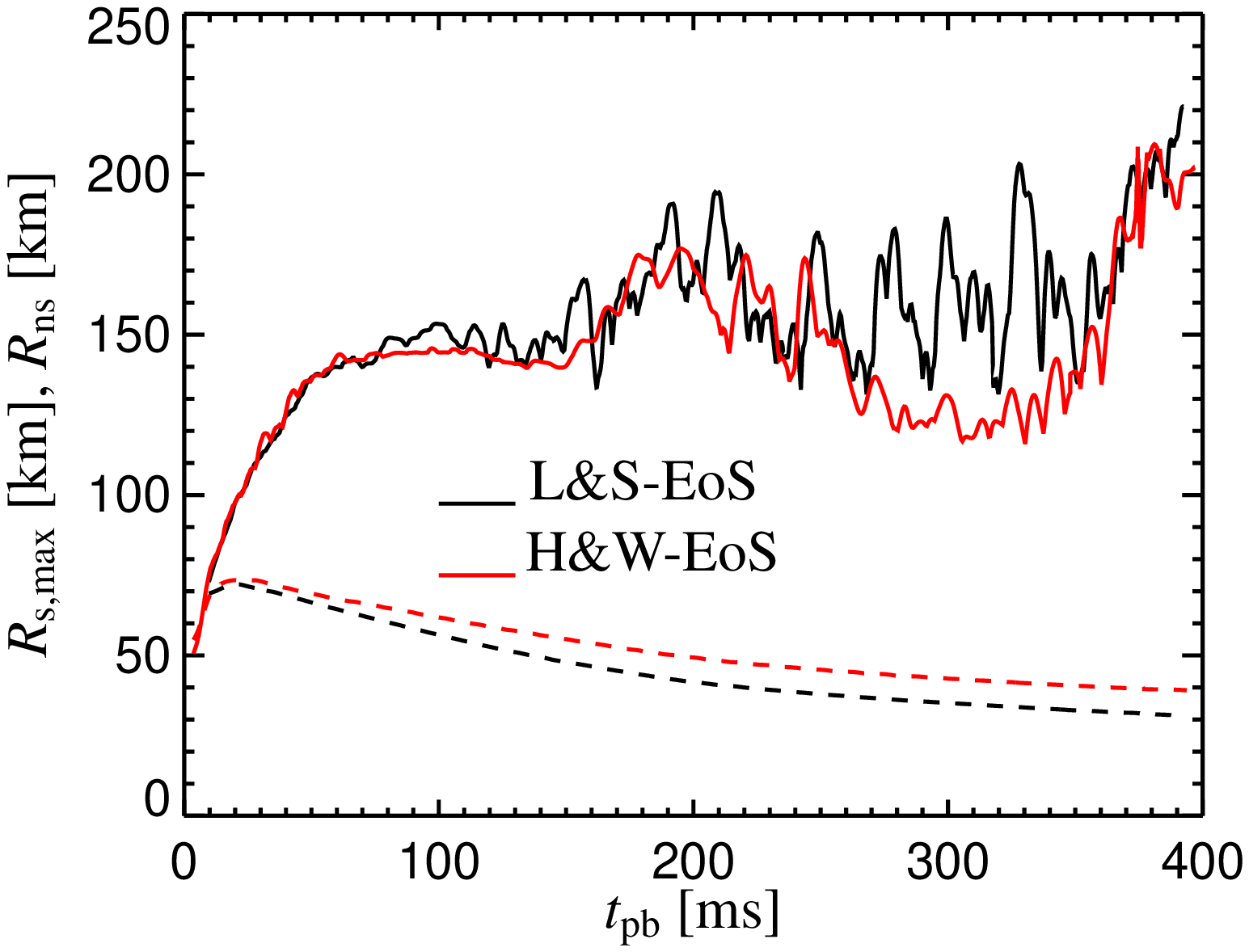}
\includegraphics[width=8.5cm]{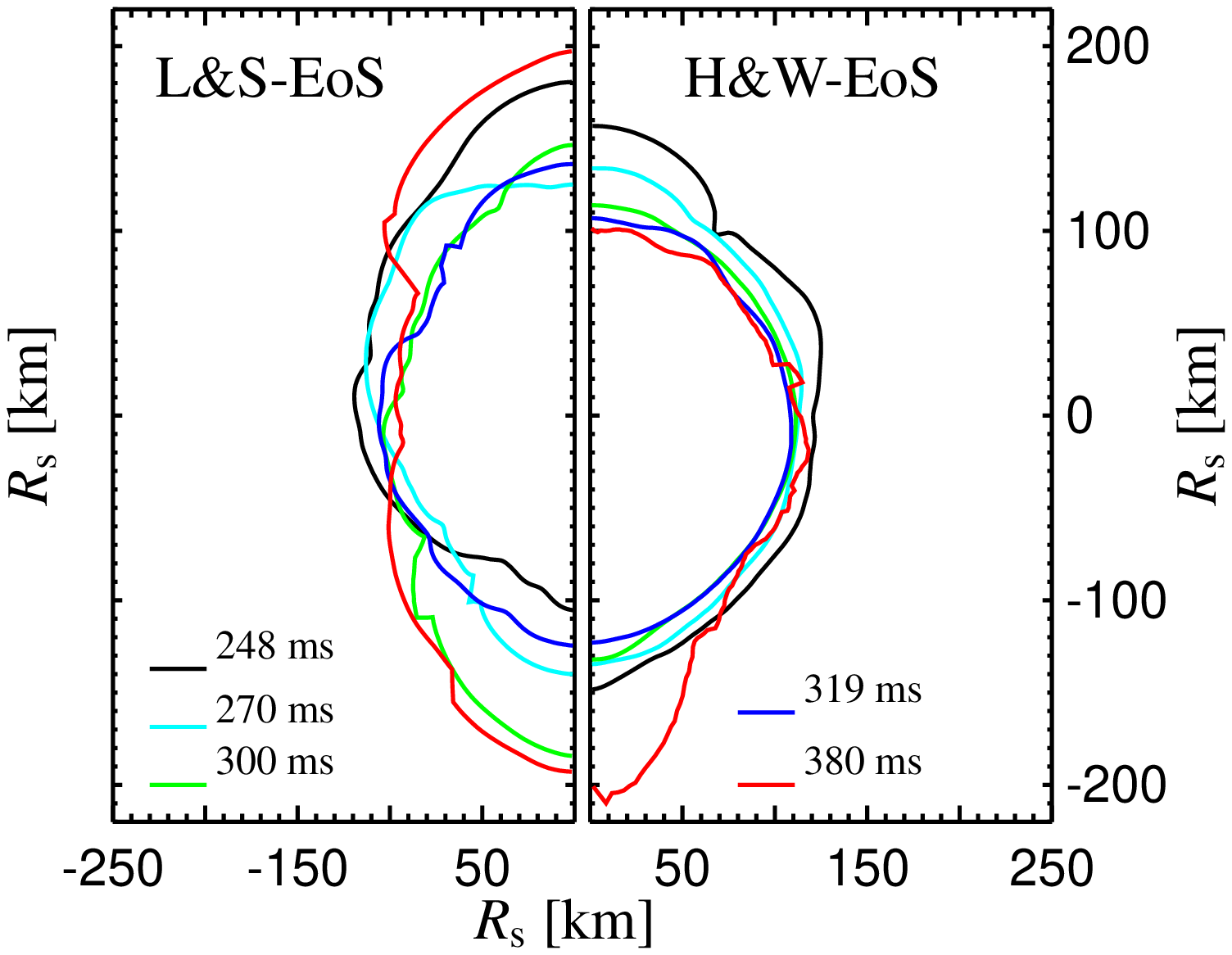}
\end{center}
\caption{{\em Left:} Maximum shock radii (solid lines) and proto-neutron
star radii (dashed lines) as functions of post-bounce time for the
2D simulations with different nuclear equations of state. The neutron
star radii are determined as the locations where the rest-mass density is
equal to $10^{11}\,$g$\,$cm$^{-3}$.
{\em Right:} Shock contours at the different post-bounce times
listed in the figure. The
vertical axis of the plot is the symmetry axis of the simulation.}
\label{fig:shockradius}
\end{figure*}

\section{Results}
\label{sec:results}

In the following we will compare our two simulations for these
equations of state in detail, focussing on the observable signals
whose properties depend on the asymmetries due to hydrodynamic 
instabilities in the supernova core after bounce.

\subsection{Hydrodynamic instabilities and shock motion}
\label{sec:shock}

Basically one can discriminate three regions and episodes
of nonradial hydrodynamic instability in the stellar core after
bounce:  (i) prompt post-shock convection (Epstein 1979,
Burrows \& Fryxell 1993,
Janka \& M\"uller 1996, M\"uller \& Janka 1997, 
Swesty \& Myra 2005), (ii) convection 
inside the nascent neutron star, i.e.\ below the neutrinosphere
of the electron neutrinos (Burrows 1987, Keil et al.\ 1996,
Buras et al.\ 2006b, Dessart et al.\ 2006), and 
(iii) convective overturn in the neutrino-heating layer between the 
gain radius and the stalled supernova shock 
(Herant et al.\ 1994; Burrows et al.\ 1995;
Janka \& M\"uller 1996; Fryer \& Warren 2002, 2004) as well
as SASI activity (Blondin et al.\ 2003, Blondin \& Mezzacappa 2006,
Ohnishi et al.\ 2006, Scheck et al.\ 2008). These regions can be
identified for both 2D simulations in 
Fig.~\ref{fig:convectionregions}.

Region (i) arises when the newly formed supernova shock
propagates outward through the infalling stellar iron core
and experiences massive energy losses by nuclear photo-disintegrations.
This weakens the shock so that it leaves behind a layer with
negative entropy gradient (before it encounters infalling
material with even lower densities and therefore produces
rising gas entropies again in spite of its further deceleration).
In addition,
a $Y_e$ minimum is established near an enclosed mass of roughly 
1$\,M_\odot$ when a large number of newly produced electron neutrinos
begins to stream away from the neutrinosphere in a luminous 
shock-breakout burst. The electron fraction forms a trough
around the radial position
where the shock makes the transition from the neutrino-opaque to
the neutrino-transparent regime (Fig.~\ref{fig:promptconvection}). 
The electron fraction decreases towards this minimum, because
electron-capture neutrinos escape more slowly from regions with
higher densities, and because the conversion of electrons to 
neutrinos proceeds more
slowly at lower densities. In both simulations the negative 
entropy gradient overlaps partly with the negative $Y_e$ gradient.

The layer where both the entropy and the $Y_e$ gradients are
negative is found to be Ledoux unstable (marked by the grey
vertical bars in Fig.~\ref{fig:promptconvection}). This region
lies between the first local maximum and the following local
minimum of the entropy profile. It
contains considerably more mass in the model computed with the
H\&W EoS, where it is bounded by the mass coordinates of 
0.63$\,M_\odot$ and 0.79$\,M_\odot$. At higher enclosed masses,
the entropy rises slowly out of its local minimum. In contrast,
in the L\&S case the boundaries are at
0.60$\,M_\odot$ and 0.71$\,M_\odot$, and the increase of the
entropy in the overlying shells is much steeper. In the H\&W model
the thicker layer in mass corresponds also to a much wider radial
domain at a larger distance from the stellar center.
Because of higher growth rates the convective activity sets in 
more immediately after the shock passage and lateron becomes 
stronger and encompasses a greater fraction of the stellar core.
This can be seen at $t \la 30\,$ms after bounce in 
Fig.~\ref{fig:convectionregions}, where the righthand panels for the 
H\&W simulation in comparison to the lefthand panels show significantly 
more violent convective overturn. While the L\&S model develops 
visible mass motions between $\sim$15$\,$km and $\sim$40$\,$km,
the H\&W model shows stronger activity at radii of $15\,\mathrm{km}\la
r\la 80\,$km. 

After some 10$\,$ms the profiles of entropy and 
electron fraction are flattened and with the disappearance of
the driving force the convective activity begins to calm down (this
can be recognized better in the two lower panels of 
Fig.~\ref{fig:convectionregions}). Now, however, region (ii) is 
being formed. Neutrino transport begins to reduce the electron 
fraction deeper inside the nascent neutron star. As a consequence
of this the lower boundary of the $Y_e$ trough and thus the
negative $Y_e$ gradient moves gradually inward. At the 
same time the energy losses by neutrinos reduce the level of
the entropy plateau at $s\sim 5\,k_\mathrm{B}$/nucleon such that
the plateau becomes wider and its inner edge also moves inward. 
These changes in the entropy and lepton
number profiles in combination lead to
the establishment (or/and maintenance) of a convectively 
unstable layer inside the proto-neutron star. 
About 50--60$\,$ms after bounce, following
a short, intermediate period of relative quiescence, both of our 
2D models indeed show the reappearance of vivid overturn activity
at densities above $10^{12}\,$g$\,$cm$^{-3}$ 
(Fig.~\ref{fig:convectionregions}). The convective
transport of lepton number and entropy supports and enhances
the mentioned structural changes so that a long-lasting zone with
convective mixing develops (compare our Fig.~\ref{fig:convectionregions}
with Figs.~9 and 10 of Dessart et al.\ 2006). 
This zone encompasses a growing mass region within the broadening 
$Y_e$ trough.

Region~(iii) with nonradial hydrodynamic mass motions develops
outside of the steep density gradient at the surface of the 
nascent neutron star, where the neutrinospheres of all neutrinos are
located. First signs of low-$\ell$ mode, low-amplitude SASI sloshing 
can be seen here already some 10$\,$ms after bounce. At about 
100$\,$ms after bounce this activity strengthens significantly 
because at this time neutrino energy deposition in the gain layer 
has begun to build up a negative entropy gradient between the gain 
radius and the
stagnant supernova shock. The neutrino-heating region is unstable
to convection, and the presence of convective motions produces
vorticity and entropy perturbations that feed back into 
the advective-acoustic cycle that is considered as an explanation of
the SASI phenomenon (see Foglizzo 2001, 2002; Foglizzo et al.\ 2007).
Thus the onset of convection may lead to an amplification of the
SASI growth. On the other hand, the violent shock oscillations that
are characteristic of the fully developed SASI produce large entropy
variations in the downstream region. These act as seeds for secondary
convection (Scheck et al.\ 2008) and have been found to aid the 
neutrino-heating mechanism for powering supernova explosions
(Scheck et al.\ 2008, Marek \& Janka 2007; 
see also Murphy \& Burrows 2008). 
In the nonlinear regime convective and SASI activity are
inseparably coupled, and a high-$\ell$ convective mode pattern
occurs superimposed on the low-$\ell$ mode SASI deformations of
the shock contour and of the post-shock region (for 
snapshots, see Fig.~\ref{fig:snapshots} and Marek \& Janka 2007, 
Scheck et al.\ 2008).

Although both simulations show the same basic features
and evolutionary stages, the differences in details are interesting.
As mentioned above, the H\&W model develops more vigorous prompt 
post-shock convection in a region with higher mass and wider 
radial extension. In contrast, the SASI activity in this model is
appreciably less strong than that in the calculation with the 
L\&S EoS until roughly 350$\,$ms after bounce. This can be seen in
Fig.~\ref{fig:shockradius}, where the maximum shock radius of 
the L\&S case exhibits significantly bigger SASI amplitudes and
correspondingly the shock contours show more extreme nonspherical 
deformation (righthand panel of Fig.~\ref{fig:shockradius}). These
differences grow during the nonlinear phase of the SASI between 
$\sim$150 and 350$\,$ms until at $t \ga 350\,$ms p.b.\ the 
SASI sloshing in the H\&W model also gains more power and the 
maximum shock radii of both simulations become more similar again.

We suspect that the more compact proto-neutron star for the softer
L\&S EoS leads to conditions that favor strong SASI activity, possibly
because of the more efficient neutrino heating (and more vigorous
post-shock convection) as a consequence of the higher neutrino
luminosities and harder neutrino spectra that are radiated from
a more compact and hotter nascent neutron star 
(see Figs.~\ref{fig:luminosities} and \ref{fig:meanenergies}
and Sect.~\ref{sec:neutrinos}). Another possible explanation in
the context of the advective-acoustic cycle scenario may be
different amplification factors of perturbations in the two
models. Due to the lack of good theoretical insight into the 
behavior of the SASI in the fully nonlinear regime, we do not 
see a way how to facilitate deeper understanding by further analysis.
For both suggested explanations, however, one might expect that
in the H\&W simulation at late post-bounce times, when the
proto-neutron star radius has contracted (Fig.~\ref{fig:shockradius})
and the neutrino-heating timescale has decreased (see Fig.~6
in Marek \& Janka 2007), the conditions for violent SASI activity
have the tendency to improve. This would be consistent with our
observed growth of the SASI amplitudes near the end of this 
simulation.

\begin{figure*}[!htp]
\begin{center}
\includegraphics[width=8.5cm]{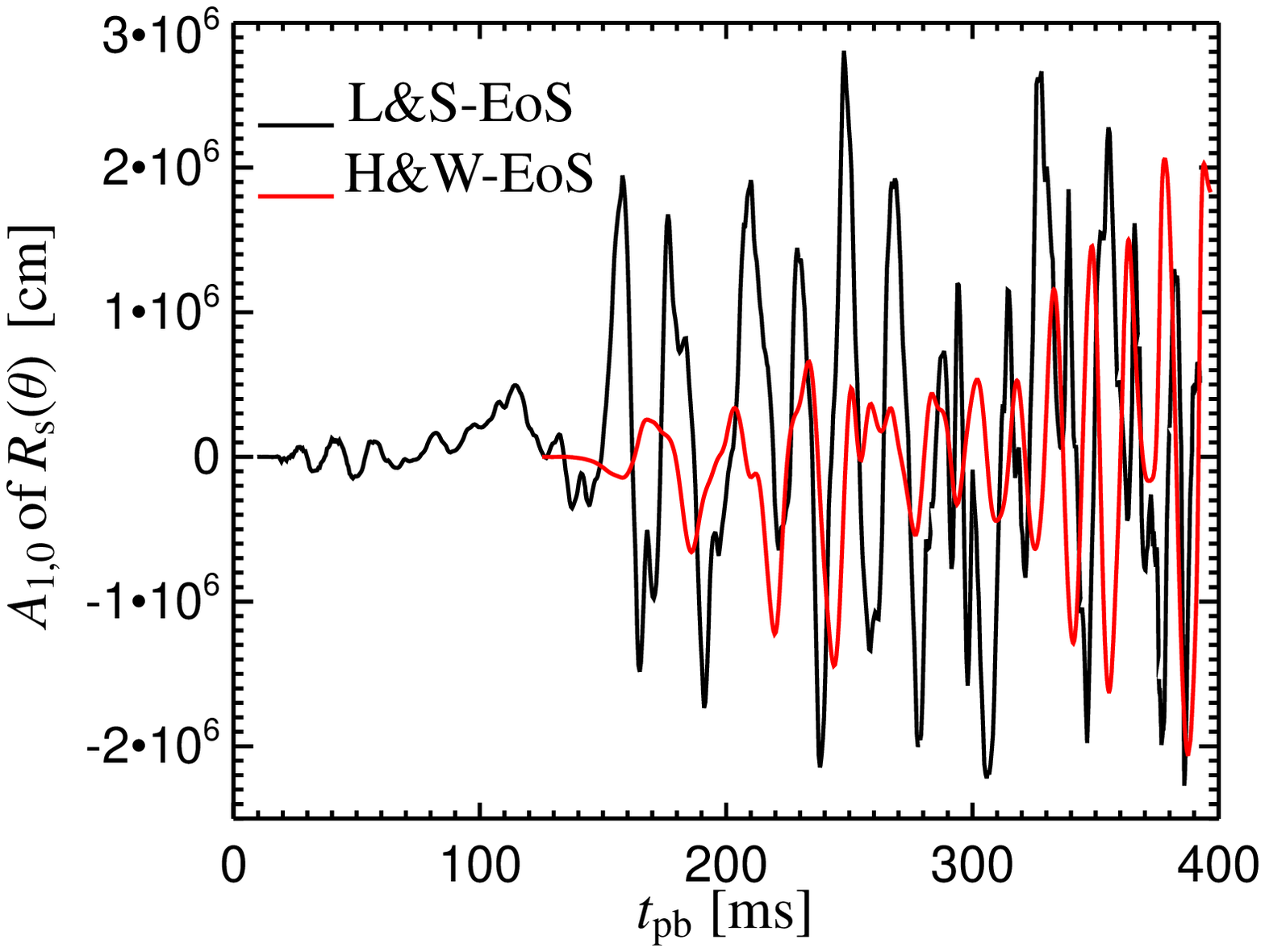}
\includegraphics[width=8.5cm]{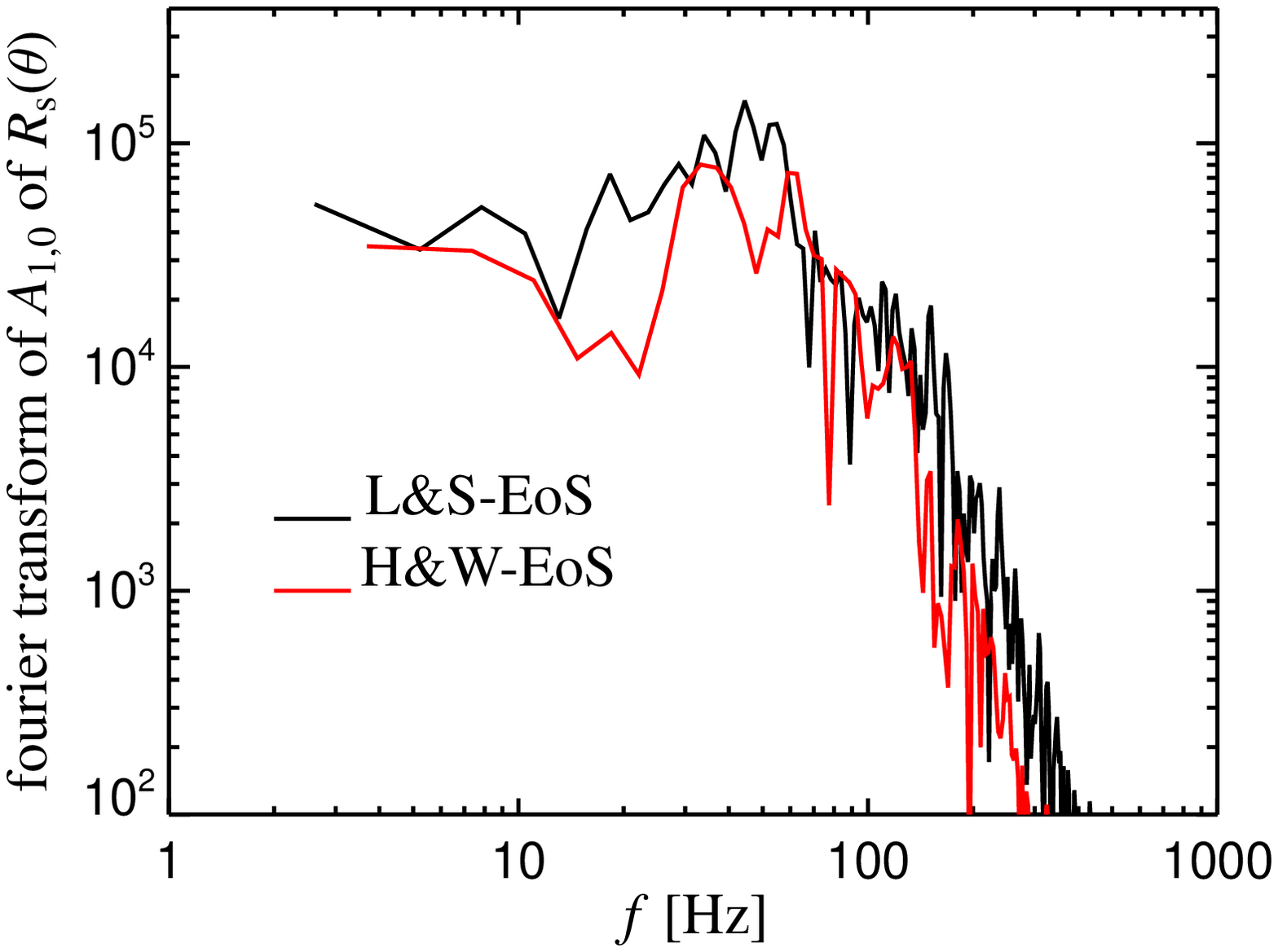}\\ \vspace{0.3cm}
\includegraphics[width=8.5cm]{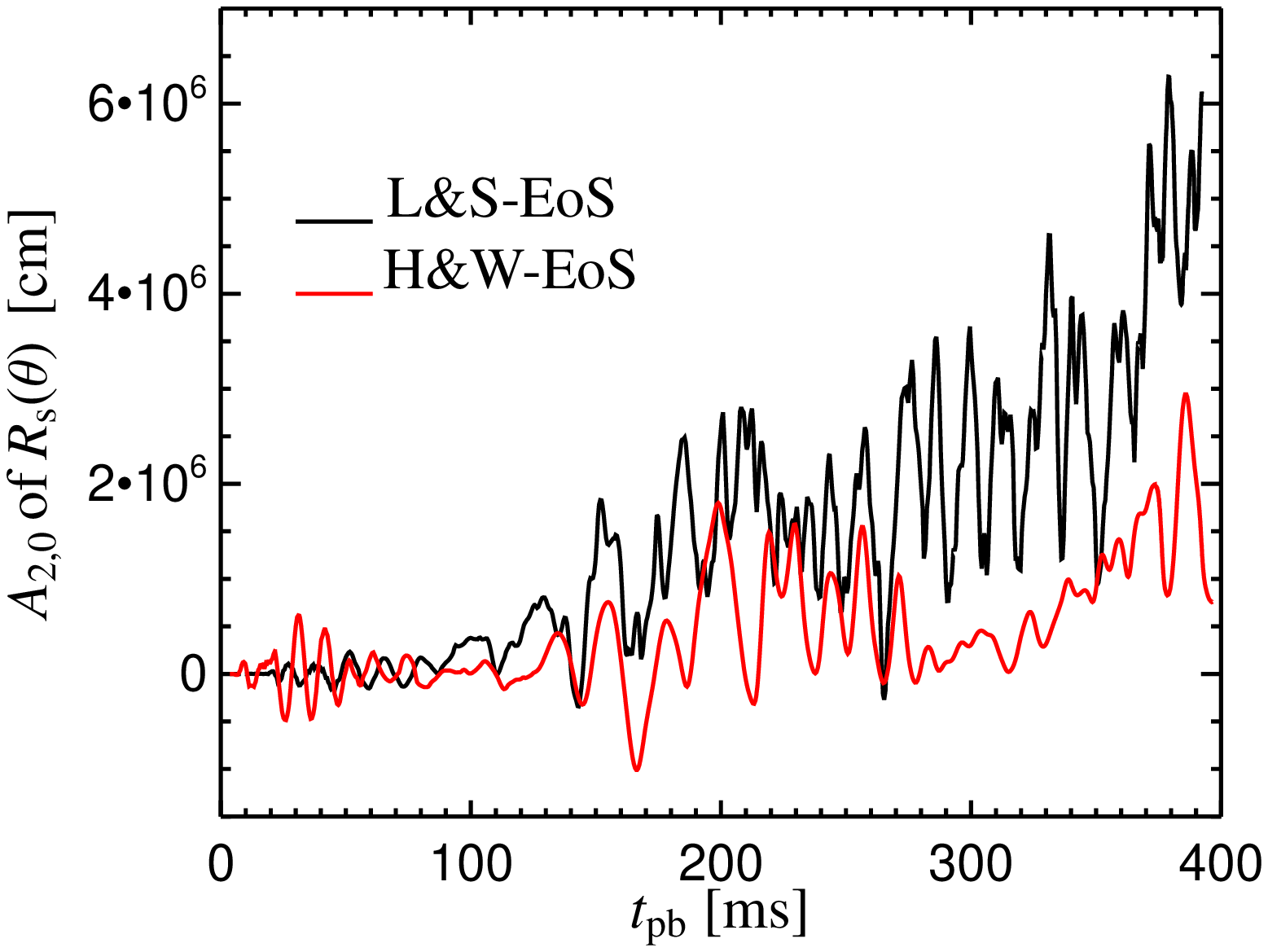}
\includegraphics[width=8.5cm]{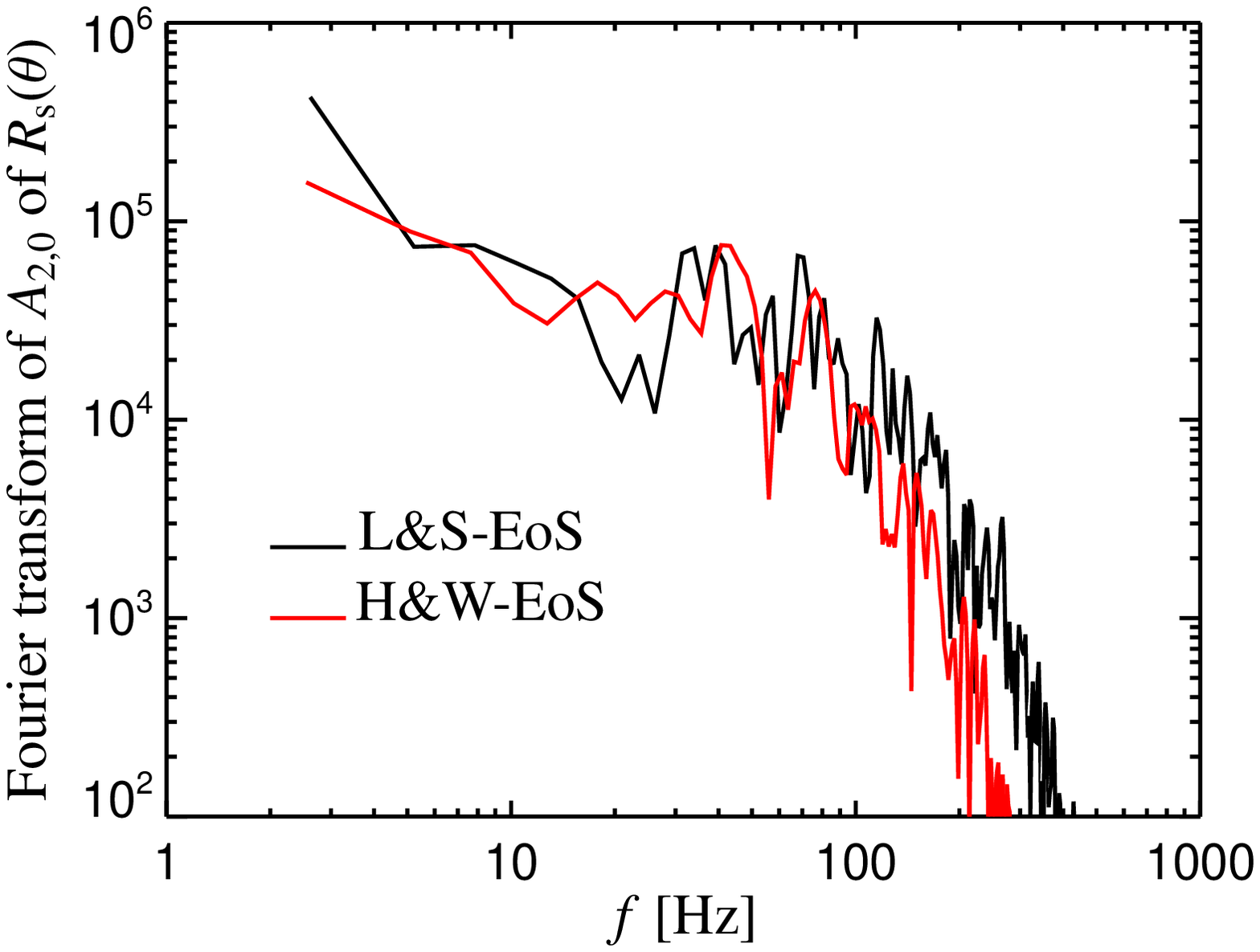}\\ \vspace{0.3cm}
\includegraphics[width=8.5cm]{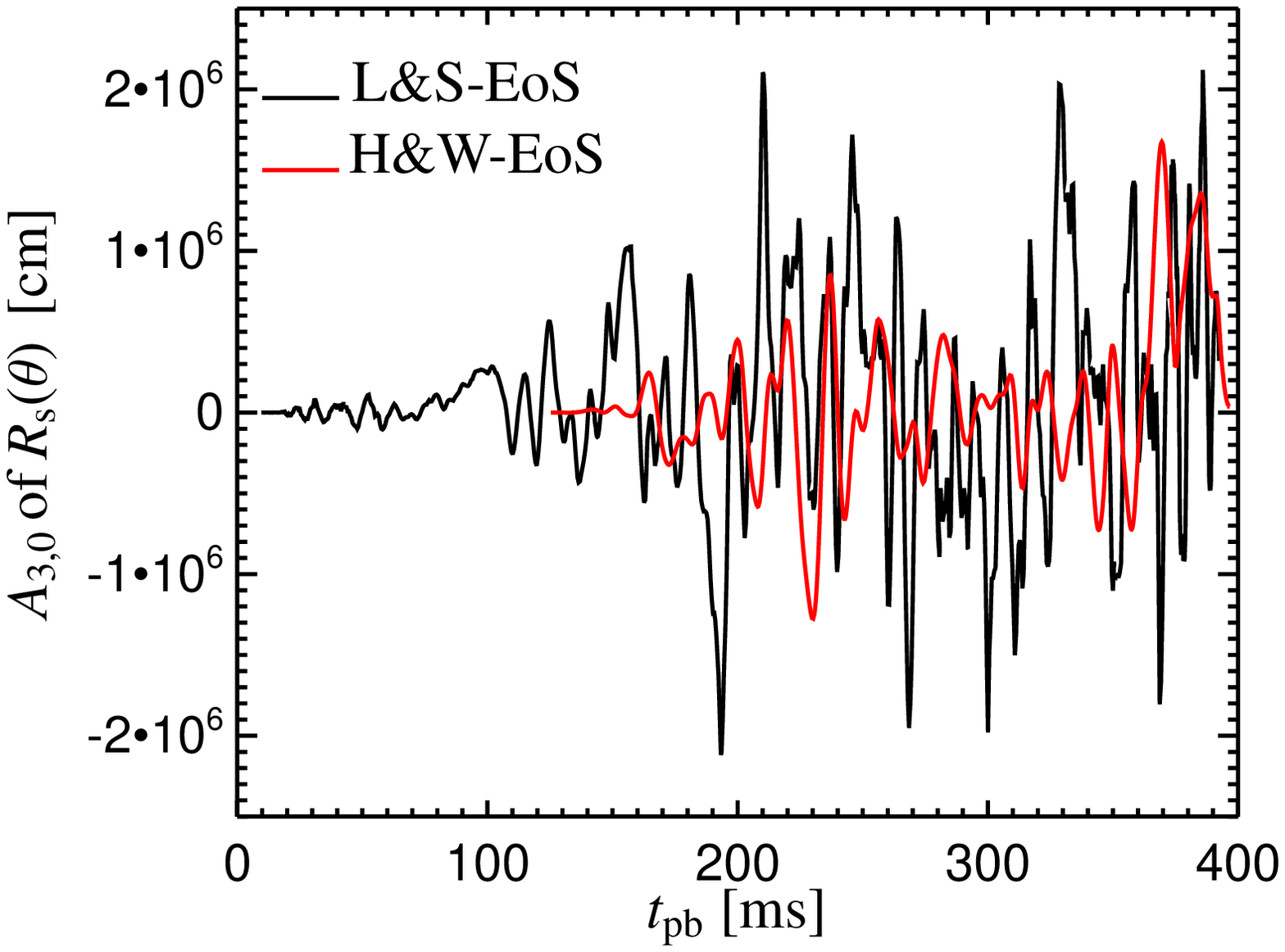}
\includegraphics[width=8.5cm]{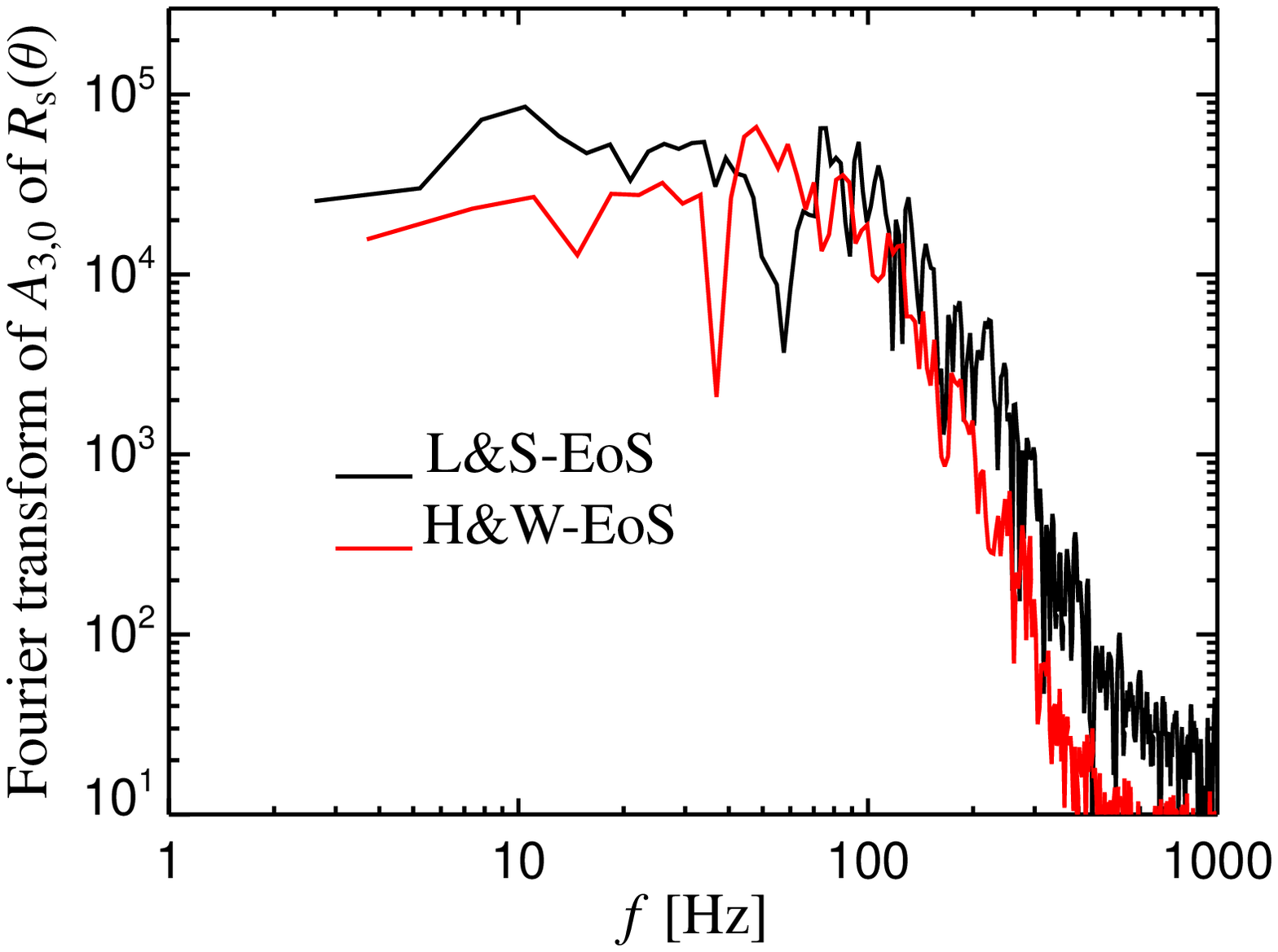}
\end{center}
\caption{{\em Left:} Amplitudes for the decomposition of the
angle-dependent position of the 
shock front, $R_\mathrm{s}(\theta,t)$, into spherical harmonic  
functions for the 2D simulation with the L\&S EoS (black lines)
and with the H\&W EoS (red lines). 
Shown are the contributions of the modes for 
$(\ell,m) = (1,0)$ (top), $(2,0)$ (middle), and $(3,0)$ (bottom). 
The underlying trend in growing amplitudes for the $l=2$ mode
towards the end of our simulations signals an increasing and
persistent prolate
deformation of the shock. Note that during the first 125.3$\,$ms
after core bounce, when the shock deformation is still
relatively small, the simulation with the H\&W EoS was performed
only with a 90$^\circ$ wedge from the pole to the equator, and 
we can therefore plot only the amplitudes with even $\ell$ for 
this equatorially symmetric phase.
{\em Right:} The Fourier transforms of the time-dependent
mode amplitudes.} 
\label{fig:shockmodes}
\end{figure*}

In Fig.~\ref{fig:shockmodes}, lefthand panels, 
we plot the time-evolution of the
amplitudes $A_{1,0}$, $A_{2,0}$, and $A_{3,0}$, of the decomposition
of the angle-dependent shock radius $R_\mathrm{s}(\theta,t)$ into 
spherical harmonics:
\begin{equation}
R_\mathrm{s}(\theta,t)\,\equiv \, \sum_{\ell = 0}^{\infty}
A_{\ell,0}(t) P_{\ell}^0(\cos\theta) \ ,
\label{eq:sphericalharmonics}
\end{equation}
where $m = 0$ because of the axial symmetry of our models and 
$P_{\ell}^0(\cos\theta)$ are the Legendre polynomials\footnote{Note 
that we have renormalized our amplitudes $A_{\ell,0}$ compared to 
the coefficients $a_{\ell}$ used by Kotake et al.\ (2007) in 
order to directly display in Fig.~\ref{fig:shockmodes} the 
radius variations associated with the different spherical 
harmonics components. The coefficient $A_{0,0}$ is identical
with the angular average of the shock radius: $A_{0,0}(t) = 
\left\langle R_{\mathrm{s}}(t)\right\rangle \equiv (4\pi)^{-1}
\int{\mathrm{d}}\Omega\,R_\mathrm{s}(\theta,t)$, whose values
and evolution are similar to those of the maximum shock radius
plotted in Fig.~\ref{fig:shockradius}.}. 
Figure~\ref{fig:shockmodes} confirms
our description above: the amplitudes for the L\&S run are
typically 2--3 times larger than those of the H\&W simulation.
Only near the end of the computed evolution (at $t \ga 350\,$ms
p.b.), $A_{1,0}$ and $A_{3,0}$ of both models reach very similar
values. This is not so for $A_{2,0}$, which exhibits oscillatory
behavior on top of a clear slope to positive values in both 
simulations, signaling a trend in growing prolate deformation of
the shock. Until the end of the computational runs, however, this 
deformation remains considerably more pronounced in the case of 
the model with the L\&S EoS. 

The righthand panels of Fig.~\ref{fig:shockmodes} provide the 
Fourier transforms of the time-dependent mode amplitudes. One
can see a broad peak of the Fourier spectra between about 10 and
roughly 100$\,$Hz, followed by a steep decline towards higher
frequencies. The $\ell = 1$ amplitude exhibits the clearest,
jagged peak structure at a frequency of 20--60$\,$Hz for 
the L\&S run, and a smoother peak at 30--40$\,$Hz with a 
secondary one around
65$\,$Hz for the H\&W model. $A_{2,0}$ shows most power
at frequencies around 30--50$\,$Hz and 70--90$\,$Hz in both 
simulations, and $A_{3,0}$ possesses a broad Fourier maximum
below $\sim$100--130$\,$Hz, with an indication of a peak at about
50$\,$Hz in the H\&W case. There is no really clear correlation
of the SASI oscillation period with the compactness of the 
nascent neutron star or with the average radius of the stalled 
shock except for, maybe, the slightly higher frequency of the
main $\ell = 1$ peak in the L\&S run. 
Because of the time-dependent structure of the 
accretor and of the whole post-shock region the appearance
of well localized and very prominent peaks in the spectrum
might not be expected. In addition, shock wobbling
associated with convective mass motions
produces a background of short-wavelength ``noise'' at frequences 
up to roughly 200$\,$Hz (Fig.~\ref{fig:snapshots}). 
Accordingly, in particular the 
amplitudes for higher $\ell$ values show power nearly evenly
distributed over a wide range of frequencies. The Fourier 
transforms for $A_{1,0}$, $A_{2,0}$, and $A_{3,0}$ plotted
in Fig.~\ref{fig:shockmodes} nevertheless confirm the 
stronger SASI activity in the simulation with the L\&S EoS:
only for a few, narrow frequency windows the Fourier amplitudes
of the H\&W model become larger than those of the L\&S run.

In the following two sections we will turn to an analysis of 
the consequences of the described
nonradial hydrodynamic instabilities, in particular also
of the SASI, for observable signals from the supernova core. One
question of interest will be, whether the differences caused by 
the use of stiff or soft neutron star equations of state manifest
themselves in some distinctive features of the signals.

\begin{figure*}[!htp]
\begin{center}
\includegraphics[width=8.5cm]{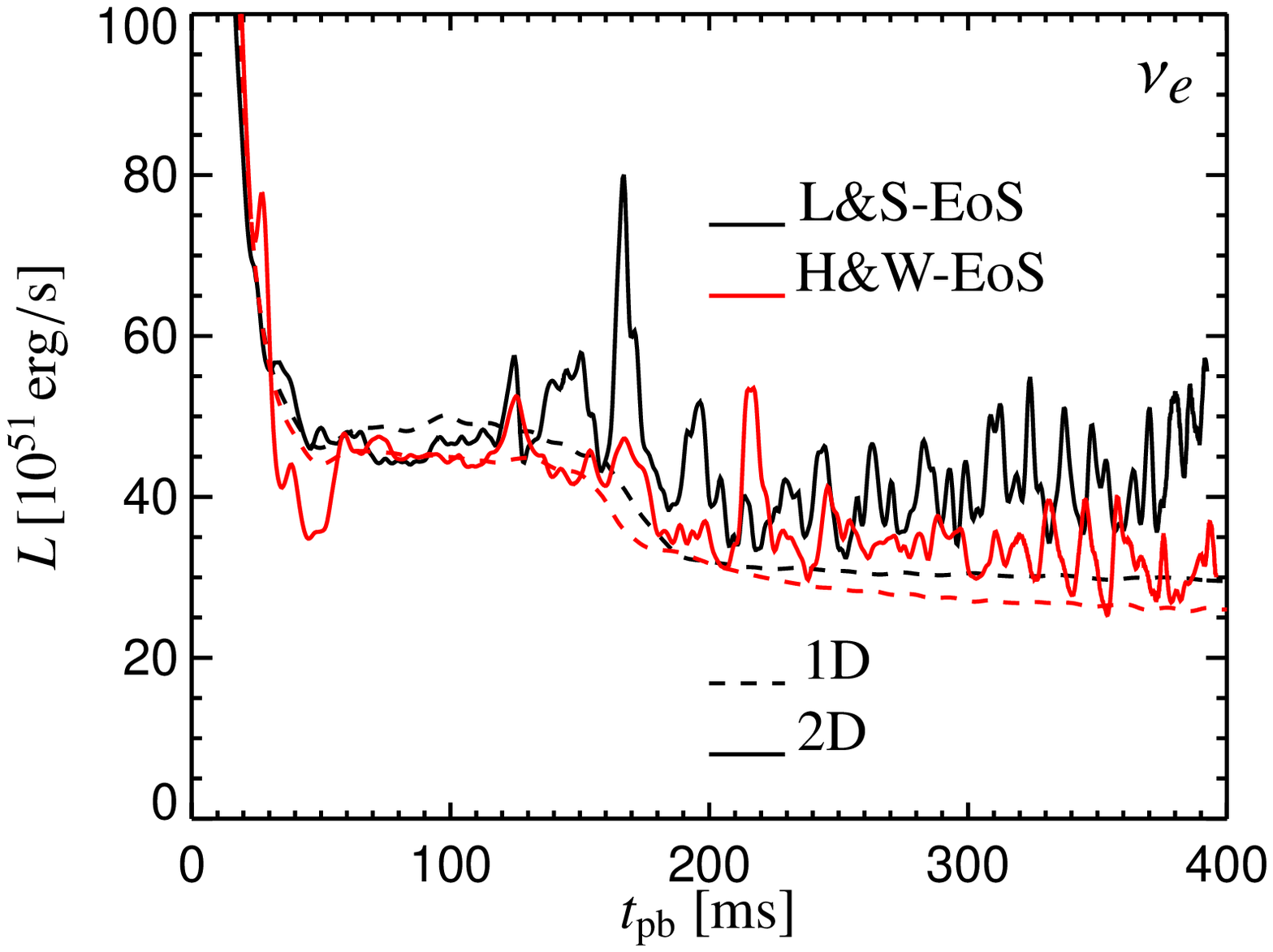}
\includegraphics[width=8.5cm]{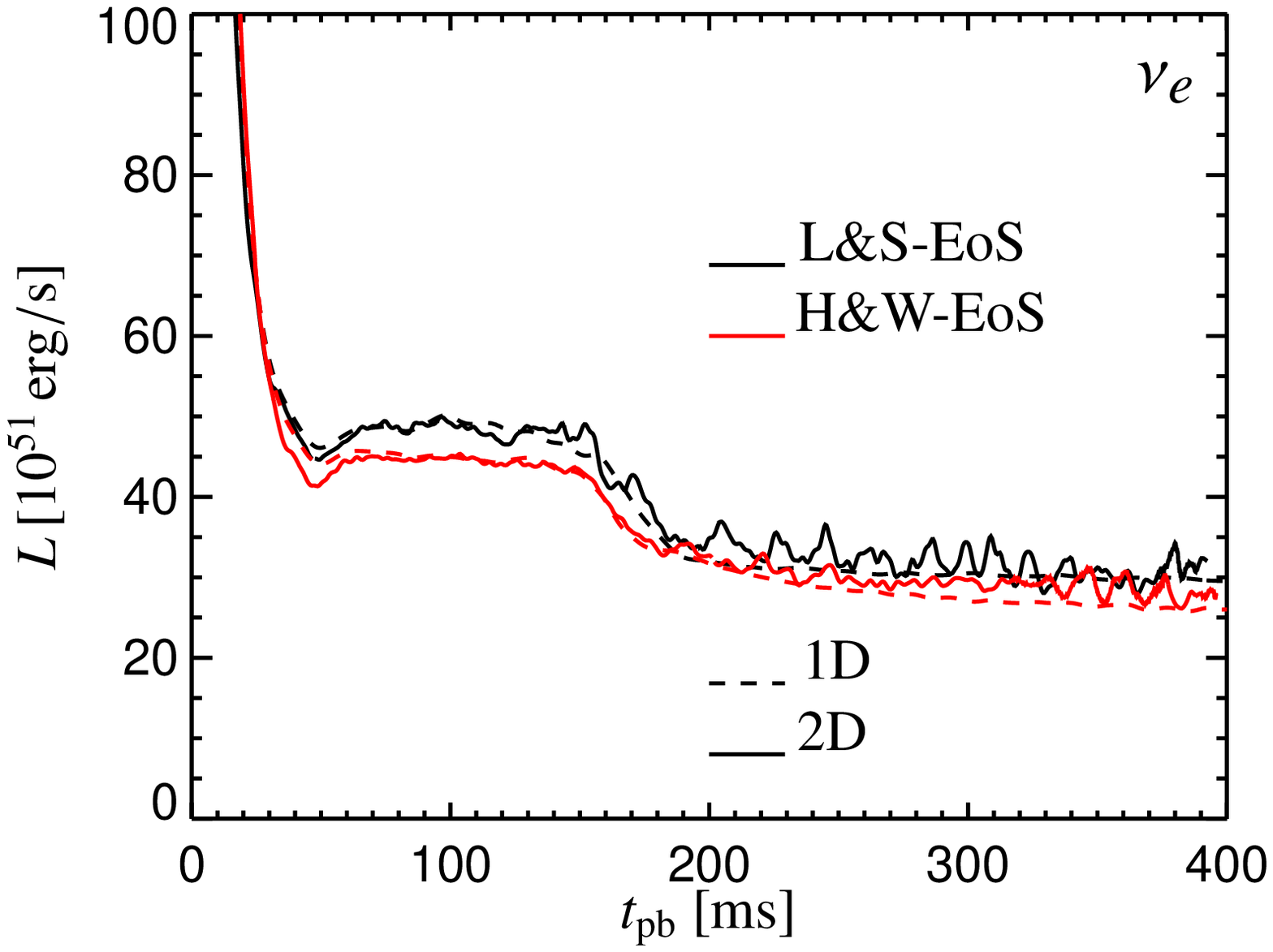}\\ \vspace{0.3cm}
\includegraphics[width=8.5cm]{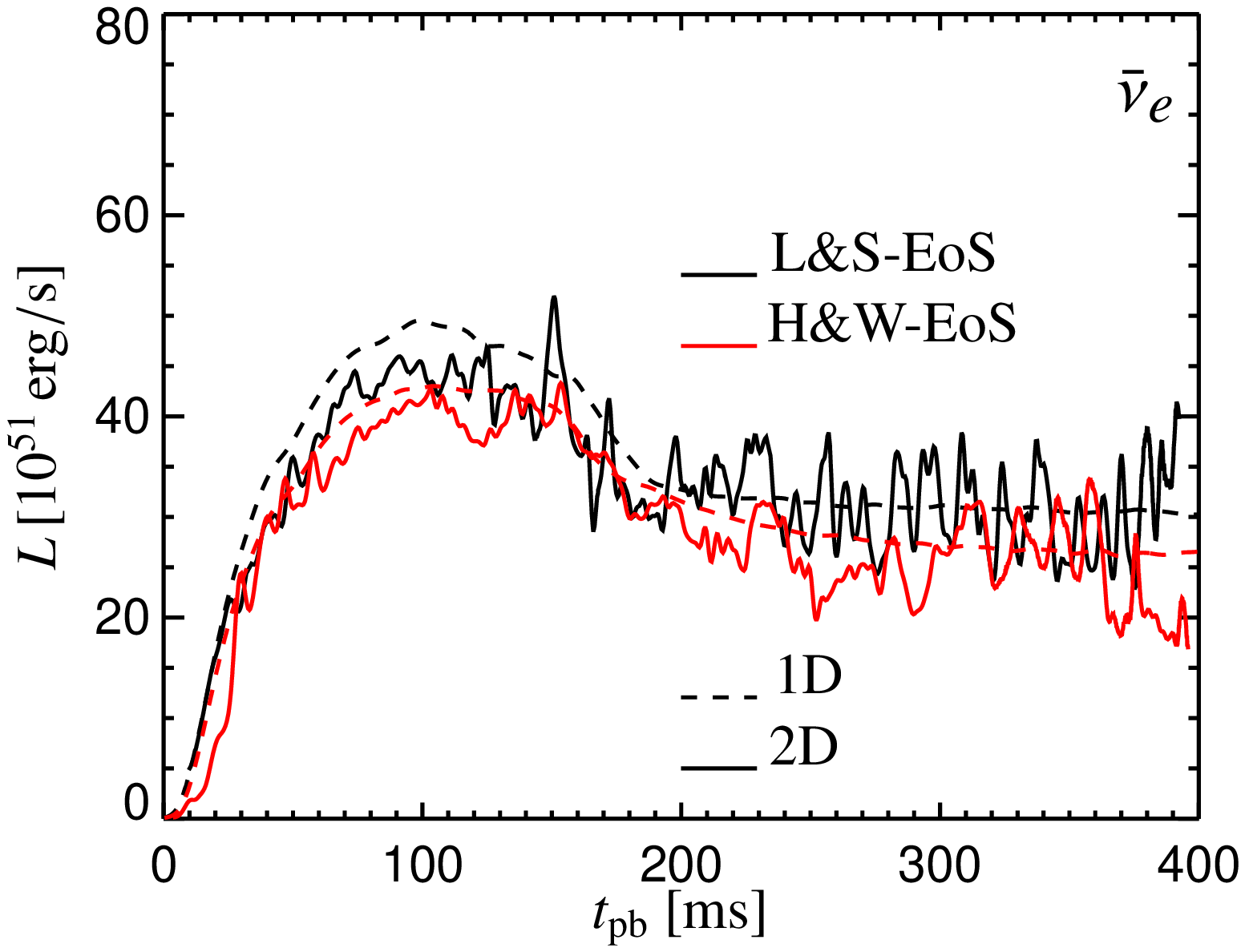}
\includegraphics[width=8.5cm]{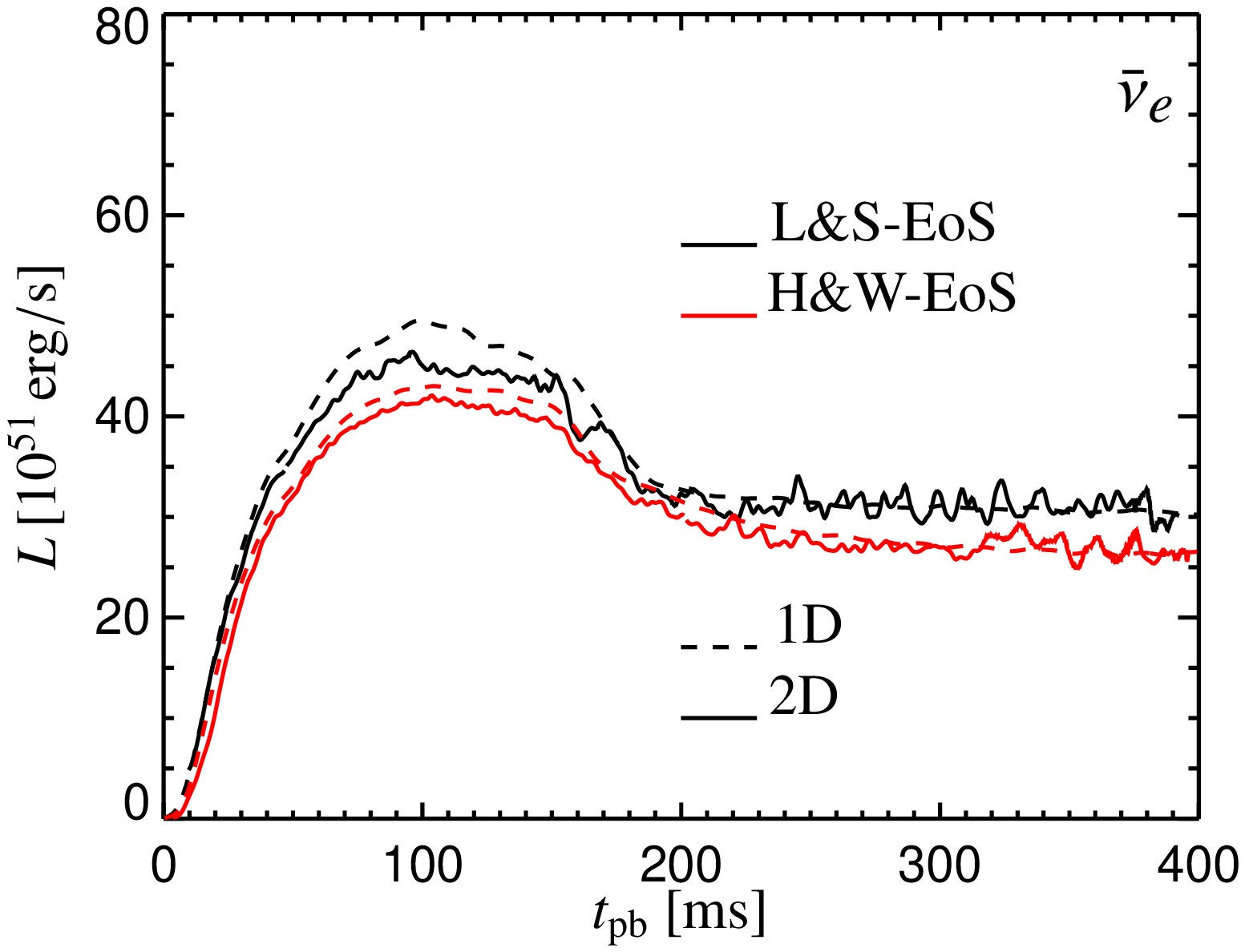}\\ \vspace{0.3cm}
\includegraphics[width=8.5cm]{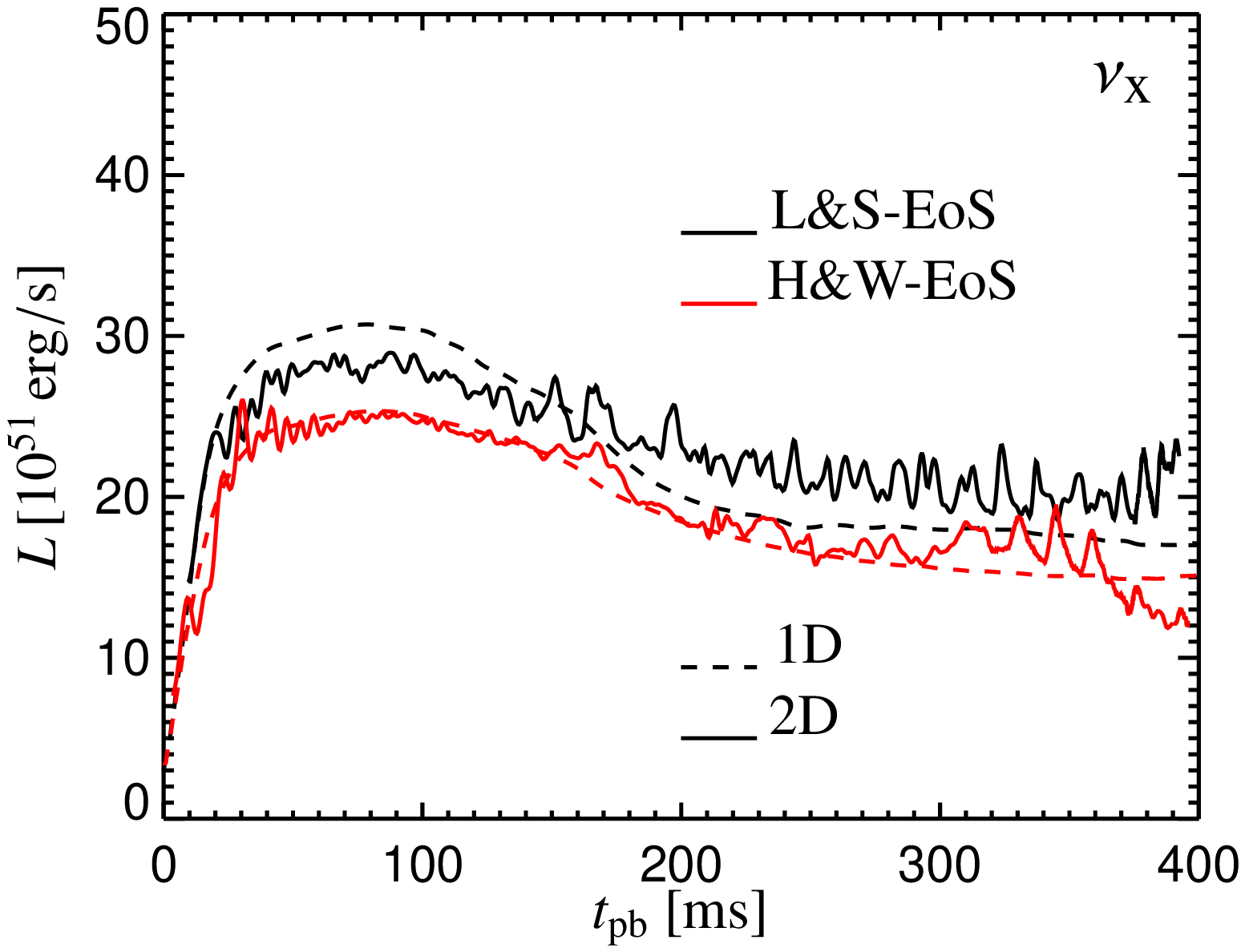}
\includegraphics[width=8.5cm]{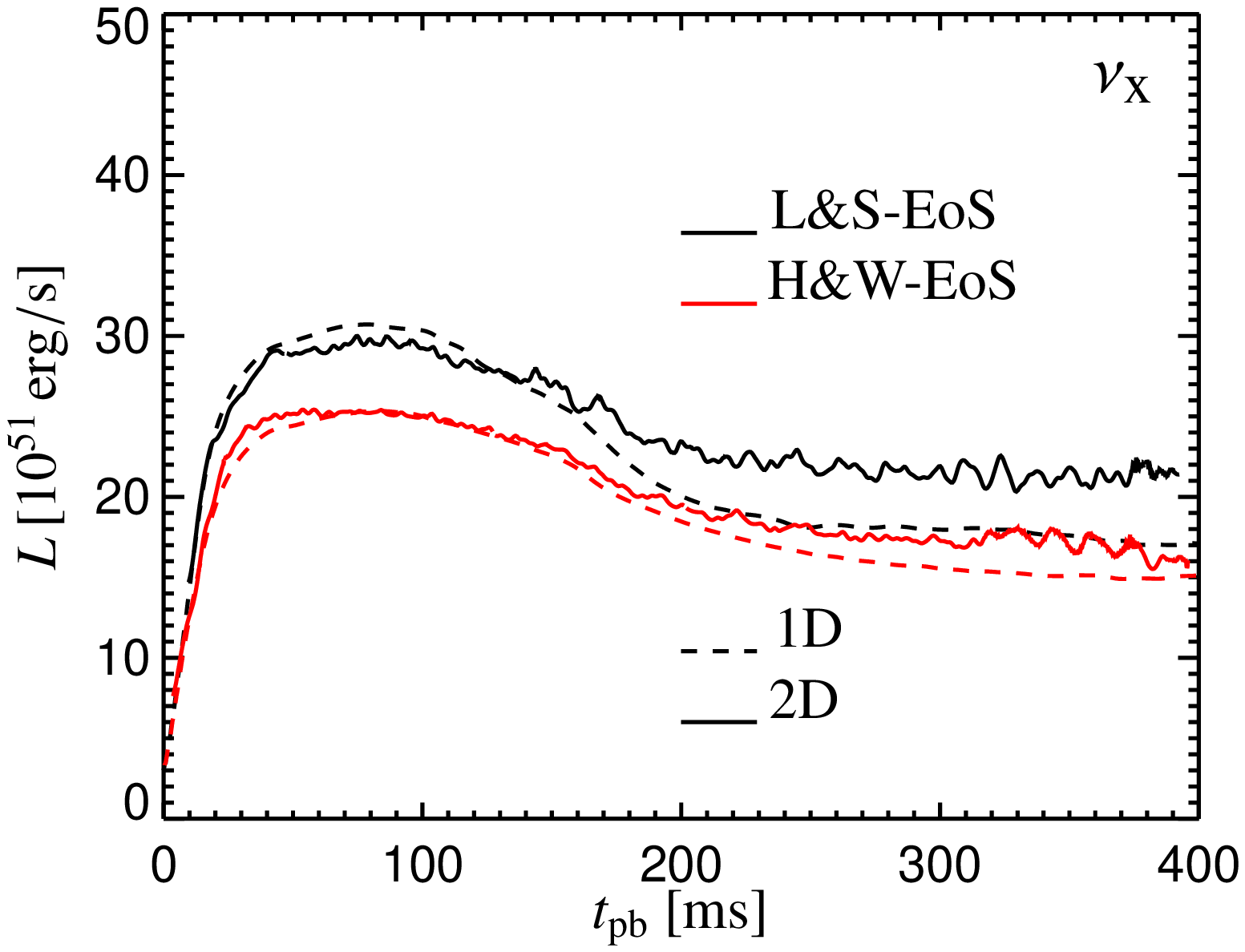}
\end{center}
\caption{Isotropic equivalent luminosities of electron neutrinos
  (top), electron antineutrinos (middle), and one kind of heavy-lepton
  neutrinos ($\nu_\mu$, $\bar\nu_\mu$, $\nu_\tau$, or $\bar\nu_\tau$; bottom)
  versus time after core bounce as measurable for a distant
  observer located along the polar axis of the 2D spherical 
  coordinate grid (solid
  lines). The dashed lines display the radiated luminosities of the 
  corresponding spherically symmetric (1D) simulations. 
  The evaluation was performed at a radius of 400$\,$km
  (from there the remaining gravitational redshifting to infinity
  is negligible) and the results are given for an observer at rest
  relative to the stellar center. While the left column shows the
  (isotropic equivalent) luminosities computed from the flux that is
  radiated away in an angular grid bin
  very close to the north pole, the right column
  displays the emitted (isotropic equivalent) luminosities when the
  neutrino fluxes are integrated over the whole northern hemisphere of
  the grid (see Eqs.~(\ref{eq:lum1}) and (\ref{eq:lum2b}), respectively).}
\label{fig:luminosities}
\end{figure*}

\begin{figure*}[!htb]
\begin{center}
\includegraphics[width=8.5cm]{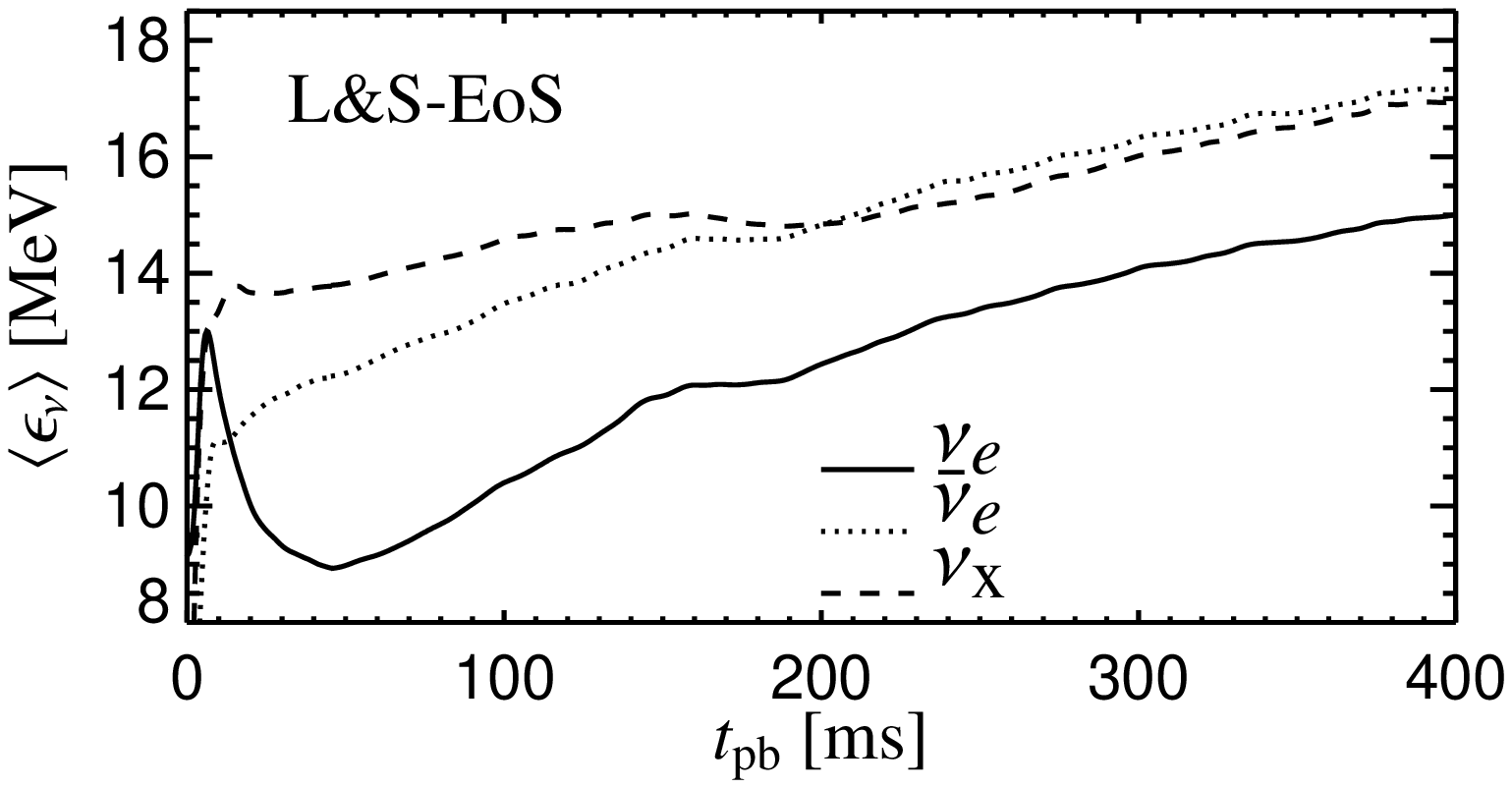}
\includegraphics[width=8.5cm]{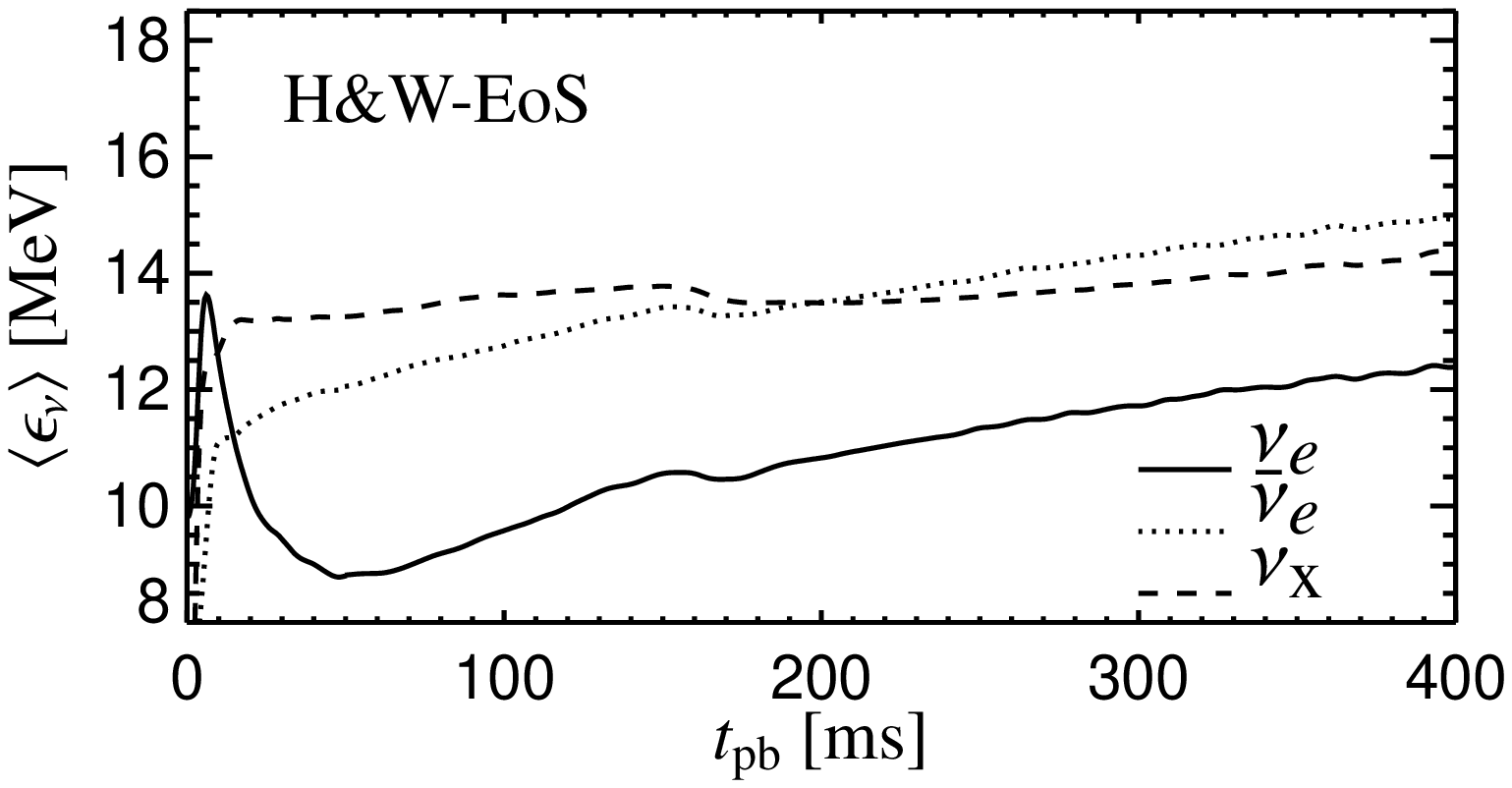}\\ \vspace{0.3cm}
\includegraphics[width=8.5cm]{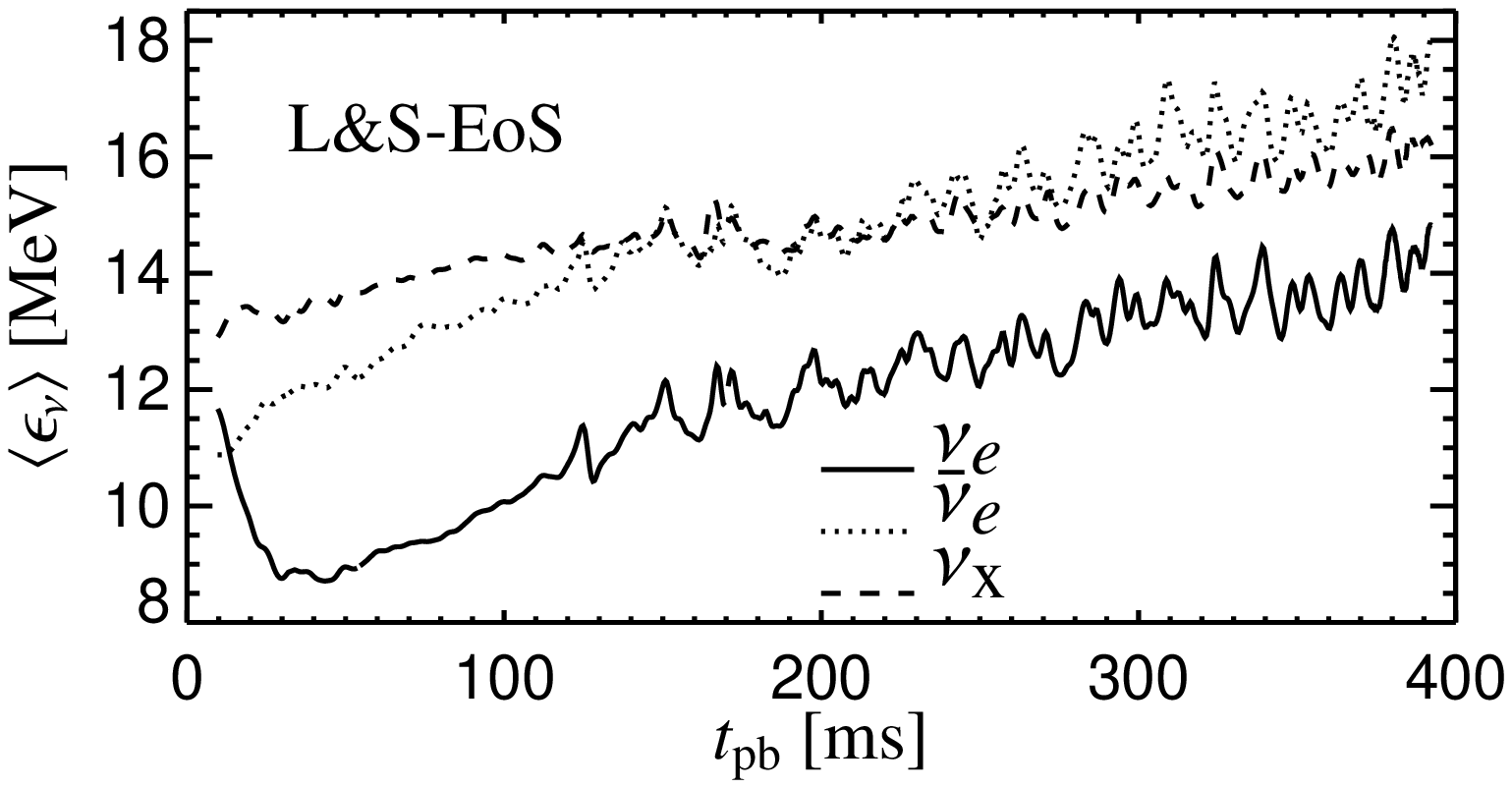}
\includegraphics[width=8.5cm]{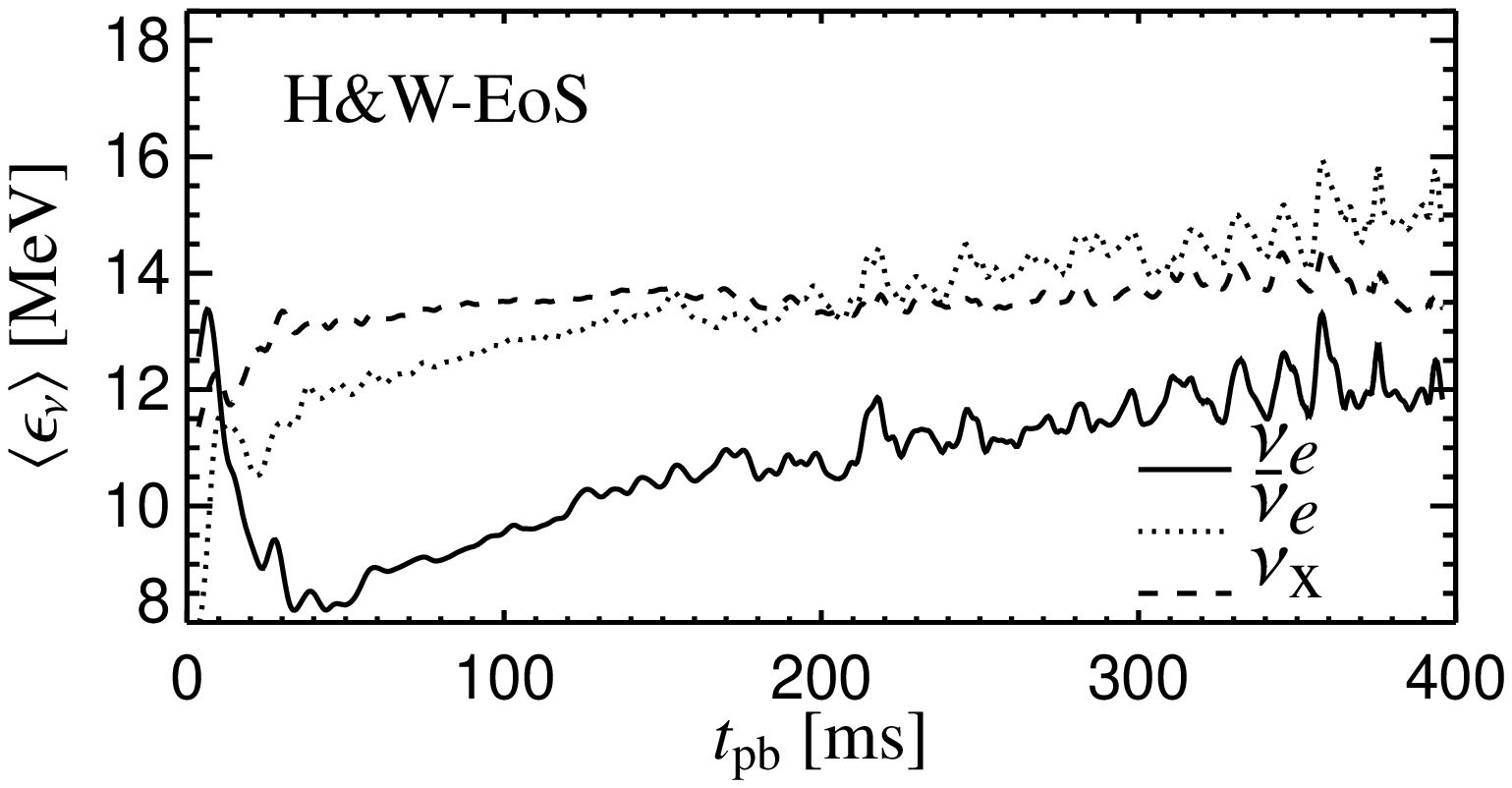}\\ \vspace{0.3cm}
\includegraphics[width=8.5cm]{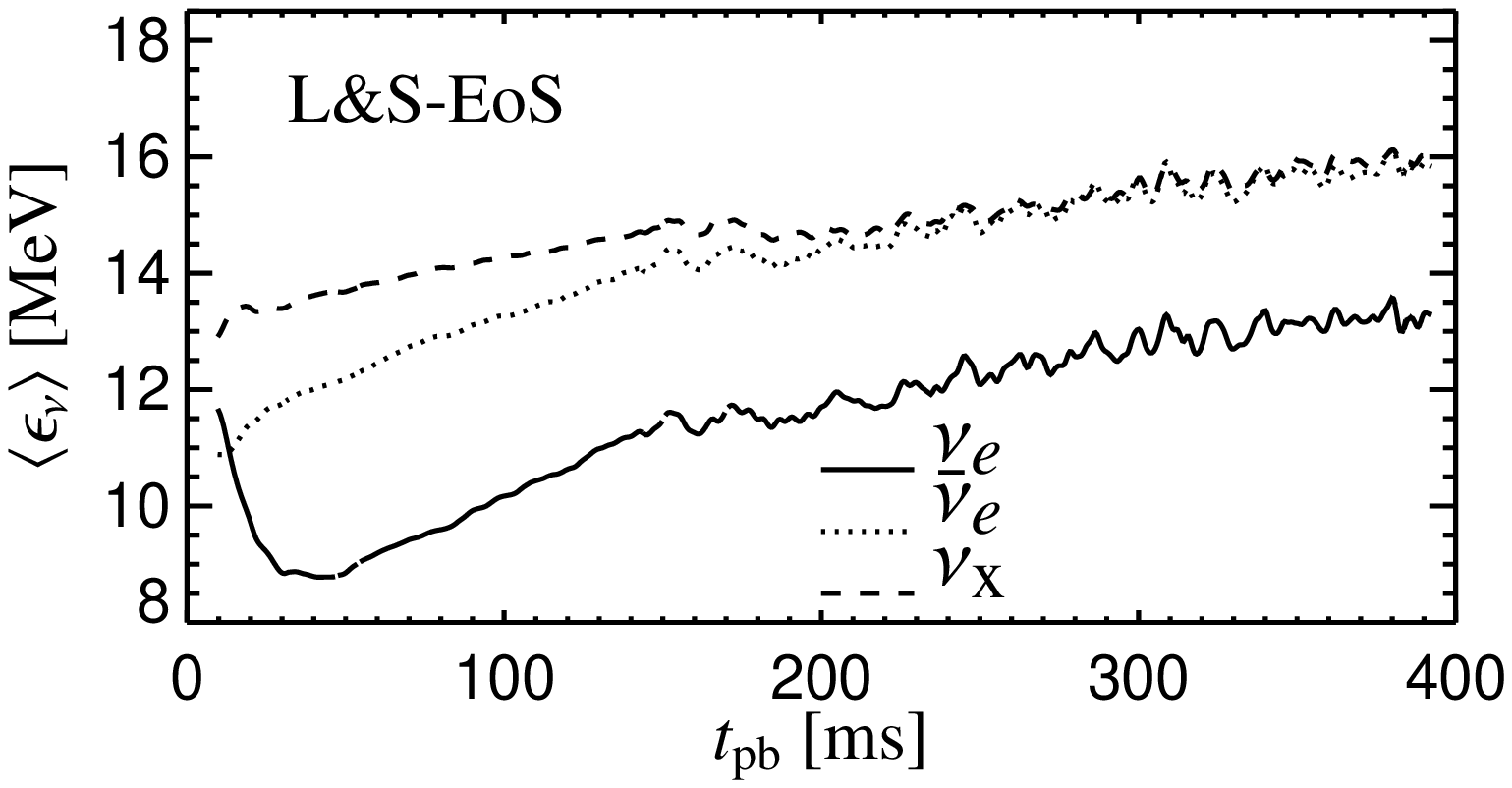}
\includegraphics[width=8.5cm]{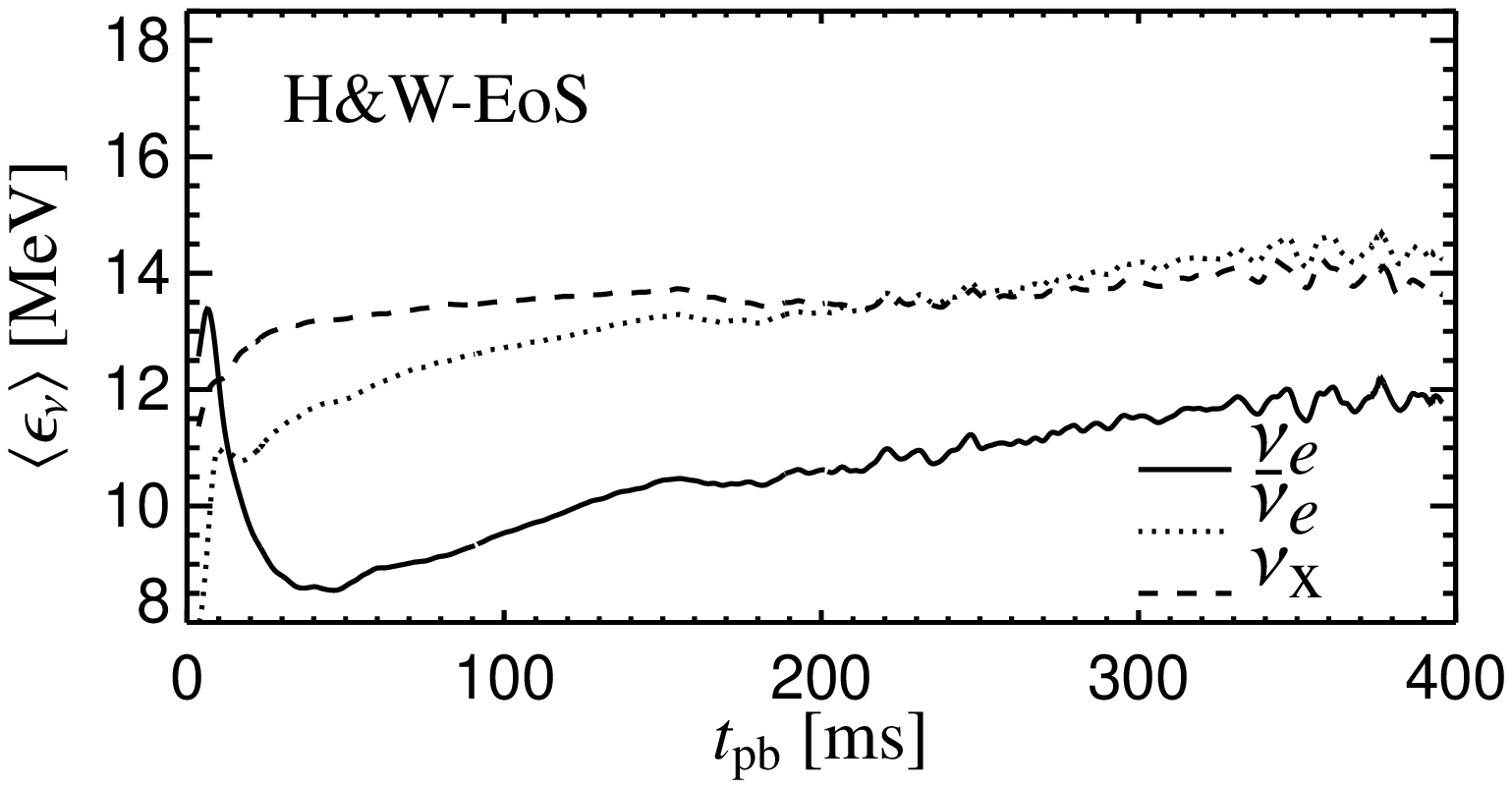}
\end{center}
\caption{Mean energies of radiated neutrinos as functions of post-bounce 
time for our 1D simulations ({\em top}) and 2D models ({\em middle} and
{\em bottom}) with both 
equations of state (the lefthand panels are for the L\&S EoS, the
right ones for the H\&W EoS). The displayed data are defined as ratios
of the energy flux to the number flux and correspond to the luminosities
plotted with dashed and solid lines in Fig.~\ref{fig:luminosities}.
The panels in the middle show results
for a lateral grid zone near the north polar axis, the bottom panels 
provide results that are averaged over the whole northern hemisphere
of the computational grid.
In all cases the evaluation has been performed in the laboratory frame
at a distance of 400$\,$km from the stellar center.}
\label{fig:meanenergies}
\end{figure*}

\subsection{Neutrino emission}
\label{sec:neutrinos}

The luminosities for neutrinos and antineutrinos of
all flavors and for both 2D simulations, compared to the results
of the corresponding 1D models, are given as functions
of post-bounce time in Fig.~\ref{fig:luminosities} (muon and tau 
neutrinos and their antiparticles, which we sometimes denote
with $\nu_x$, are treated in the same way, 
and the plots show the luminosities of one individual type of
these heavy-lepton neutrinos). The luminosities
displayed there are actually not defined as the total rate of energy 
loss of the supernova core through neutrinos radiated in all directions.
Instead, they represent the isotropic
equivalent luminosities inferred by an observer at rest relative to
the stellar center, viewing the source from the direction of the
polar axis of the grid and from a large distance above the north pole
(i.e., gravitational redshifting is taken into account). The results
for an observer above the south pole look qualitatively and 
quantitatively very similar and lead to the same conclusions.

One of the goals of the present work is an analysis of luminosity
fluctuations that are associated with modulations of the mass
accretion rate of the nascent star due to SASI oscillations and
convective overturn in the post-shock layer,
which become very vigorous and create nonstationary conditions
at later post-bounce times (see Sect.~\ref{sec:shock} and
Figs.~\ref{fig:shockradius} and \ref{fig:shockmodes}).
Strong shock retraction leads to a transient increase of the gas 
flow towards the neutron star and to the compression and enhanced
cooling of the matter near the neutron star surface. In contrast,
shock expansion has the opposite
effect because it causes a deceleration of the infall or even outward
acceleration of material that is accreted through the shock. Thus
shock expansion stretches the time this matter remains in the gain
layer and does not cool by neutrino emission 
(Marek \& Janka 2007, Scheck et al.\ 2008, Murphy \& Burrows 
2008). Corresponding quasi-periodic fluctuations of the neutrino
luminosities, strongest for electron neutrinos 
$\nu_e$ and antineutrinos
$\bar\nu_e$ (which are more abundantly produced in the layer of 
freshly accreted, hot matter near the neutron star surface) can be
seen in all panels of Fig.~\ref{fig:luminosities}.

In the lefthand panels of Fig.~\ref{fig:luminosities} we have constructed
the isotropic equivalent luminosities from the neutrino radiation
leaving the stellar core in one angular grid bin near the north pole
of the mesh, i.e.,
\begin{equation}
L_{\nu,1}(\theta_{j+{1\over 2}},t)\,=\, 4\pi r^2 
F_{\nu}(r,\theta_{j+{1\over 2}},t)
\label{eq:lum1}
\end{equation}
for $F_{\nu}(r,\theta_{j+{1\over 2}},t)$ being the energy-integrated 
neutrino energy flux in the rest frame at the cell-centered
(pole-near) latitudinal angle $\theta_{j+{1\over 2}}$ of the polar
grid of the simulation,
at time $t$ and a radius $r$ that is large enough
such that the luminosity is conserved at greater radial distances
(for the numerical evaluation we have chosen $r = 400\,$km).
In contrast, for the righthand panels of this figure we have integrated
the neutrino emission over the whole northern hemisphere and then 
rescaled the result to the full sphere, i.e.,
\begin{eqnarray}
L_{\nu,2}(t) &=& 2 r^2 \int_{{\mathrm{hemisphere}}}\mathrm{d}\Omega\,
F_{\nu}(r,\theta,t) \label{eq:lum2a} \\  
&=& 4\pi r^2 \sum_{j = 1}^{N_{\theta}/2}   
(\cos\theta_{j-1} - \cos\theta_j)F_{\nu}(r,\theta_{j-{1\over 2}},t)\ ,
\label{eq:lum2b}
\end{eqnarray}
where the second line exploits axial symmetry and the 
sum is performed over the angular zones of one hemisphere.
The former approach means that 
one assumes that an observer at a position near the polar axis receives 
the radiation emitted only from one pole-near lateral zone (in our 
treatment this means that the received radiation flux has the properties
calculated in exactly one angular grid bin or angular ``ray''),
whereas the second evaluation implies the assumption that each 
unit of area on the
hemisphere oriented towards the observer contributes 
to the observable luminosity with a weight defined by
the (radial) flux calculated
for the corresponding angular bin (``ray'') of the polar grid 
in our ray-by-ray transport treatment. This assumption 
would be correct, of course, if a spherical source were radiating 
uniformly and isotropically in all directions, but in general the 
surface parts oriented with smaller angles to the observer direction
contribute more strongly.
For a sphere with a sharp radiating surface 
(in contrast to a radially extended neutrino-decoupling layer)
that emits neutrinos with a (locally)
isotropic intensity towards an observer at great distance on the
axis of symmetry of the source, Eq.~(\ref{eq:lum2a}) would
contain a factor $(2\cos\theta)$ in the integrand\footnote{The factor 
$\cos\theta$
accounts for the projection of the surface elements on the radiating 
sphere perpendicular to the line connecting source and observer.}. 
If the intensity was a function
of the latitudinal angle $\theta$ of the stellar grid, for example,
this factor would enhance the relative
contributions of surface parts that are closer to the observer. 
Because of limb darkening the importance of the observer-near 
regions of the radiating neutrinosphere is further enhanced.

Not solving the full multi-angle problem of neutrino transport in
the 2D geometry but using our ray-by-ray transport approximation
does not allow us to exactly compute the direction
dependence of the neutrino intensity radiated from the nascent
neutron star. The ray-by-ray treatment essentially implies
the construction of a 1D transport solution in every lateral bin
of the computational polar grid, assuming a spherically symmetric
transport problem for the stellar conditions that are present in 
this angular bin. Therefore the ray-by-ray approach 
tends to underestimate
the directional smearing of luminosity features that result from
local and time-dependent emission increase in the semi-transparent 
accretion layer near the neutron star surface. Such features are,
for example, caused by the
SASI and convection modulated mass inflow and must be
expected to be more prominent in the plots showing the measurable 
luminosity $L_{\nu,1}$, because this quantity is the isotropic 
equivalent luminosity
resulting from neutrino emission occurring in one angular grid bin
(see lefthand panels of Fig.~\ref{fig:luminosities}). In contrast, 
spatially localized effects will show up only with 
significant damping in the luminosity $L_{\nu,2}$, where neutrinos
coming from the whole hemisphere facing the observer are added up
with equal weighting (righthand panels of Fig.~\ref{fig:luminosities}).
We consider these two evaluated cases
as the extrema that define the theoretical limits of the true 
result. We suspect that $L_{\nu,1}$ yields the better approximation 
of the real situation, but we also provide $L_{\nu,2}$ for a 
pessimistic estimate of the magnitude of the luminosity 
fluctuations. It is important to note that the total energy loss
of the supernova core in neutrinos (or one sort of neutrinos), 
which is usually defined as neutrino luminosity and computed by
integrating the neutrino flux over all observer directions, 
shows SASI induced variability only on a much reduced level (see 
the corresponding plots in the Marek \& Janka 2007 paper). This
can be understood from the fact that the SASI fluctuations in the
northern and southern hemispheres are phase shifted by roughly 
half a SASI period so that the emission maxima on one side fill
the emission minima on the other. The combined signal is therefore
much smoother than the neutrino flux that can be received by an 
observer in any chosen direction.

The modulations of the apparent luminosity due to SASI and convective 
variations of the neutron star accretion rate in the 2D simulations
are superimposed on the general luminosity evolution that is 
characteristic of the employed progenitor star. These modulations
add another feature on top of the effects that discriminate 1D from
2D results and models based on the use of the L\&S EoS from those
with the H\&W EoS. The 15$\,M_\odot$ stellar model used in the
present simulations shows a distinct drop in the mass accretion rate
by roughly a factor of two when the steep density decline at the interface
between the silicon shell and the oxygen-enriched Si layer falls 
through the stalled shock (see Appendices A and B of Buras et al.\ 2006b).
This happens between 160$\,$ms and 180$\,$ms
and leads to the pronounced decrease of the luminosities of all neutrinos
and antineutrinos at around this time.
 
In Figs.~\ref{fig:luminosities} and \ref{fig:meanenergies} we see 
an effect already mentioned in Sect.~\ref{sec:shock}, namely that the 
proto-neutron star in the simulations with the softer L\&S EoS is
more compact and hotter and therefore radiates higher luminosities
and significantly higher mean energies of all kinds of neutrinos 
and antineutrinos (the mean neutrino energies $\left\langle\epsilon_\nu
\right\rangle$ are defined here as
the ratio of the energy flux to the number flux).
This holds for 1D as well as 2D models. Besides the quasi-periodic
fluctuations, the 2D results exhibit the characteristic differences 
compared to the 1D case that were discussed before in much detail by
Buras et al.\ (2006b): convective transport below the neutrinospheres 
leads to an accelerated loss of lepton number and energy, which 
manifests itself on the one hand in a slightly enhanced electron neutrino
flux relative to the electron antineutrino flux, and on the other hand in 
significantly increased muon and tau neutrino luminosities compared
to the 1D results. These differences are visible in both the polar
luminosities, $L_{\nu_i,1}$, as well as in the hemispherically 
integrated data, $L_{\nu_i,2}$, of Fig.~\ref{fig:luminosities}.
On the other hand, the convective neutron star is less compact
than its 1D counterpart, for which reason the mean energies of the
radiated neutrinos are slightly reduced in the 2D simulations (compare
the upper two panels of Fig.~\ref{fig:meanenergies} with the bottom
ones). This
phenomenon is particularly strong for the L\&S EoS, in which case the
more compact neutron star structure seems to react more sensitively to 
the redistribution of lepton number and entropy associated with the
convective activity in its interior.

\begin{figure*}[!htp]
\begin{center}
\includegraphics[width=8.5cm]{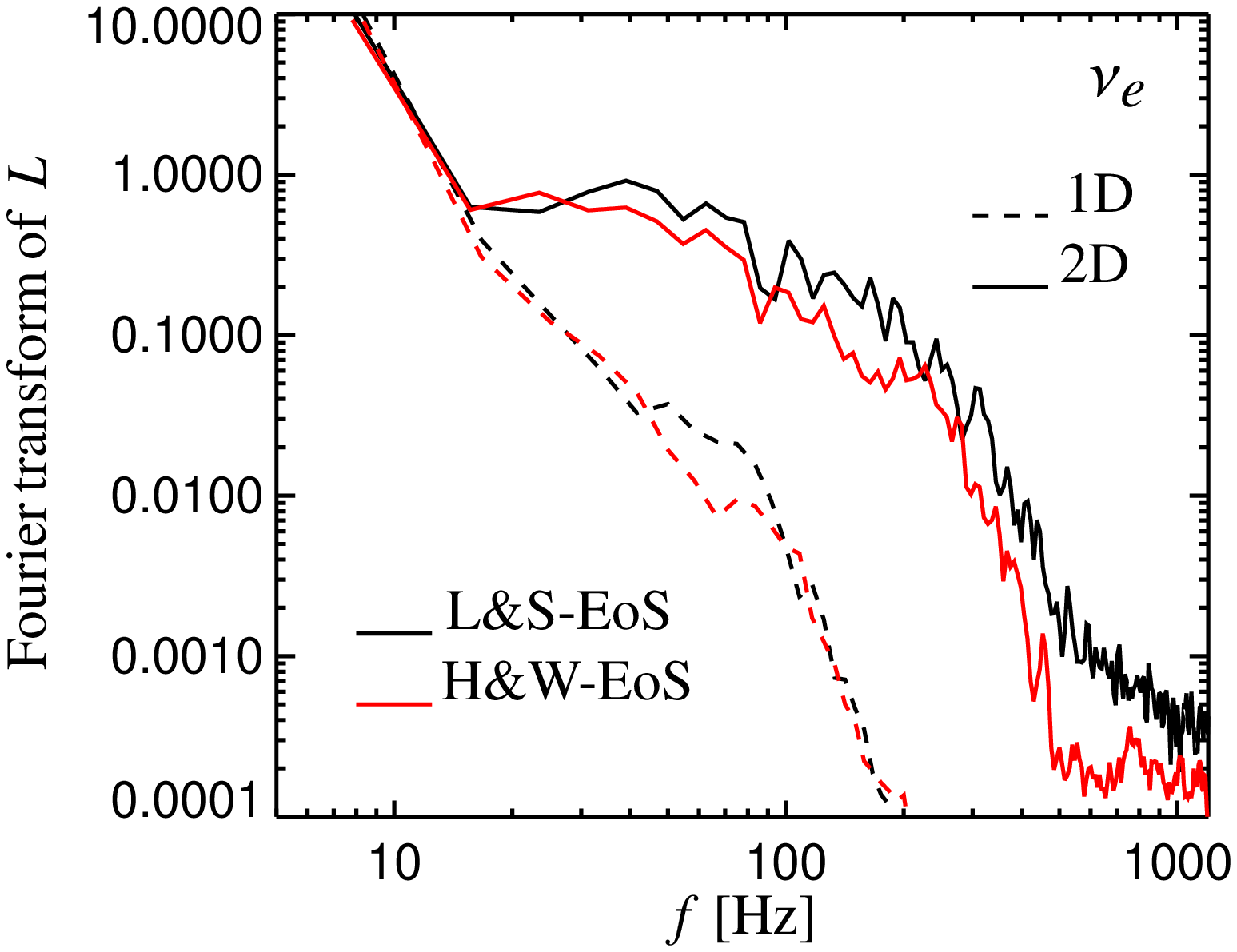}
\includegraphics[width=8.5cm]{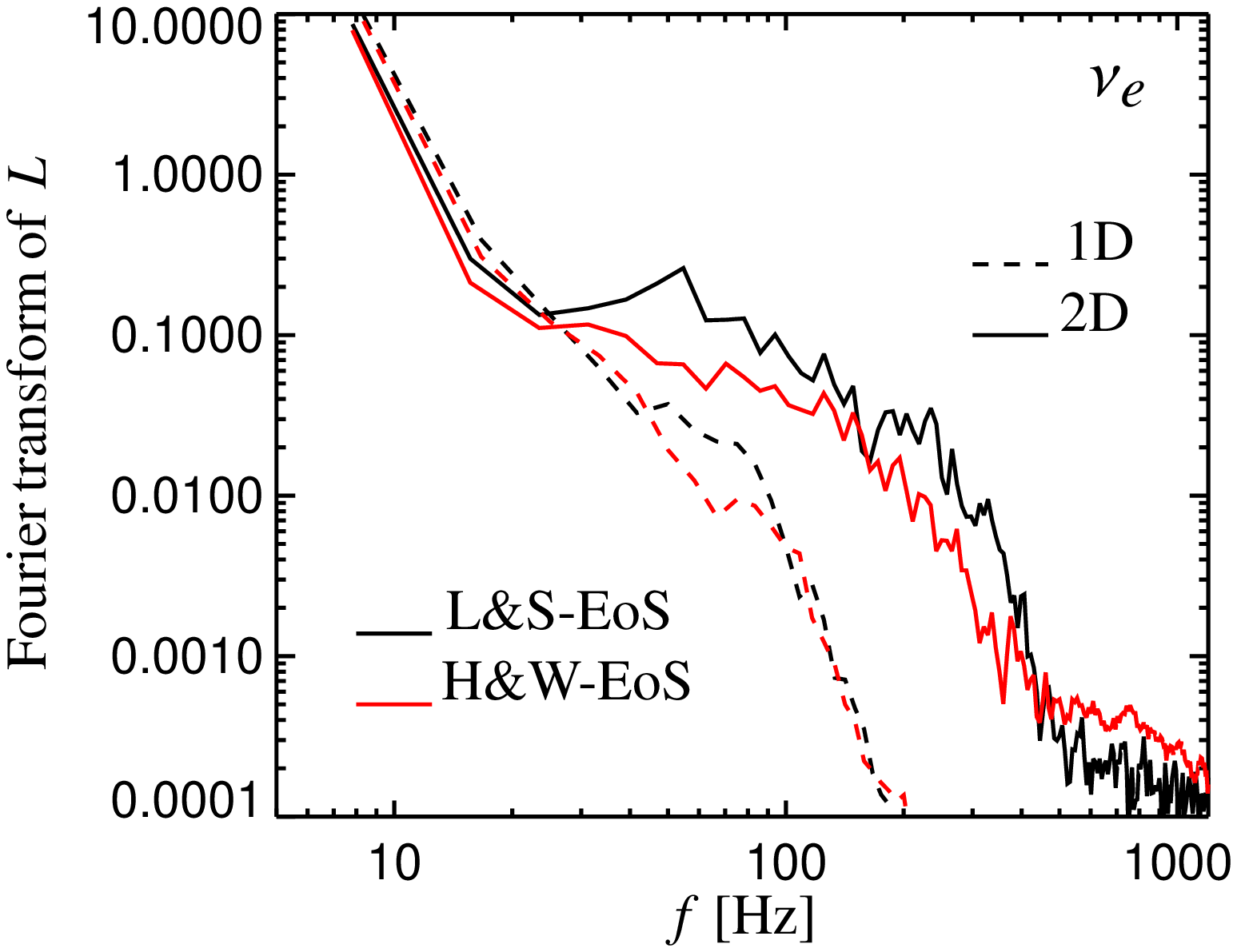}\\ \vspace{0.3cm}
\includegraphics[width=8.5cm]{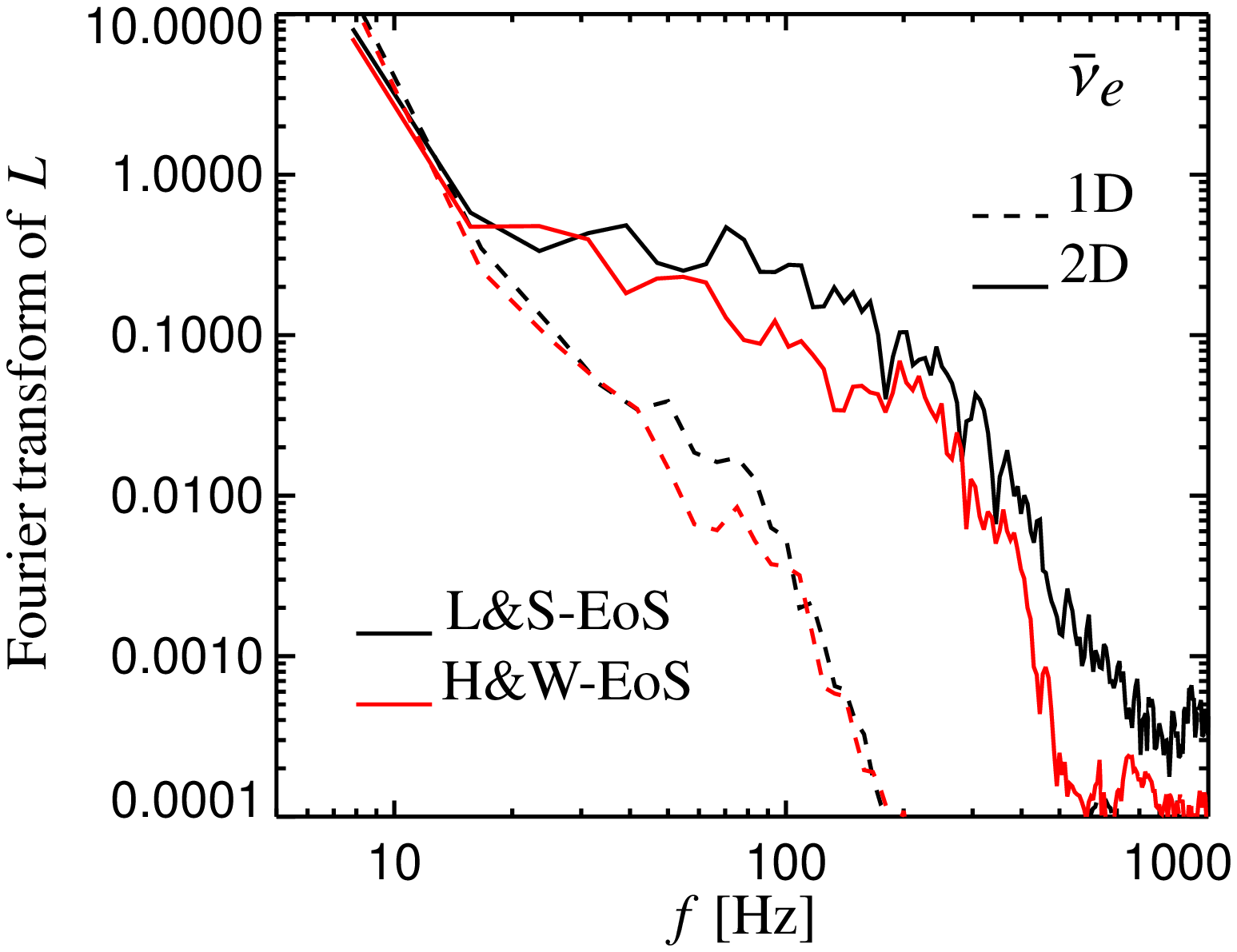}
\includegraphics[width=8.5cm]{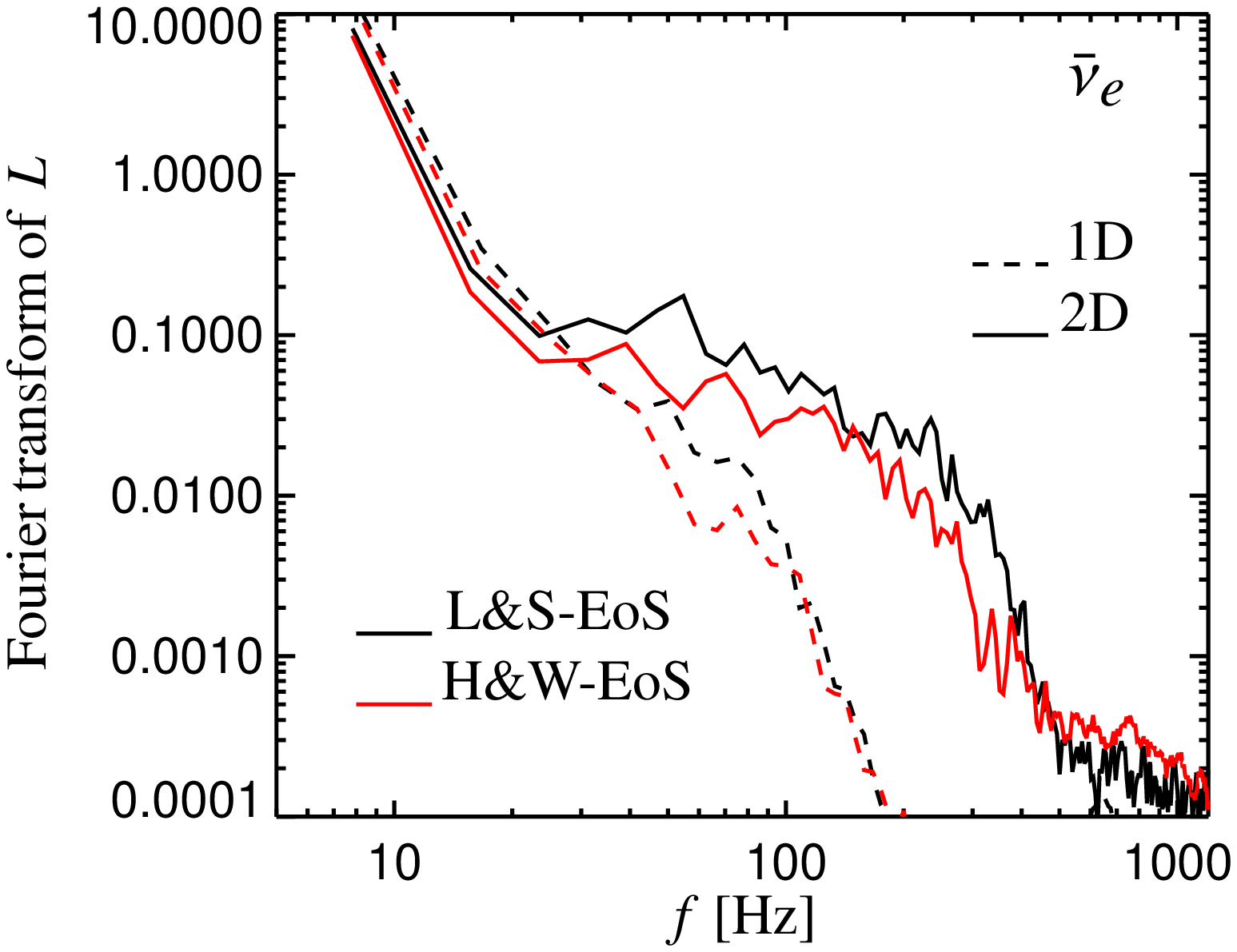}\\ \vspace{0.3cm}
\includegraphics[width=8.5cm]{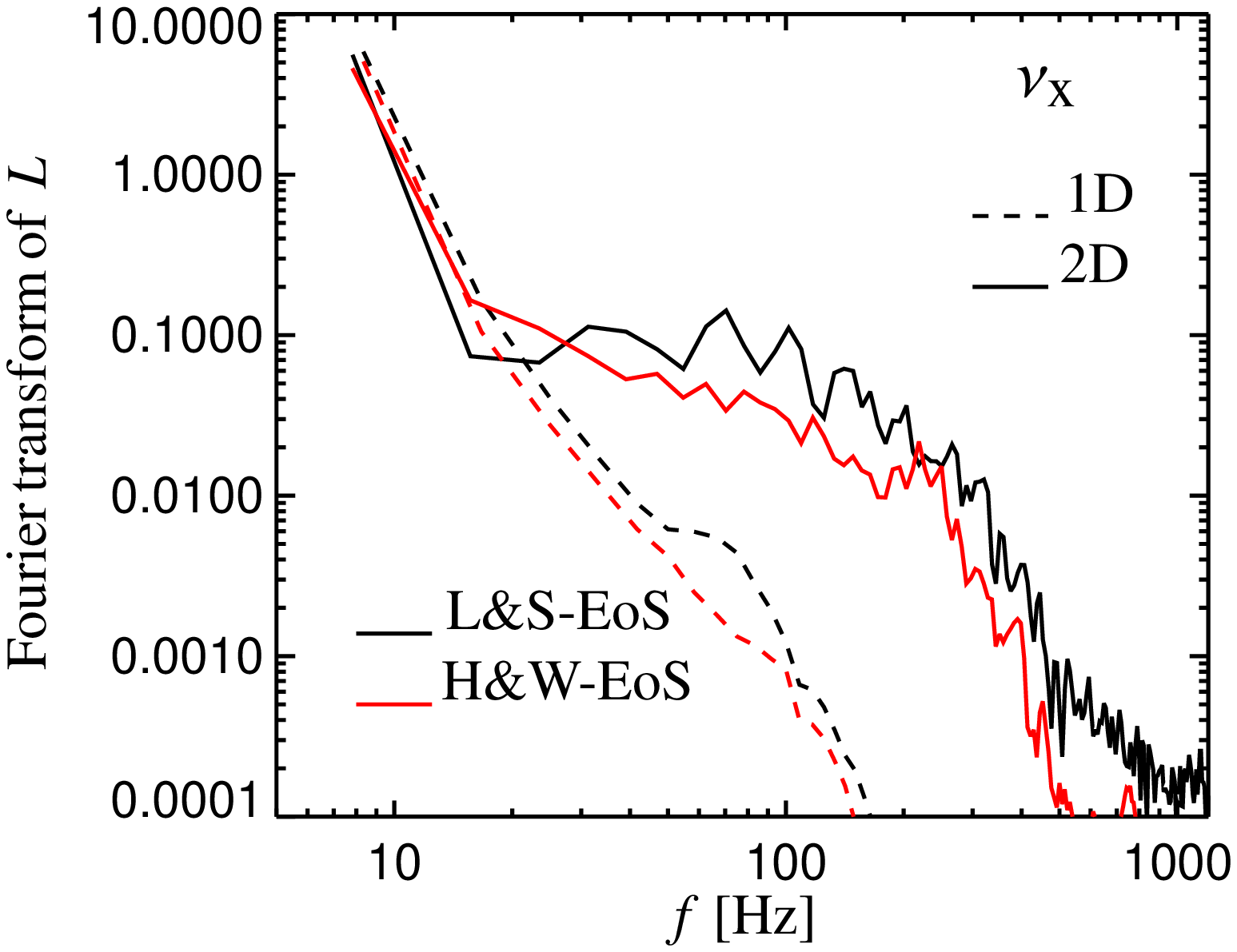}
\includegraphics[width=8.5cm]{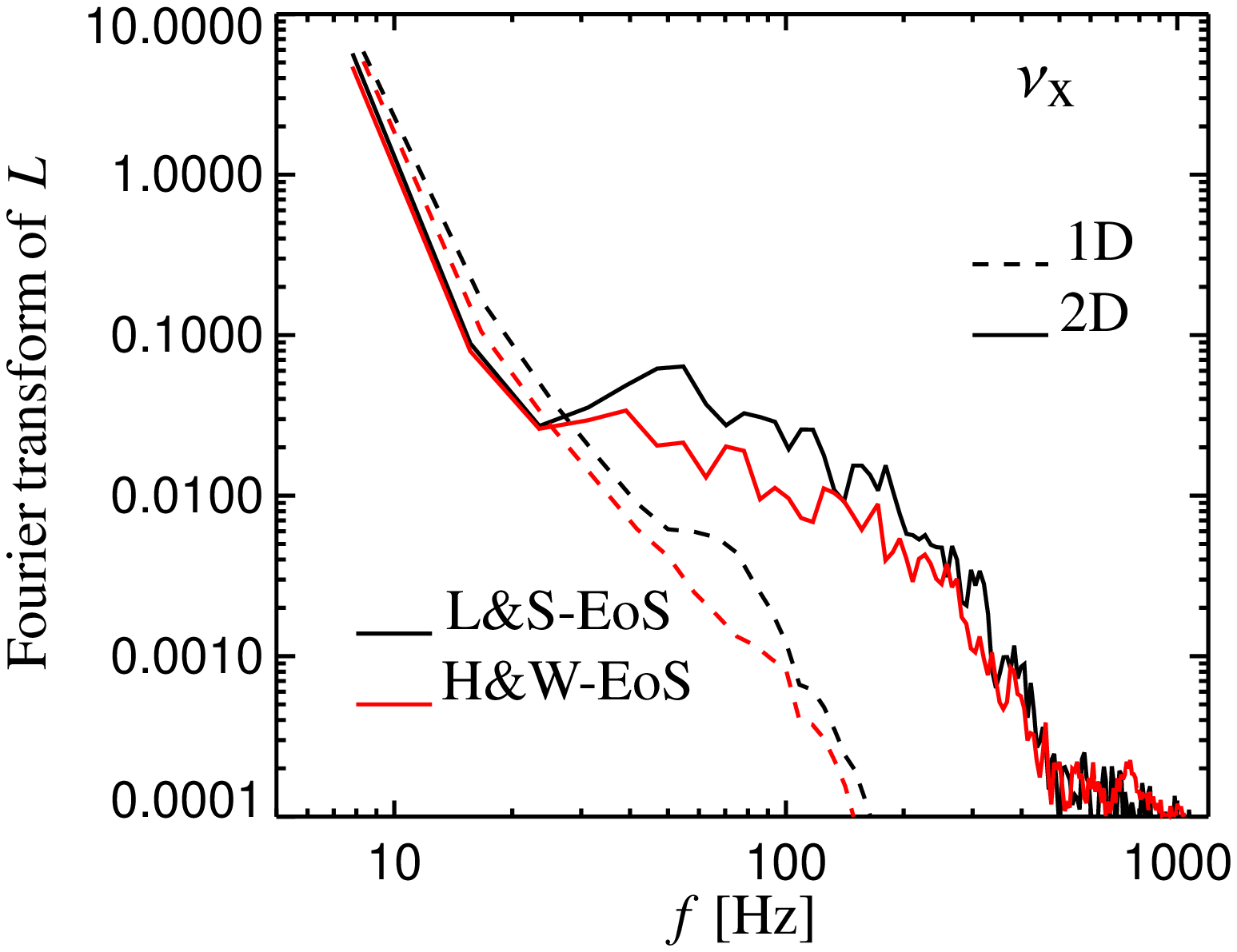}
\end{center}
\caption{Fourier amplitudes of the time-dependent neutrino luminosities
shown in Fig.~\ref{fig:luminosities} (the left/right column there corresponds
to the left/right column in the present figure). 
In order to reduce the level of
background noise, the evaluation was performed for post-bounce times
$t > 100\,$ms, for which the SASI modulation is clearly visible in the
neutrino luminosities, and the signals were split in 17 overlapping segments 
of 128$\,$ms, advanced in steps of 10$\,$ms. The Fourier spectra of the
1D models show a weak contribution from low-amplitude numerical 
fluctuations of the transport results.} 
\label{fig:fourierlumis} 
\end{figure*}

An interesting effect is the crossing of the mean energies 
of electron antineutrinos and heavy-lepton neutrinos ($\left\langle
\epsilon_{\bar\nu_e} \right\rangle$ becoming higher than 
$\left\langle \epsilon_{\nu_x} \right\rangle$)
that occurs in five of the six panels of Fig.~\ref{fig:meanenergies} at
about 200$\,$ms after bounce, but which {\em cannot} be observed until
this post-bounce time for other energy moments; for example it is 
absent in the case of the rms energies of the energy
spectra. These rms energies are ordered in the standard way at
all times, with heavy-lepton neutrinos $\nu_x$ being more energetic
on average than $\bar\nu_e$, and the latter being more energetic 
than $\nu_e$: $\left\langle \epsilon_{\nu_e} \right\rangle_{\mathrm{rms}} 
< \left\langle \epsilon_{\bar\nu_e} \right\rangle_{\mathrm{rms}}
< \left\langle \epsilon_{\nu_x} \right\rangle_{\mathrm{rms}}$
(this was pointed out already by Marek \& Janka 2007).
The energy crossing therefore is a consequence of a subtle difference
in the spectral shape of $\bar\nu_e$ and $\nu_x$. A close inspection
of our results reveals that this effect is connected with the 
modification of the radiated neutrino fluxes in the accretion layer
around a successively more compact, contracting proto-neutron star.
In contrast to the mean energies at a radius of 400$\,$km, i.e.\ 
far outside of the neutrino source (see Fig.~\ref{fig:meanenergies}), 
the corresponding data at a fixed density of
$10^{11}\,$g$\,$cm$^{-3}$ always follow the relation $\left\langle
\epsilon_{\bar\nu_e} \right\rangle \la \left\langle
\epsilon_{\nu_x} \right\rangle$. This holds for our 1D results and
for the time-dependent unidirectional as well as (hemi)spherically 
averaged 2D results with both equations of state; only during 
short phases of a strong increase of the accretion rate can
the mean energy of electron antineutrinos exceed the one of 
heavy-lepton neutrinos also at $\rho = 10^{11}\,$g$\,$cm$^{-3}$. 
Similarly, accretion around the poles of
the 2D models also causes the polar rms energies of $\bar\nu_e$ and 
$\nu_x$ --- measured outside of the accretion layer --- to converge
($\left\langle \epsilon_{\bar\nu_e} \right\rangle_{\mathrm{rms}}
\la \left\langle \epsilon_{\nu_x} \right\rangle_{\mathrm{rms}}$),
whereas the laterally averaged values are 
always well separated, in agreement with the hierarchy of the rms 
energies found in the corresponding 1D simulations. Moreover, 
the neutrino radiating ``naked'' neutron star in 
spherically symmetric models with artifically initiated explosions
does not produce the inversion of the mean $\bar\nu_e$ and $\nu_x$ 
energies, whereas the accreting neutron star in a corresponding 
nonexploding model does (see Fig.~2 in Janka et al.\ 2007).

All these findings point to the importance of accretion
for an explanation of the phenomenon. Accretion modifies the
density profile outside of the neutrinosphere and thus is 
responsible for 
a subtle change in the $\bar\nu_e$ spectrum relative to the 
$\nu_x$ spectrum. Two effects seem to contribute. On the one hand,
with the inclusion of $\nu_x\bar\nu_x$-pair production by 
nucleon-nucleon bremsstrahlung and by $\nu_e\bar\nu_e$ annihilation
in state-of-the-art supernova models,
the energysphere of heavy-lepton neutrinos is shifted to
lower temperatures than in older simulations where these processes
were ignored. The heavy-lepton neutrinos that leave
their average neutrino-energysphere also experience a
``spectral filter effect'' in being down-graded in energy space
by neutrino-nucleon and neutrino-electron collisions when diffusing
through the cooler, outer layers of the neutron star (modern
multi-energy transport calculations follow the corresponding
spectral changes with great accuracy). 
For both reasons the spectra of $\nu_x$ and $\bar\nu_e$ 
are now much more similar than in the older models (for a detailed
discussion, see Raffelt 2001 and Keil et al.\ 2003). On the other
hand, electron neutrinos and antineutrinos, which are emitted in 
rapid charged-current reactions of electrons and positrons with protons 
and neutrons, are more copiously produced in the hot accretion layer
than muon and tau neutrinos. This leads to an enhancement of the 
$\bar\nu_e$ flux at medium and high energies ($\ga$10$\,$MeV) compared
to the $\nu_x$ flux at these energies. In combination, these effects
can raise the average energy of electron antineutrinos radiated from
an accreting neutron star above that of heavy-lepton neutrinos.

The described relative spectral changes become stronger at later 
times after core bounce so that a crossing of 
$\left\langle \epsilon_{\bar\nu_e} \right\rangle$
and $\left\langle \epsilon_{\nu_x} \right\rangle$
occurs. As the post-bounce time advances,
the average energies of muon and tau neutrinos grow slowly because the 
neutrinospheric temperature increases in the shrinking neutron star.
This increase is stronger for the soft L\&S EoS than for the
stiff H\&W EoS (see Fig.~\ref{fig:meanenergies}). In contrast, the
$\nu_e$ and $\bar\nu_e$ created in the semitransparent accretion layer
escape with mean energies that rise relatively faster, because 
accretion onto a more compact proto-neutron star releases more 
gravitational binding energy. This leads to more extreme heating
of the infalling and compressed gas and more abundant production of
$\nu_e$ and $\bar\nu_e$ with higher energies. This reasoning also 
explains why the polar-near data in Fig.~\ref{fig:meanenergies}
show the effect more clearly than the 
hemispherically averaged results: the SASI sloshing motions of the 
shock along the direction of the polar grid axis with alternating
periods of shock inflation and contraction lead to successive phases
of gas expansion and mass accumulation on the one hand, and gas
compression on the other, in particular around the north
and south poles of the neutron star (compare Figs.~\ref{fig:snapshots}
and \ref{fig:luminosities}). 

At first sight this argument seems in 
conflict with the observation that the accretion 
funnels in the sequence of snapshots of Fig.~\ref{fig:snapshots}, in
particular the most prominent and persistent ones that carry most
mass, reach down to the neutron star typically at intermediate 
latitudinal angles, i.e., around the equator; still the fluctuations of
the neutrino emission are highest close to the poles. We understand 
this as a consequence of the following two facts. Firstly, because
of the extremely large infall velocities only little neutrino loss
occurs within the downdrafts. Instead, most of the binding energy
is radiated when the accreted gas spreads around the neutron star and
begins to settle onto the neutron star surface (for a discussion of
this aspect, see also Scheck et al.\ 2006). Secondly, due to the
SASI sloshing of the whole accretion layer, much of the gas accreted
through the shock
is redirected alternatingly towards the north pole or the south pole.
There the gas is decelerated near the grid (and sloshing) 
axis and deflected so that its duration of stay is longest in the polar 
regions. The gas that is sucked to a pole at an enhanced rate during the
SASI half-cycle of shock expansion is compressed there during the next
half-cycle of shock contraction. This is the reason
why most of the losses of accretion energy happen around the poles.
Future three-dimensional (3D) simulations with neutrino transport will
have to show whether this behavior is a consequence of the assumption
of axial symmetry or whether a similar effect can also be found in 
connection with the SASI in three dimensions.

Concerning the possibility of measuring the SASI (and convection) 
induced luminosity variations by the detection of neutrinos from a future 
Galactic supernova it is important to note that the strong compressional 
heating of accreted gas during the SASI half-cycles with shock
contraction does not only boost the neutrino emission at the poles
but also leads to the production of more energetic
neutrinos. Therefore the variations of the neutrino luminosities 
correlate and are in phase with modulations of the mean neutrino energies
(see in particular the polar data in Figs.~\ref{fig:luminosities} and          
\ref{fig:meanenergies}). While (for the polar emission) the luminosity 
differences between maxima
and minima can be several 10\% up to about 50\% of the values in 
low-emission episodes and the mean energies show variations up to 
roughly 1$\,$MeV or 10\%, the observationally relevant energy moments
(approximately scaling like $L\left\langle\epsilon\right\rangle^2$ 
or $L\left\langle\epsilon\right\rangle$, depending on the neutrino
detector) fluctuate with an amplitude of typically $\sim$50\% 
and extrema up to nearly 100\% of the values in the minima.
The fluctuations are similarly strong for $\nu_e$ and 
$\bar\nu_e$, which are both abundantly created in the accretion layer,
and somewhat reduced for heavy-lepton neutrinos, whose emission is
more indirectly affected by the perturbations of their decoupling
region near the neutron star surface, e.g.\ through the surface gravity
modes caused by the SASI activity and through changes in the density 
structure and optical depth associated with the mass motions in the
SASI region.

In Fig.~\ref{fig:fourierlumis} we provide the Fourier spectra of
the neutrino luminosities displayed in Fig.~\ref{fig:luminosities};
the lefthand panels belong to the polar luminosities, the righthand 
panels to the hemispherically averaged ones.
The 2D simulations exhibit a broad peak between about 20$\,$Hz and
roughly 400$\,$Hz with a clear power excess compared to 1D results.
Instead of the expected monotonic decline the Fourier transforms of 
the 1D luminosities possess a flat shoulder at frequencies between 
$\sim$50$\,$Hz and $\sim$100$\,$Hz. This feature is a consequence
of low-amplitude numerical noise produced by the transport scheme,
which is so small that it is physically insignificant for the 
simulations. As expected from
the time evolution of the shock radii and corresponding luminosity
variations, which both are larger in the case of the L\&S run,
the Fourier spectra of the 2D simulation with the L\&S 
EoS dominate the result with the H\&W EoS by a factor of 2--5 for
all kinds of neutrinos. The flat local maxima in the 
$\sim$25--80$\,$Hz range, whose location roughly 
agrees with the peak of the 
Fourier amplitude of the $\ell = 1$ shock oscillation mode 
(Fig.~\ref{fig:shockmodes}) and also shows the tendency of slightly
higher frequencies for the L\&S case, are superimposed by a 
high-frequency component that originates from the faster convective 
modulations of the accretion flow and neutrino emission.

\begin{figure*}[!htp]
\begin{center}
\includegraphics[width=8.5cm]{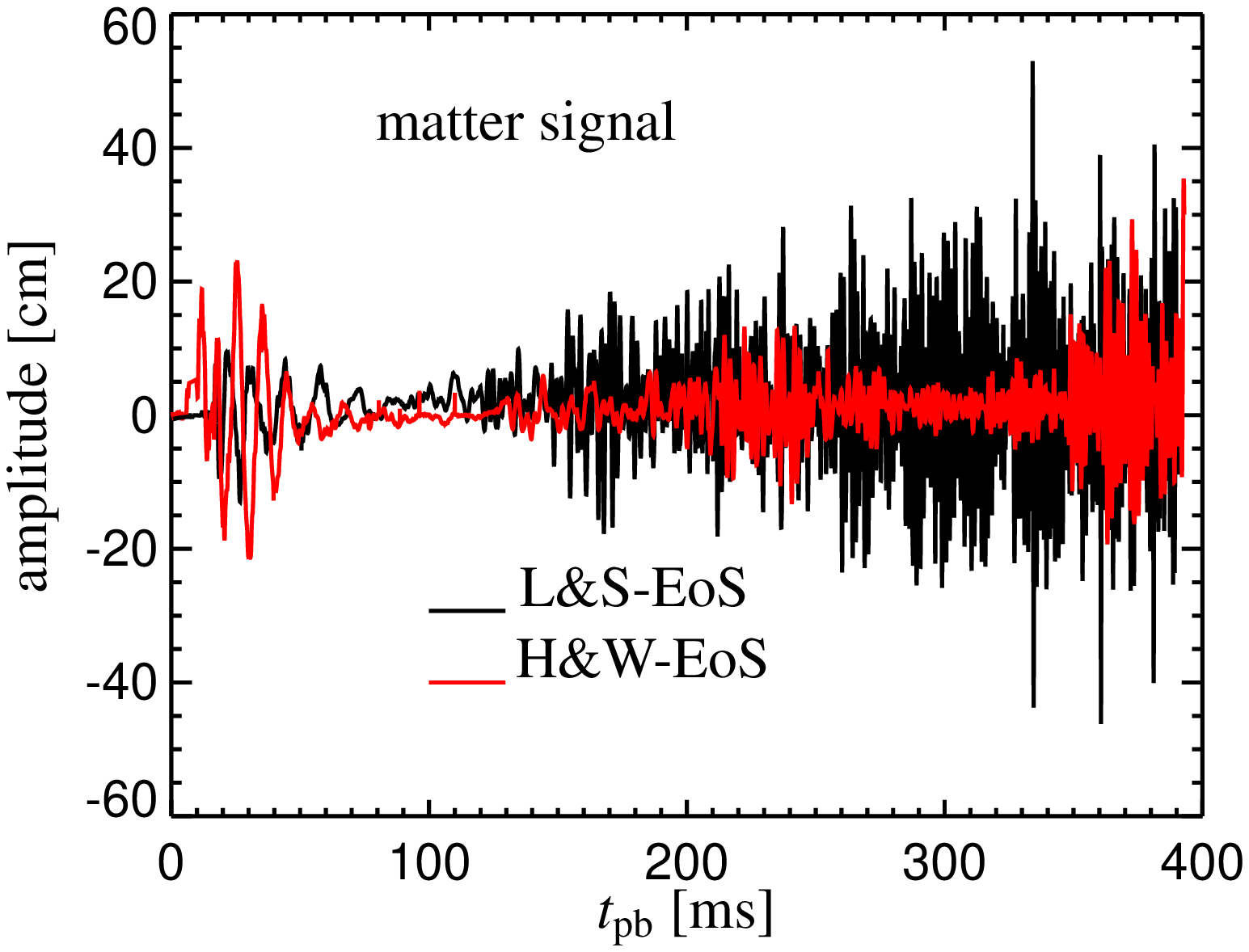}
\includegraphics[width=8.5cm]{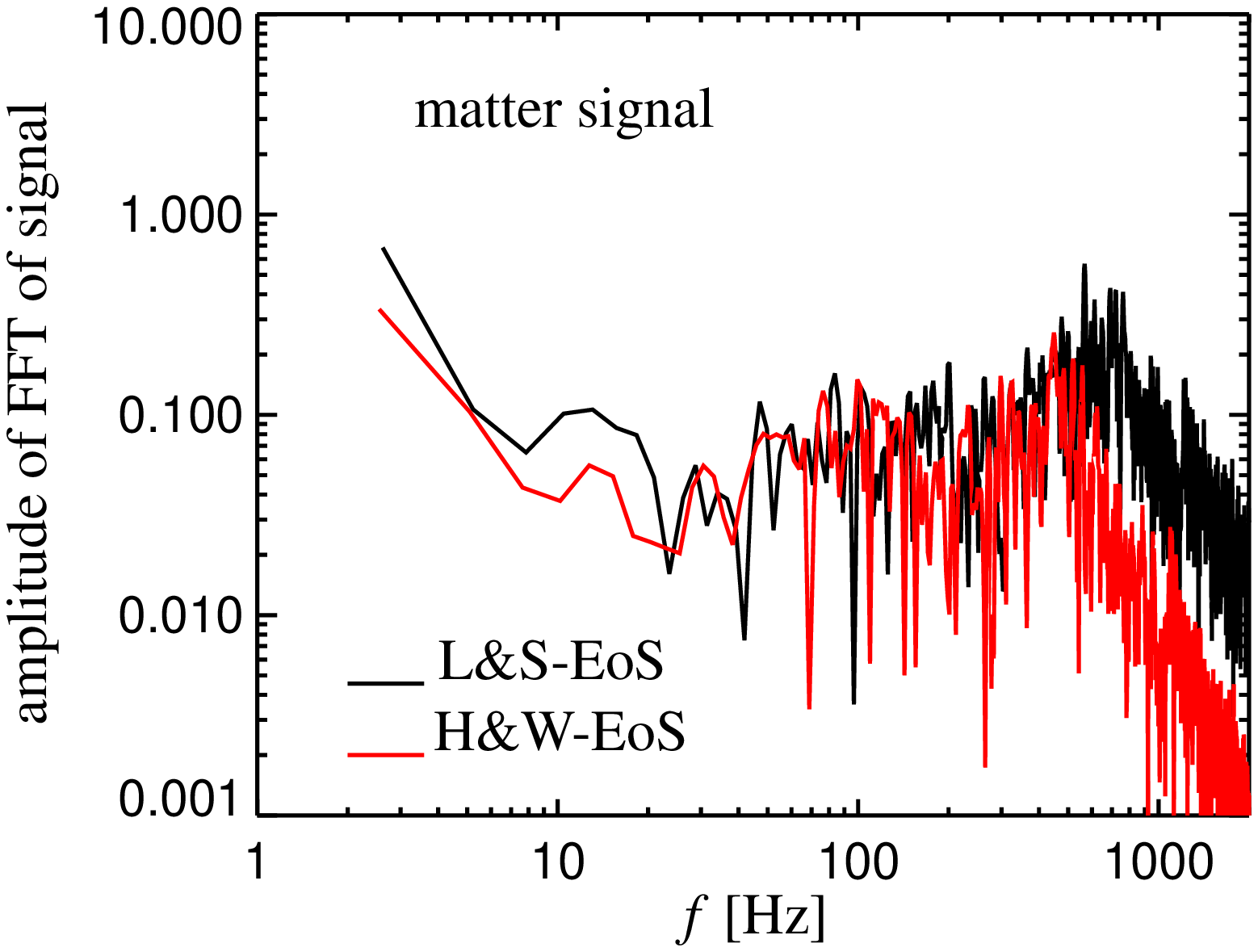}\\ \vspace{0.3cm}
\includegraphics[width=8.5cm]{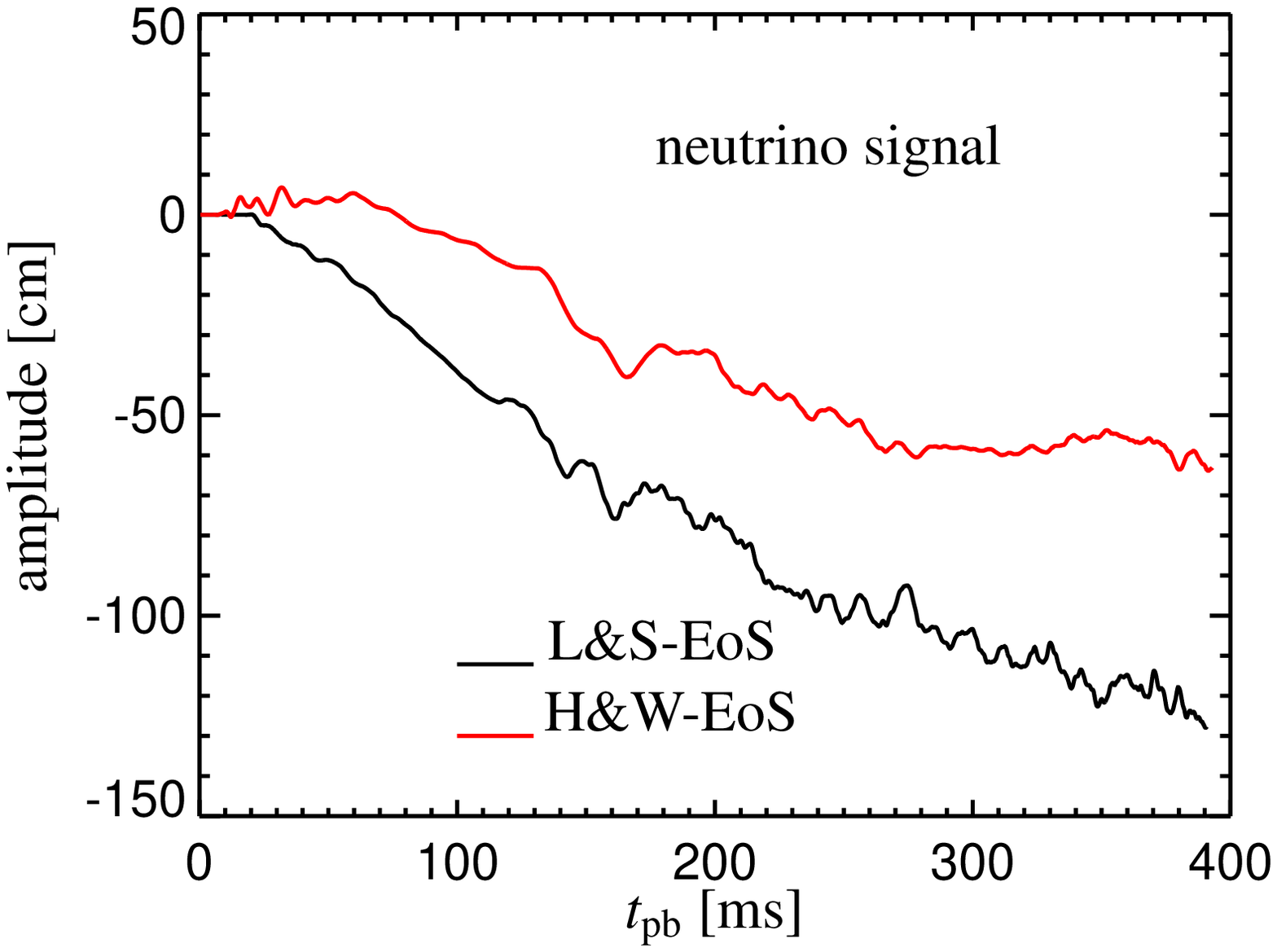}
\includegraphics[width=8.5cm]{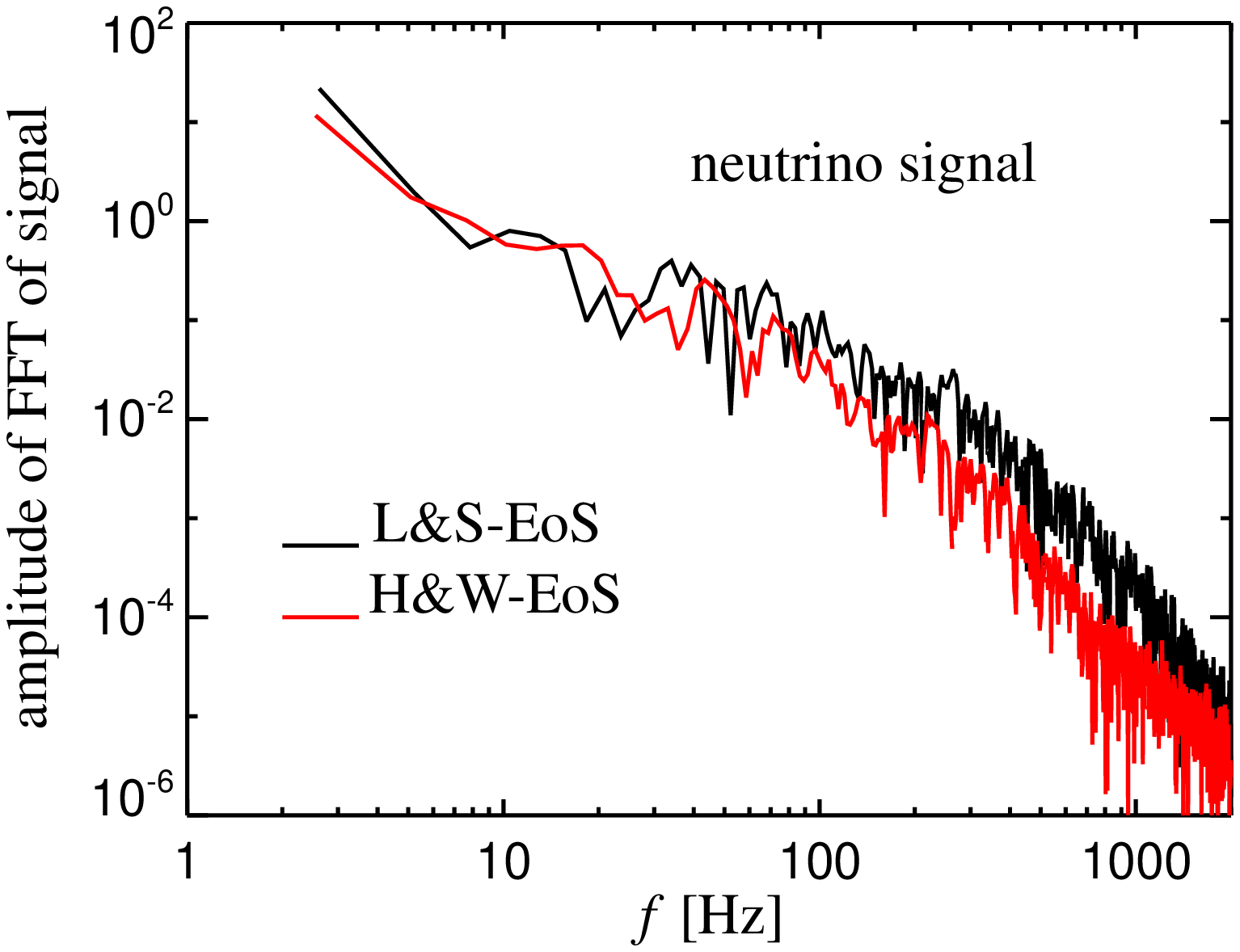}
\end{center}
\caption{Gravitational-wave quadrupole amplitudes $A^{E2}_{20}$
as functions of post-bounce time ({\em left}) and
corresponding Fourier spectra ({\em right}) associated with mass
motions ({\em top}) and anisotropic neutrino ($\nu_e + \bar\nu_e$ plus
all heavy lepton neutrinos and antineutrinos) emission ({\em bottom})
for a distant observer in the equatorial plane of the axisymmetric
source. Note that despite the trend toward a prolate shock deformation
at late post-bounce times (see Fig.~\ref{fig:snapshots} and the lefthand
panels of
Fig.~\ref{fig:shockmodes}) the neutrino amplitude develops negative
values, suggesting that the signal is not determined by a dominant
polar neutrino
emission in this case. The prominent peaks of the matter signals
at Fourier frequencies around 500$\,$Hz and 700$\,$Hz, respectively,
are caused by mass motions in the proto-neutron star surface at
densities between $10^{11}\,$g$\,$cm$^{-3}$ and $10^{13}\,$g$\,$cm$^{-3}$.}
\label{fig:graviwaves}
\end{figure*}

\subsection{Gravitational waves}
\label{sec:graviwaves}

Nonradial mass motions in the three regions of hydrodynamic 
instabilities as well as anisotropic neutrino emission are sources
of gravitational-wave signals from stellar core-collapse events
(e.g., M\"uller et al.\ 2004 and references therein). 
We will analyze our data for EoS-specific properties of these
signals now.

The (quadrupole) gravitational-wave amplitudes, energy spectra, and
spectrograms resulting from anisotropic mass motion can be computed 
for axisymmetric models as described in 
M\"uller \& Janka (1997; Eqs.~(10)--(12)), using the Einstein
quadrupole formula in the numerically convenient form derived
by Blanchet et al.\ (1990), and by standard FFT techniques.  
Assuming an observer that is located at an angle $\theta$
with respect to the symmetry axis of the source, the dimensionless
gravitational wave amplitude $h(t)$ is related to the quadrupole wave
amplitude $A^{E2}_{20}$ (measured in units of cm), the lowest-order
non-vanishing term of a multipole expansion of the radiation field
into pure-spin tensor harmonics (see Eq.~(9) of M\"uller 1997),
according to
\begin{equation}
 h\, =\, \frac{1}{8} \sqrt{\frac{15}{\pi}} \sin^2\theta
     \,\frac{A^{E2}_{20}}{R} ,
\label{eq:qwa}
\end{equation}
where $R$ is the distance to the source. In the following we will 
always assume $\sin^2\theta = 1$, i.e., the observer is positioned in
the equatorial plane of the polar grid. The wave 
amplitudes due to the anisotropic neutrino emission are obtained with
the formalism given in M\"uller \& Janka (1997; Eqs.~(28)--(31)),
which is based on the work of Epstein (1978). According to these
references, the dimensionless gravitational-wave
amplitude for the radiated energy in neutrinos can be written as
convolution of the time-dependent neutrino luminosity $L_\nu (t)$ of the
supernova core and the emission anisotropy $\alpha_\nu(t)$,
\begin{equation}
h_\nu^{\mathrm{TT}}\,=\, {2G\over c^4 R}\int_0^t {\mathrm{d}}t'\,
L_\nu(t')\,\alpha_\nu(t') 
\label{eq:nuamplitude}
\end{equation}
(the quadrupole wave amplitude $A^{E2}_{20}$ for neutrinos then results
from inversion of Eq.~(\ref{eq:qwa})), with
\begin{equation}
\alpha_\nu(t) \,\equiv\, {1\over L_\nu (t)}\,\int_{4\pi}{\mathrm{d}}\Omega\,
\Psi(\theta,\phi)\,
{{\mathrm{d}}L_\nu({\bf \Omega},t)\over \mathrm{d}\Omega}\ ,
\label{eq:alpha}
\end{equation}
where ${\mathrm{d}}L_\nu({\bf \Omega},t)/\mathrm{d}\Omega$ is the 
neutrino energy radiated at time $t$ per unit of time and per unit
of solid angle into direction $\mathrm{d}\Omega$, and 
${\mathrm{d}}\Omega = -{\mathrm{d}}\cos\theta {\mathrm{d}}\phi$ 
defines the solid angle element in the polar coordinate system of
the source. The angle-dependent function $\Psi(\theta,\phi)$ is
given in Eq.~(27) of  M\"uller \& Janka (1997). For the case of an 
axially symmetric source, Kotake et al.\ (2007) derived a
compact expression (see their Eq.~(8) and note that our $\Psi$,
following the notation of M\"uller \& Janka 1997, is
their $\Phi$), which is visualized
as a function of the polar angle $\theta$ in their Fig.~1. For
the further discussion it is important to note that $\Psi$ has
positive values in the polar cap regions between $\theta = 0$ and
$\theta = 60^\circ$ and between $\theta = 120^\circ$ and 
$\theta = 180^\circ$, but adopts negative values in the equatorial
belt for $60^\circ < \theta < 120^\circ$. The combined polar caps 
and the equatorial belt have equally large surface areas.

The quadrupole wave amplitudes $A^{E2}_{20}$ of our two 2D models and the 
corresponding Fourier spectra are shown in Fig.~\ref{fig:graviwaves},
in the upper panels for the signals from aspherical matter flow,
and in the lower panels for those from anisotropic neutrino emission.

\begin{figure}[!htb]
\begin{center}
\includegraphics[width=8.5cm]{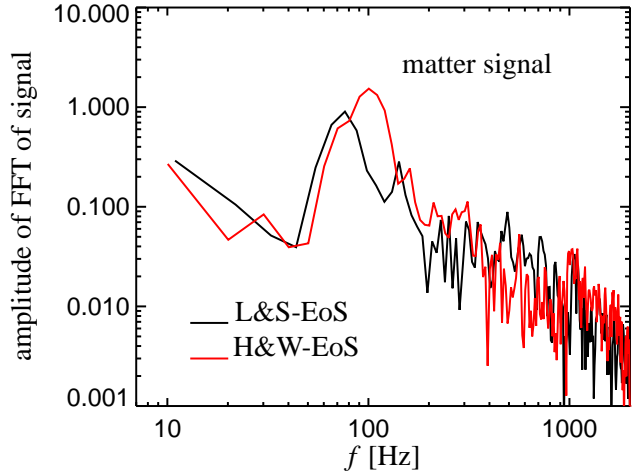}
\end{center}
\caption{Fourier spectra of the gravitational wave quadrupole amplitudes
of the mass motions caused by prompt post-shock convection during the
first 100$\,$ms after core bounce (see upper lefthand panel of
Fig.~\ref{fig:graviwaves}).}
\label{fig:gwpromptconvection} 
\end{figure}

The matter signal consists of a superposition of
quasi-periodic variations
on timescales ranging from a few to several ten milliseconds.
Two phases of enhanced activity can be distinguished. The first one
occurs within tens of milliseconds after core bounce and
is followed by a more quiescent episode before a long-lasting period
with an overall trend in growing amplitudes begins. The transient, 
early phase of strong gravitational-wave emission is a consequence of 
the prompt post-shock convection (cf.\ Sect.~\ref{sec:shock}) and 
exhibits larger amplitudes, slightly higher frequencies (spectral
peak at about 100$\,$Hz instead of $\sim$70$\,$Hz), 
and more power in our simulation with the 
stiffer H\&W EoS (see Fig.~\ref{fig:gwpromptconvection}). The reason 
for these differences is the
fact that the region of prompt convection involves more mass and
extends to larger radii in the H\&W case (Figs.~\ref{fig:convectionregions} 
and \ref{fig:promptconvection}), where also the growth rates
of the instability are higher and the mass motions become more violent
and faster. The prominent peaks of the Fourier spectra  
in Fig.~\ref{fig:gwpromptconvection} correspond to a clearly dominant
frequency that can be identified in the wave train during the first 
$\sim$80$\,$ms after bounce.

The later phase of intense gravitational-wave production is associated
with the development of convective overturn and large-amplitude SASI
oscillations in the post-shock layer as well as mass flows and
convective activity in the 
proto-neutron star. Accordingly, the wave amplitudes at $t \ga 100\,$ms
show variability with different frequencies. It is interesting to note
that the enveloping curves of the wave amplitudes at this time
follow roughly the evolution of the maximum shock radii in  
Fig.~\ref{fig:shockradius}. This means that more active phases and
stronger gravitational-wave emission correlate with larger shock 
radii. The nonradial gas motions in the post-shock
region contribute mostly to the power radiated at
frequencies up to about 200$\,$Hz in accordance with the SASI
Fourier spectra shown in Fig.~\ref{fig:shockmodes}. The main peak
of the matter gravitational-wave spectrum at 300--600$\,$Hz (H\&W EoS)
and 600--800$\,$Hz (L\&S EoS) as well as most of the lower ``background''
above $f\sim 200\,$Hz, however, originates from changes
in the mass-quadrupole moment in deeper regions, namely at densities
between $10^{11}\,$g$\,$cm$^{-3}$ and $10^{13}\,$g$\,$cm$^{-3}$,
where more mass is involved than in the post-shock layer.
The main source of this signal are gravity modes and gas flows 
instigated by the violent impact of accretion funnels
in the surface layers of the nascent neutron star, which 
perturbs regions at densities up to
$\rho \sim 10^{12}\,$g$\,$cm$^{-3}$. A substantial part of the
emission also comes from the convective zone inside the neutron 
star at $\rho > 10^{12}\,$g$\,$cm$^{-3}$. The importance of the 
interaction of the SASI and accretion flows with the proto-neutron
star surface can be concluded from a closer inspection of
the upper lefthand panel of Fig.~\ref{fig:graviwaves} in comparison with 
Fig.~\ref{fig:convectionregions}. The gravitational-wave amplitudes
of the L\&S run at $t > 100\,$ms p.b.\ are significantly larger than 
those of the H\&W 
model until near the end of our simulations. Only then the amplifying
SASI and accretion activity begins to affect shells at
$\rho \sim 10^{12}\,$g$\,$cm$^{-3}$ also in the H\&W case,
while earlier the involved regions in this model possess clearly lower
densities.

The maximum values of the dimensionless gravitational-wave strain 
$h(t)$ for the matter signal are of the order of several $10^{-22}$
for a source at a distance of 10$\,$kpc, and the
dominant peak may be marginally observable with LIGO~I in the case
of a Galactic supernova (for a detailed discussion of measurement
aspects, see M\"uller et al.\ 2004). While the spectral peak is
slightly higher if the nuclear EoS is soft and the proto-neutron
star more compact, the stiffer H\&W EoS leads to somewhat lower
peak frequencies and thus to a shifting of the main emission closer
to the region of highest sensitivity of the LIGO instrument. It is
therefore not clear that a soft EoS is more favorable for a 
measurement of such gravitational-wave signals.

Significantly larger amplitudes are generated by the anisotropic
neutrino emission, however most power of this component of the
gravitational-wave signal comes at frequencies below 100$\,$Hz
(Fig.~\ref{fig:graviwaves}).
The lower lefthand panel of Fig.~\ref{fig:graviwaves}
shows negative amplitudes with the continuous trend in growing 
absolute values and superimposed high-frequency modulations for both
2D simulations discussed in this paper.
The negative values indicate a time-dependent quadrupole moment that
is determined by an excess of neutrino emission from regions near the
equatorial plane, a fact that can be concluded easily from the
latitudinal variation of the function $\Psi(\theta)$ in
the angular integral of the anisotropy parameter $\alpha_\nu(t)$ of
Eq.~(\ref{eq:alpha}). This function becomes negative in the equatorial 
belt between $60^\circ$ and $120^\circ$,
see Fig.~1 of Kotake et al.\ (2007).

A continuous decrease of the neutrino gravitational-wave amplitude to
negative values is no common feature of all 2D simulations. This
behavior depends strongly on the particular emission asymmetry that
develops within the nascent neutron star and in its accretion layer. 
M\"uller et al.\ (2004) found long-period variations between positive
and negative values, for example,
for an explosion simulation of an 11.2$\,M_\odot$
star (cf.\ their Fig.~4) and for a 2D neutrino-cooling simulation of a 
convective proto-neutron star (cf.\ their Fig.~5), and they obtained a 
positive gravitational-wave amplitude with an overall trend in
growth and only shorter periods of decrease for the neutrino emission
of a centrifugally flattened, accreting neutron star at the center
of a collapsing, rotating 15$\,M_\odot$ supernova progenitor (cf.\
their Fig.~3). 

This diversity of possible behaviors is in conflict
with results obtained by Kotake et al.\ (2007), who reported almost
monotonically rising, positive wave amplitudes for all of their
2D simulations of SASI unstable accretion shocks in collapsing
supernova cores. They explained their finding by the particular
properties of the SASI and the corresponding post-shock accretion
flow in models that are constrained by the assumption of axial
symmetry. In this case the SASI oscillations must occur along the
polar grid axis, and Kotake et al.\ (2007) argued that the influence
of this on the accretion flow makes the neutrino emission biased
towards the regions around the symmetry axis 
(see also Sect.~\ref{sec:neutrinos}), thus leading to positive
and growing wave amplitudes.

Why do we come to different conclusions about the neutrino 
emission asymmetry than Kotake et al.\ (2007) in spite of
our apparent agreement about the role of the SASI for the 
$\nu_e$ and $\bar\nu_e$ production in the accretion layer?
In the first place one must note that the modeling approach taken
by Kotake et al.\ (2007) is fundamentally different from ours. 
Besides many other differences, they, for example, did not simulate
the collapse of a stellar progenitor model but started from
a steady-state solution of an accretion shock with
a mass accretion rate of 1$\,M_\odot\,$s$^{-1}$; they
excised the neutron star interior at a constant radius chosen as
the inner boundary of the computational grid, where they assumed a
fixed density of $10^{11}\,$g$\,$cm$^{-3}$; and they kept the
fairly high
mass accretion rate and the accretor mass constant during all
of the simulation (time-independent boundary conditions at the
outer grid boundary), using a spherically symmetric Newtonian
gravitational potential and ignoring self-gravity of the matter
in the accretion flow. Moreover and very important in the context
of the present discussion of gravitational-wave signals associated
with anisotropic neutrino emission, they did not solve the 
energy-dependent, multi-flavor neutrino
transport problem but employed a light-bulb description with 
prescribed, time-independent neutrino luminosities and spectra
from the nascent neutron
star and local source rates for lepton number and energy due to
neutrino reactions in the accretion flow.

\begin{figure*}[!htb]
\begin{center}
\includegraphics[width=8.5cm]{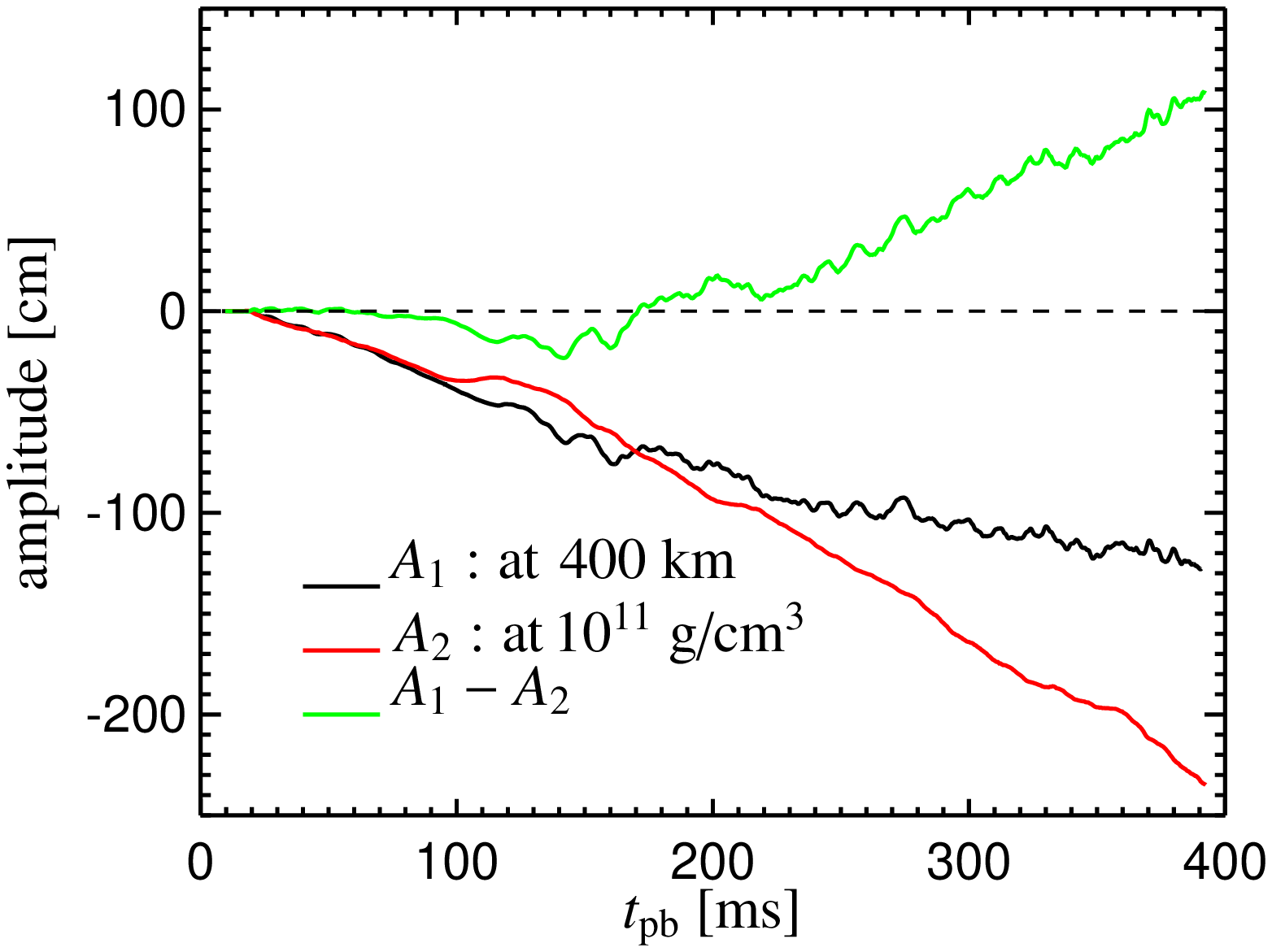}
\includegraphics[width=8.5cm]{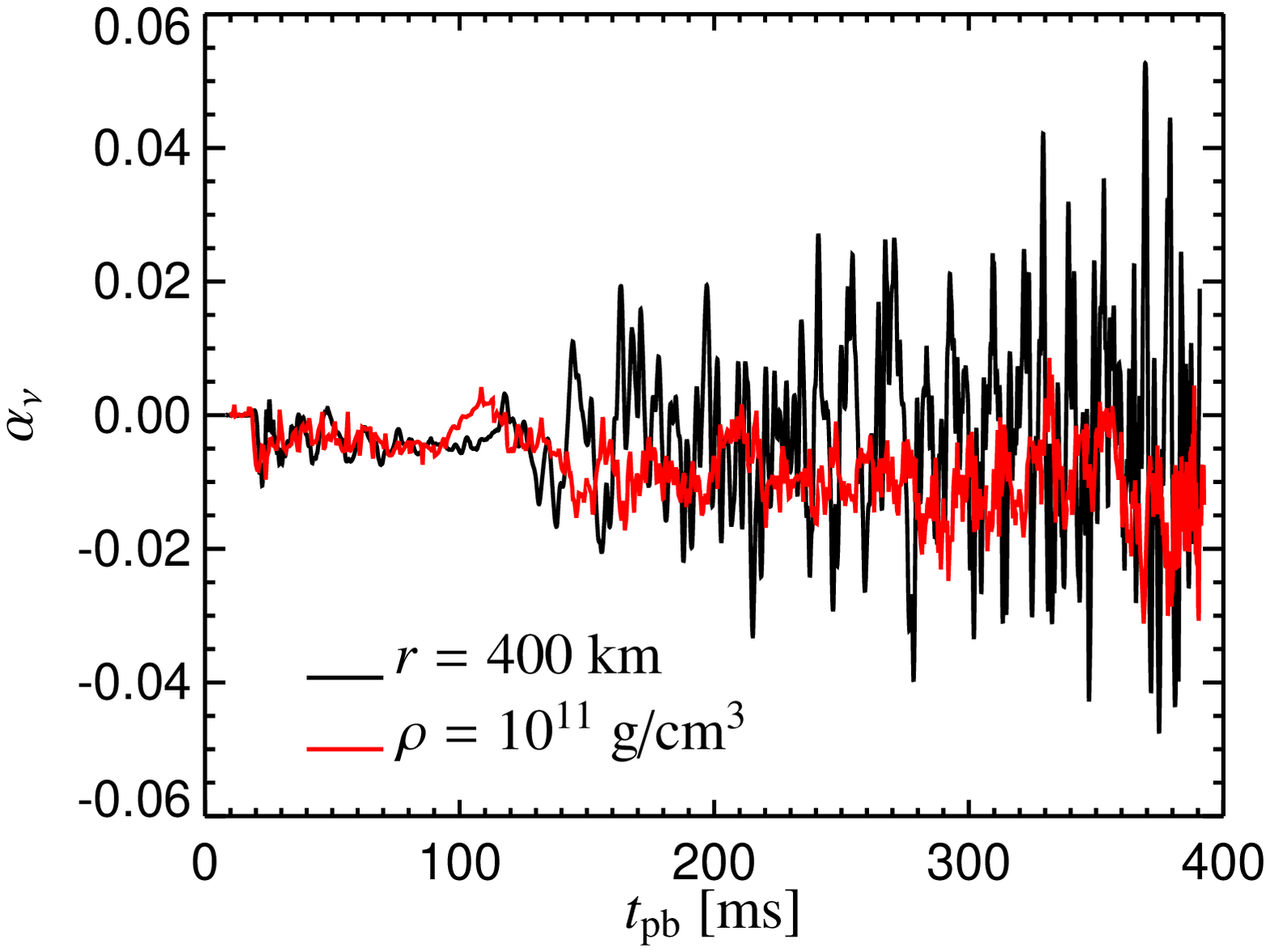}\\ \vspace{0.3cm}
\includegraphics[width=8.5cm]{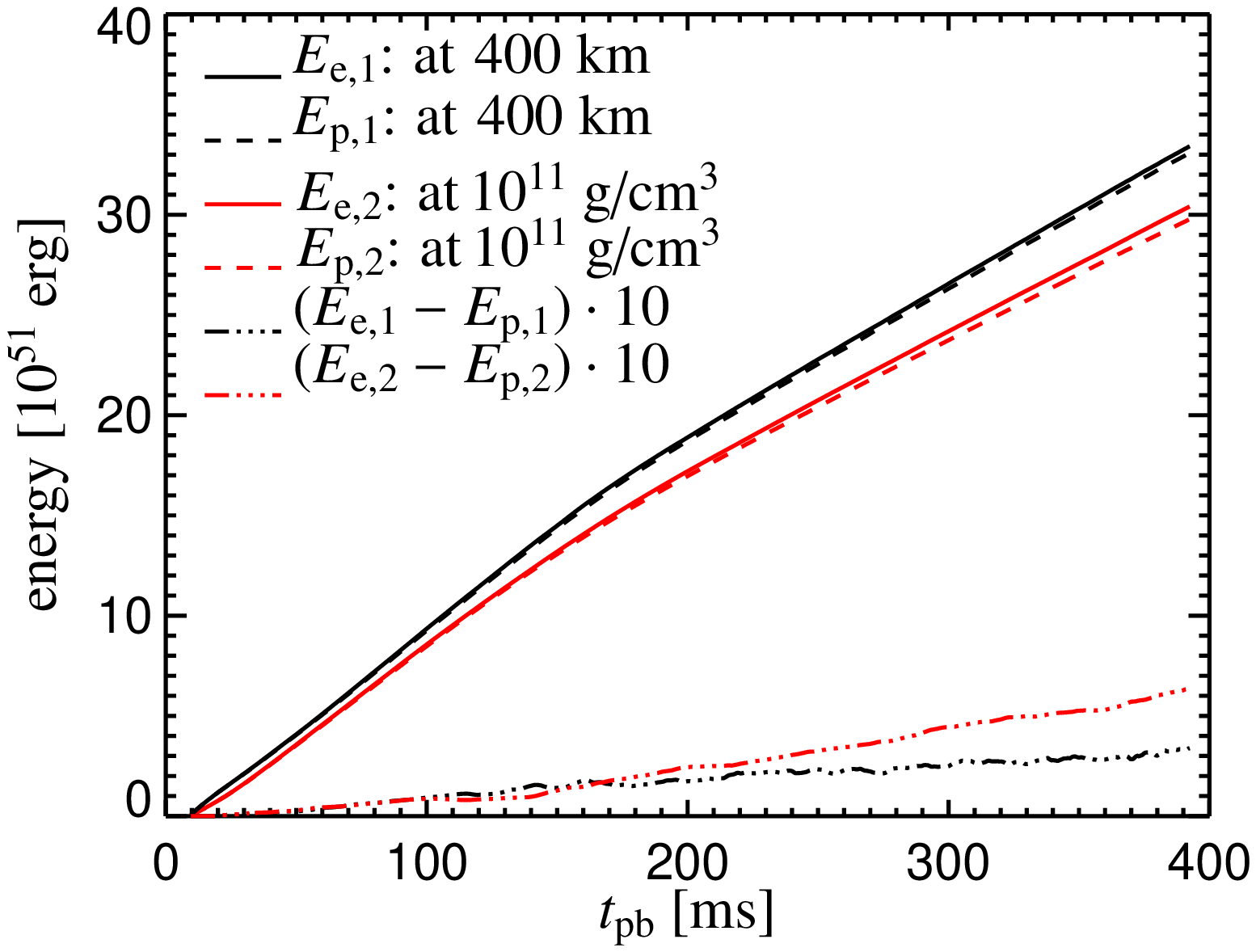}
\includegraphics[width=8.5cm]{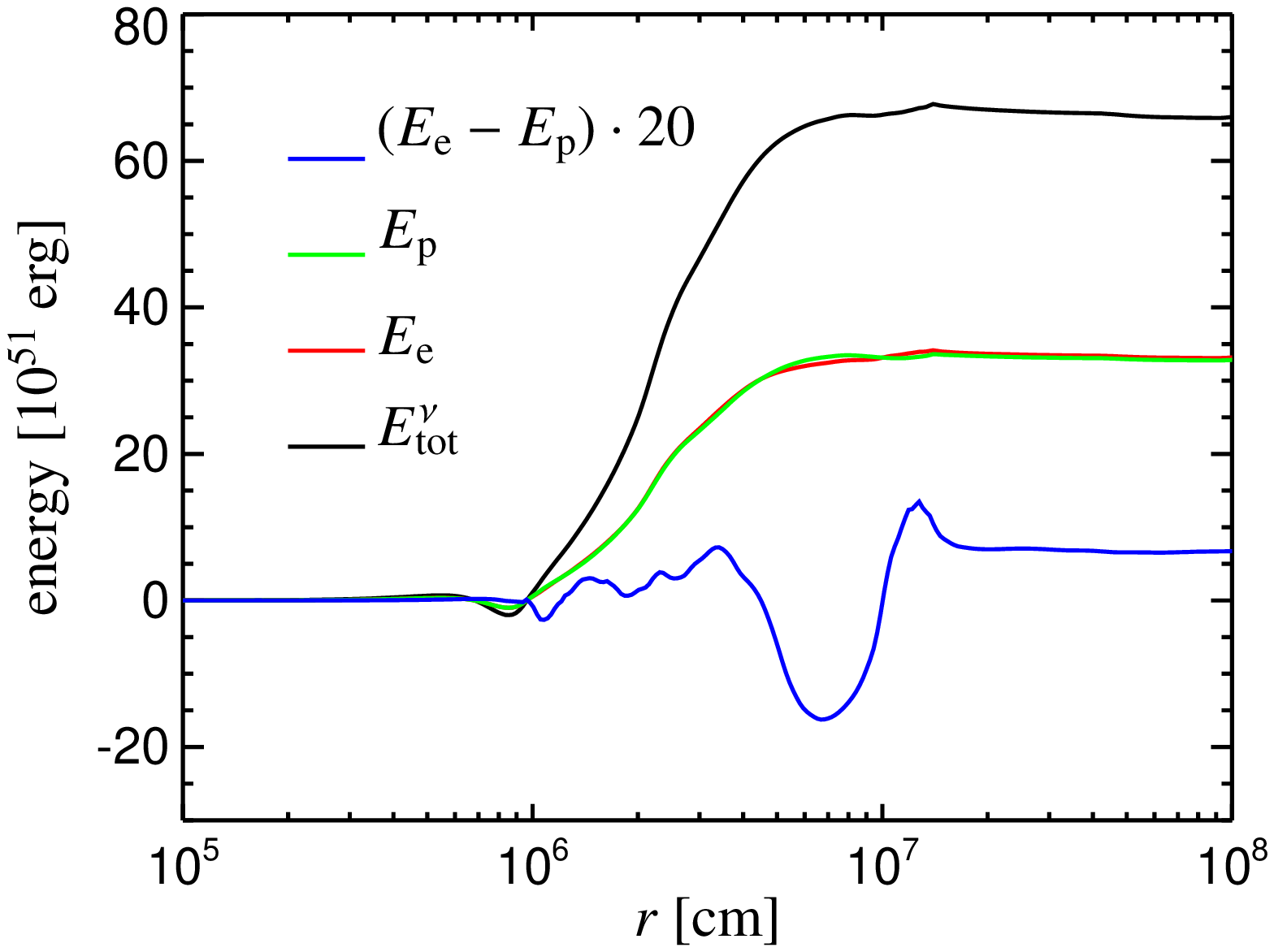}
\end{center}
\caption{Quantities that are relevant for the gravitational-wave
signal associated with the anisotropic neutrino emission in our 2D model
using the L\&S EoS.
{\em Top left:} Gravitational-wave amplitudes as functions of time
after core bounce. The black curve was obtained by considering the
neutrino radiation at a large distance from the source (identical with
the corresponding line in
the lower lefthand panel of Fig.~\ref{fig:graviwaves}, where, as in
Figs.~\ref{fig:luminosities} and \ref{fig:meanenergies}, the evaluation
was performed in the laboratory frame at a radius of 400$\,$km).
The red line shows the wave amplitude computed from the neutrino 
luminosities that leave the neutrinospheric region
(rather, a sphere where the density is $10^{11}\,$g$\,$cm$^{-3}$).
The green line displays the difference between the black and red curves.
{\em Top right:} Anisotropy parameter $\alpha_\nu$ of the neutrino 
emission as function of post-bounce time.
The color coding of the curves is the same as in the previous plot.
{\em Bottom left:} Energy radiated in all neutrinos through
the ``polar cap regions'' ($E_\mathrm{p}$) and through the
``equatorial belt'' ($E_\mathrm{e}$; see text for
details), and the difference in these results (scaled by a factor
of 10), integrated from the beginning of the 2D simulation shortly
after bounce until the time $t_\mathrm{pb}$ on the horizontal axis.
As in the two panels before, the color coding of the curves
corresponds to the two different locations where this analysis was
performed.
{\em Bottom right:} Total energy radiated in neutrinos through
the ``polar cap regions'' ($E_\mathrm{p}$), through the
``equatorial belt'' ($E_\mathrm{e}$), difference in both
scaled by a factor of 20, and sum of both ($E_\mathrm{tot}^\nu$)
as functions of radius. The time integration was performed from the
beginning of the 2D simulation shortly after bounce until the end
of the model run at about 400$\,$ms after bounce.
}
\label{fig:gwneutrinos}
\end{figure*}

Based on the possibilities available to us for comparison
we cannot make a reasonable judgement whether any, and
if so, which of these simplifications are causal for the discrepant
gravitational-wave results obtained by us and by Kotake
et al.\ (2007). However, we strongly suspect 
that the approach of the latter work is too simplistic for reliably 
estimating the emission asymmetry of neutrinos that is responsible
for the gravitational-wave generation. This conclusion is supported
by a close analysis of our neutrino transport data and the insight 
they yield into the formation of the neutrino emission asymmetry.  
Figure~\ref{fig:gwneutrinos} provides the corresponding information
for our 2D simulation with the L\&S EoS.

In the upper lefthand panel of Fig.~\ref{fig:gwneutrinos} we display 
the time evolution of the neutrino gravitational-wave amplitude for
this model. Different from
Fig.~\ref{fig:graviwaves} we have performed the evaluation now
not only at a radius of 400$\,$km, where the neutrino
luminosity has reached its asymptotic level, but also at the 
(time-dependent) radius where the density has a value of
$10^{11}\,$g$\,$cm$^{-3}$, i.e.\ in the vicinity of the
electron neutrinosphere and thus below the accretion layer where
neutrino cooling still enhances the outgoing $\nu_e$ and $\bar\nu_e$
fluxes. The upper righthand panel of Fig.~\ref{fig:gwneutrinos}
contains the corresponding parameters $\alpha_\nu(t)$ of the
neutrino emission anisotropy, which exhibit rapid variations 
between positive and negative values on timescales of milliseconds
to $\ga$10$\,$ms and growing amplitudes at later times.
Both upper panels show an even clearer long-time
trend to negative gravitational-wave amplitudes for the neutrino
luminosity leaving the neutrinospheric layer: $\alpha_\nu(t)$ 
measured at $10^{11}\,$g$\,$cm$^{-3}$ attains positive values only 
during the very short time intervals when the high-frequency 
modulations reach their largest amplitudes,
while this parameter at a radius of 400$\,$km enters the positive
side for significantly more numerous and longer time intervals,
although $\int_0^t{\mathrm{d}}t'\alpha_\nu(t') < 0$ for all times
at both locations.
Correspondingly, the gravitational-wave amplitude at
$\rho = 10^{11}\,$g$\,$cm$^{-3}$ is even more negative than the one
at $r = 400\,$km, and their difference (green line in the upper 
lefthand
panel of Fig.~\ref{fig:gwneutrinos}) suggests that the amplitude
associated with the neutrino emission in the accretion layer is
positive and reduces the neutrinospheric emission asymmetry. 
This seems to mean that the final emission anisotropy of the neutrinos
is determined already by the flux that leaks out from the 
neutrinospheric region and in which the equatorial region has
obtained a slight flux excess. Below
we will see, however, that this interpretation is not completely
correct and the true situation is more complex. A comparion of the
different lines in the upper lefthand panel of Fig.~\ref{fig:gwneutrinos}
also reveals that the fluctuations of the radiated neutrino
gravitational-wave signal with a period of 10--20$\,$ms 
(frequency $\sim$50--100$\,$Hz)
originate from the accretion layer and are therefore connected 
to the SASI and convective activity in the post-shock region and
its influence on the neutrino-cooling zone as described in 
Sect.~\ref{sec:neutrinos}.

\subsubsection{Neutrino emission anisotropy}
\label{sec:neutrinoasymmetry}

In order to obtain deeper insight into the origin of the neutrino
emission anisotropy, we have calculated the cumulative loss of 
energy in neutrinos (i.e., the time- and surface-integrated 
neutrino flux) once for the polar cap regions and another time
for the equatorial belt, again at $r = 400\,$km and at 
$\rho = 10^{11}\,$g$\,$cm$^{-3}$. The results are shown in
the lower lefthand panel of Fig.~\ref{fig:gwneutrinos}. The small
differences $E_\mathrm{e}-E_\mathrm{p}$ between equatorial and
polar neutrino energy losses, scaled by a factor of 10, are 
plotted with
dash-dotted lines. Again we see that at large distances as 
well as near the neutrinospheres the equatorial emission 
exceeds slightly the polar one during the whole evolution after
$t\sim 40\,$ms post bounce, and
this excess is reduced by the neutrino production that occurs
in the cooling layer outside of a density of 
$10^{11}\,$g$\,$cm$^{-3}$. Separating the analysis for the
different neutrino kinds, it becomes obvious that only the 
$\nu_e$ and $\bar\nu_e$ luminosities grow in the cooling layer
above the neutrinospheres, and that these neutrinos are more
abundantly produced in the polar regions where the loss of accretion
energy mainly occurs, as discussed in Sect.~\ref{sec:neutrinos}. 
For electron neutrinos the emission in the accretion 
layer near the poles is so strong that their polar energy
loss at large distances can become bigger than the equatorial
one. In contrast, the flux of heavy-lepton neutrinos, which 
are radiated from the neutrinospheric region with a persistent 
surplus of the equatorial energy loss, gets a bit reduced in
the accretion and postshock layers between 
$\rho = 10^{11}\,$g$\,$cm$^{-3}$ and $r = 400\,$km, whereby
their equatorial emission excess grows to roughly twice the
value near the neutrinospheres. Their asymmetry
determines the final emission asymmetry, because they contribute
the major part ($\sim$56\%) of the total neutrino energy loss
(in fact, the pole-equator asymmetries of $\nu_e$ and $\bar\nu_e$ 
at large distances have opposite signs and nearly cancel
each other). These findings are in clear contradiction to the
results obtained by Kotake et al.\ (2007), who reported that the
neutrino gravitational-wave signals in their models were dominated 
by the $\nu_e$ emission and the other neutrinos did only contribute
very small amplitudes with the same signs (see their Fig.~3).

What is the reason that slightly (2--3\%) more energy is
radiated from the neutrinospheric layer near the equator than 
near the poles? We find 
that the emergence of this asymmetry coincides
with the development of the stable convective shell inside 
the neutron star. This might suggest that convective transport
near the equatorial plane is slightly more effective than in
the polar regions. We have no satisfactory explanation for
this phenomenon, because one would expect convection to 
operate similarly in all directions and not to produce a global
asymmetry on angular scales much larger than the size of the 
(nonstationary) convective cells. This, however, is a statement
that is correct in three dimensions but might not hold for convection 
in the 2D environment of our simulations. Here convective plumes
and downflows are toroidal features because of the axial 
symmetry, and the polar grid has a preferred direction along
the polar axis. We suspect --- without having performed a detailed
analysis --- that the operation of convective overturn with cells 
of such a toroidal structure near
and along the polar axis might be slightly different from the
in- and outflows near the equatorial plane with fluid motions
perpendicular to the
axis. If this were indeed the origin of the neutrino emission 
asymmetry, it would imply a serious warning that the computation 
of such anisotropies for estimating gravitational-wave signals (and
also pulsar kicks based on the recoil associated with anisotropic
neutrino emission) requires 3D modeling, because in 2D the discussed
effects may be just a consequence of the symmetry constraints.

But is this fishy directional asymmetry really responsible for 
our prediction of the neutrino gravitational-wave signal? Fortunately,
it is not! To see this, we have evaluated our results in yet another
way. In the lower righthand panel of Fig.~\ref{fig:gwneutrinos} we 
display the cumulative --- i.e.\ time-integrated from the beginning
until the end of our 2D simulation --- neutrino energy loss (again
summed up for neutrinos and antineutrinos of all flavors) as a
function of radius for the polar, equatorial, and total emission.
The fourth (blue) curve represents the difference ($E_\mathrm{e}-
E_\mathrm{p}$) between equatorial
and polar terms, scaled by a factor of 20. Consistent with what
we discussed above there is an excess of the equatorial losses until
about 45$\,$km. This radius encompasses well the convection zone
inside of the nascent neutron star during the whole post-bounce
evolution of the 2D model with L\&S EoS 
(Fig.~\ref{fig:convectionregions}) and is larger than the location
where $\rho = 10^{11}\,$g$\,$cm$^{-3}$ for post-bounce times 
$t \ga 170\,$ms (lefthand panel of Fig.~\ref{fig:shockradius}).
At greater radii when one enters the accretion and SASI layers, 
however, the polar energy loss first dominates before the equatorial
one wins again at $r \ga 100\,$km. After a local maximum at a 
radius slightly larger than 100$\,$km the equator-to-pole 
asymmetry settles to its final value. This radial variation is
found to be qualitatively very similar for all kinds of neutrinos,
although with different values of the local maxima and minima and
of the final, asymptotic level at large radii.

From the lower righthand panel of
Fig.~\ref{fig:gwneutrinos} we learn that the emission anisotropy
carried by the outgoing neutrinos is not imposed near the 
neutrinospheres and preserved outside, but it is completely determined 
by the accretion region behind the supernova shock and the SASI and 
convective inhomogeneities in this region. The final emission anisotropy
just by chance gets close to the value near the neutrinospheres.
Moreover, the separate analysis for the different kinds of neutrinos
reveals that it is the muon and tau neutrinos that govern
the final anisotropy of the radiated energy flux. In the two
models presented here the energy carried away by these neutrinos
in the equatorial belt is slightly
higher than the one escaping from the polar cap areas.
This as well as the total asymmetry, however, is likely to be a 
time- and model-dependent result and to
depend strongly on the detailed structure of the 
inhomogeneities that evolve behind the deformed accretion shock.
It could also depend 
on other model aspects such as the value of the mass accretion 
rate, the direction-dependent stagnation radius of the shock, 
or the relative sizes of the fluxes of the different neutrinos.
Such conjectures are not only supported by the H\&W model in 
comparison to the L\&S run: the neutrino gravitational-wave 
amplitude of the latter model exhibits a significantly clearer trend 
in a monotonic decrease to negative values, while in the H\&W case
one can see long phases where the amplitude or its time derivative
become positive (Fig.~\ref{fig:graviwaves}).
These conjectures are also in agreement with the results of other
models described in the M\"uller et al.\ (2004) paper, and, last
but not least, they are suggested by the
fact that the quantity $E_\mathrm{e}-E_\mathrm{p}$ in the lower
righthand panel of Fig.~\ref{fig:gwneutrinos} is a quantity with 
large radial variations in both positive and negative directions.
It is rather unlikely that such a complex radial behavior can
be generic for models in two dimensions so that in all cases and
at all times the same effects are recovered that ultimately determine
the emission asymmetry in our models, namely (i) a delicate 
almost-cancellation of the combined asymmetry of $\nu_e$ and
$\bar\nu_e$ due to the production and absorption processes of these
neutrinos in the neutrino-cooling and heating regions, and (ii) a 
final dominance of the anisotropy of the radiated muon and tau 
neutrinos as a consequence of their asymmetric transport through
the semi-transparent accretion layer, in which the total heavy-lepton
neutrino luminosity hardly changes but the equator-to-pole 
asymmetry grows.

Our analysis therefore reveals that the gravitational-wave results
for neutrinos depend on physical effects that are either
not included or treated in a much simplified manner in previous
works. It further reveals that a reliable
and quantitatively meaningful computation of the tiny neutrino
emission asymmetry is extremely difficult and does not only
require a full transport treatment and good numerical resolution
in all relevant regions, but ultimately must be done on the basis
of 3D models.

\section{Summary and conclusions}
\label{sec:conclusions}

We have presented results of two stellar core-collapse simulations
performed with the {\sc Prometheus-Vertex} code for a nonrotating
15$\,M_\odot$ progenitor star in axial symmetry. One of these models
was computed with a soft version of the Lattimer \& Swesty (1991)
nuclear equation of state, and the other one with the stiffer 
Hillebrandt \& Wolff (1985) EoS, which produces a significantly 
less compact proto-neutron star. For these simulations we have 
compared the neutrino and gravitational-wave signals between core
bounce and 400$\,$ms later. In particular we were interested in
signal features that may yield information about the action of the
standing accretion shock instability (SASI) in the supernova core
during the long phase of post-bounce accretion that precedes the
onset of the supernova explosion in all recent multi-dimensional
calculations of stellar core collapse.

Our main results can be summarized as follows:
\begin{itemize}
\item[(1)]
In both simulations we found that the weakened bounce shock 
produces a negative entropy gradient that partly overlaps with
the negative lepton number gradient left behind by the 
escaping shock breakout burst of electron neutrinos. 
This region is Ledoux unstable and develops prompt post-shock 
convection, which leads to an early burst of gravitational-wave
emission lasting several ten milliseconds after core bounce.
Its frequency spectrum shows a prominent peak around 70--100$\,$Hz.
\item[(2)]
A later, long-lasting phase of strong gravitational-wave emission 
with a long-time trend in growing wave amplitudes sets in at 
$t \ga 100\,$ms after bounce. It is produced by nonradial mass
motions and anisotropic neutrino emission that are caused by
convective overturn inside the proto-neutron star and in the 
neutrino-heating layer behind the stalled shock, and in particular
by vigorous sloshing of the accretion shock due to the development
of low-$\ell$ mode SASI activity. The wave train for the matter signal
consists of a superposition of quasi-periodic variations on 
timescales from a few milliseconds to several ten milliseconds.
The spectra of the dominant $\ell = 1,\,$2 SASI modes peak between
about 20 and roughly 100$\,$Hz,
and the gravitational-wave emission also shows significant
power at frequencies up to $\sim$200$\,$Hz. 
The most prominent maxima of the gravitational-wave spectra, however,
are located at higher frequencies of 300--800$\,$Hz and originate
mainly from the outer layers of the neutron star where rapid gas
flows are triggered by the supersonic impact and deceleration of 
accretion funnels.
\item[(3)]
The high-frequency peak of the gravitational-wave spectrum and
less clearly also the secondary maximum near the SASI frequency
exhibit a dependence on the 
compactness of the nascent neutron star and thus on the properties
of the high-density EoS: the more compact neutron star leads to more
powerful shock oscillations and significantly larger gravitational-wave 
amplitudes earlier after core bounce. Also the characteristic frequencies
of the low-$\ell$ SASI modes and of the gravitational-wave signal 
are higher. While the main
peak of the wave spectrum is located at 300--600$\,$Hz for the stiffer 
EoS, it can be found at 600--800$\,$Hz for the softer EoS with the 
more compact remnant.
\item[(4)]
The more compact neutron star also radiates neutrinos with higher 
luminosities and greater mean energies. Anisotropic neutrino emission
produces a low-frequency component of the gravitational-wave signal 
at $f \la 100$--200$\,$Hz with an amplitude that dominates the matter
signal by up to a factor of two. The emission anisotropy is mostly
established in the cooling and heating regions of the accretion
layer outside of the
neutrinospheres and is determined by the anisotropic transport 
of muon and tau neutrinos through this layer, which leads
to a slightly higher flux of muon and tau neutrinos near the 
equatorial plane of the polar grid, i.e., perpendicular to the 
direction of the SASI oscillations of the supernova shock.
Despite the stronger emission of $\nu_e$ and $\bar\nu_e$ from the 
accretion layer in the polar regions (i.e.\ in the direction of the 
SASI shock expansion and contraction), the gravitational-wave 
amplitude associated with the anisotropic energy loss in neutrinos
exhibits a nearly monotonic trend to negative values.
\item[(5)]
The neutrino luminosities and mean energies, in particular the
ones that can be received by an observer in the polar (and thus
SASI) direction, show a quasi-periodic time variability with an 
amplitude of several ten percent (up to about 50\%) of the minimum
values for the luminosities and of roughly 1$\,$MeV (up to 10\%)
for the mean energies. The luminosity fluctuations are correlated
and in phase with the energy variations and are somewhat bigger
for $\nu_e$ and $\bar\nu_e$ than for heavy-lepton neutrinos. They
originate from alternating phases of gas accumulation and compression
in the accretion layer in particular around the poles, caused by 
the expansion and contraction of the shock in the course of the 
SASI oscillations. 
The frequency spectra of the neutrino luminosities
show most power between about 20 and 200$\,$Hz and significant
power in a decaying tail up to about 400$\,$Hz as a consequence
of emission variability associated with faster, convective 
modulations of the accretion flow between supernova shock and
nascent neutron star. Since SASI and convective activity occur
inseparably in the gain layer and stir the cooling layer below, 
power is found to be distributed over a wide range of frequencies 
in the Fourier spectra of the luminosities.
\end{itemize}
 
The SASI induced temporal modulations of the luminosities and
mean energies are features by which neutrinos can provide direct
evidence of the dynamical processes that occur in the supernova
core. Only few other cases are known where neutrinos carry such
information, for example the prompt flash of electron neutrinos
as a signature of the shock breakout from the neutrinosphere, or
a possible second neutrino burst or abrupt termination of the 
neutrino emission, which are expected if the neutron star collapses
to a more compact object or a black hole.
A detection of the predicted characteristic, quasi-periodic 
variation in the neutrino event rate
from a Galactic supernova would confirm the action of the SASI
in the supernova core and would thus provide extremely valuable 
support of our present understanding of the
core-collapse physics: the SASI is discussed as
possible origin of global supernova asymmetries and pulsar kicks
and as a potentially crucial agent on the route to the 
explosion.

Moreover, nonradial mass motions such as convection and the SASI
in the supernova core after bounce produce gravitational-wave 
emission with significant intensity. We found that the wave
spectrum exhibits a broad hump between about 20 and 200$\,$Hz 
and a main maximum with 
clearly more power at considerably higher frequencies. The
latter is associated with SASI and accretion induced gas flows
in the neutron star surface layers, where much more
mass is involved than in the fairly dilute post-shock region.
Both the primary and the secondary maximum of the spectrum 
are sensitive to the structure of the neutron star: larger
gravitational-wave amplitudes and correspondingly higher 
spectral maxima as well as higher frequencies 
are obtained with a softer equation of state and
more compact remnant. The relative SASI variations of the 
neutrino luminosities do not appear to be very 
sensitive to the nuclear equation of state, and a clear
EoS-dependent difference in the modulation frequencies is
also not visible, but the absolute values of the luminosities
and mean energies of the neutrinos radiated by the more compact 
and hotter, accreting proto-neutron star in
our simulations are 10--20\% higher.

Anisotropic neutrino emission is also an important source of
gravitational waves with amplitudes that can dominate
the matter signal. The corresponding spectrum
possesses most power at frequencies below 100--200$\,$Hz.
The emission anisotropy is a function with rapid short-time
variability (with periods of milliseconds to $\ga$10$\,$ms)
and with an overlying, nearly monotonic long-time trend toward
slightly stronger emission from equatorial regions, i.e.,
perpendicular to the direction of the SASI shock oscillations. 
This trend decides about the sign of the neutrino 
gravitational-wave amplitude. 
A detailed analysis of our models reveals that it is
the result of complex transport physics
in which the neutrinospheric flux, which has a low
pole-to-equator anisotropy, is modified strongly in the 
inhomogeneous environment of the accretion layer
around the forming neutron star. The
pole-to-equator asymmetry changes with radius several times
in both directions and the 
contributions from $\nu_e$, $\bar\nu_e$, and heavy-lepton 
neutrinos partly cancel each other. A reasonable
determination of these transport asymmetries, which is not
only important for gravitational-wave calculations but also
for pulsar kick estimates based on neutrino recoil, requires 
detailed transport calculations. Even our sophisticated
ray-by-ray approach may not be sufficient for accurate 
predictions of measurable effects, but a full multi-angle
treatment of the transport may be necessary. 

From this discussion it is clear that any reliable assessment
of the detectability of SASI effects in the 
neutrino and gravitational-wave emission
will ultimately require results from three-dimensional models.
The growth of SASI modes with nonvanishing azimuthal order
$m$ was shown to be of 
potential importance, and the presence of such nonaxisymmetric 
asymmetries can have a big influence on the structure and the
dynamics of the accretion layer 
(Blondin \& Mezzacappa 2007, Yamasaki \& Foglizzo 2008, 
Iwakami et al.\ 2008). It is probable that also the emission
asymmetries and time modulations of the emitted signals 
exhibit significant differences compared to 2D models,
in particular because in 3D the SASI axis is not forced to
coincide with an axis of symmetry and with the axis of the
numerical grid.

The SASI and convective activity, its duration as well as
strength, must be expected to vary with the progenitor star
and to depend on the length of the postbounce accretion phase 
until the onset of the explosion (for hints to that, see the
results for 11.2$\,M_\odot$ and 15$\,M_\odot$ explosions in
Marek \& Janka 2007). In the nonlinear stage, the SASI and
convective motions are of chaotic nature and therefore
stochastic differences also in the large-scale pattern and
global asymmetry are
likely to develop from small initial differences even in 
progenitor stars (models) of the same kind (see, e.g., the
different explosion asymmetries obtained in a large set of 2D
simulations performed by Scheck et al.\ 2006). Nevertheless, 
basic features of the instabilites, in particular their mere 
presence and long-time growth behavior and the sizable 
modulation they impose on the
emitted signals over a period of possibly hundreds of 
milliseconds, should be common to different realizations.
Judged from our present understanding, however, a
template-like uniformity of the signal structure with a 
well defined frequency (or frequency range) and amplitude
that exhibit a regular dependence on the explosion 
and progenitor properties, appears not to be very likely. More
model calculations are therefore needed to find out which 
signal properties such as the modulation frequency,
are sufficiently generic to yield information about the
neutron star equation of state.

\subsection{Detectability of SASI modulations}

Should the SASI signatures and their magnitude discussed in this
work be confirmed by 3D results, the question arises whether 
they will be measurable with existing or emerging neutrino and 
gravitational-wave experiments in the case of the next Galactic 
supernova. A detailed prediction of supernova neutrino
signals in terrestrial detectors is a highly complex problem,
because the neutrinos radiated by the nascent neutron star
undergo flavor conversions so that the flavor arriving at 
the Earth is not necessarily the same as the one leaving the 
source. On the one hand, the flavor evolution is affected by
neutrino-matter oscillations during the neutrino propagation 
through the density gradients in the supernova mantle and 
envelope (where the H and L resonances are encountered at 
densities around 10$^3\,$g$\,$cm$^{-3}$ and 
10$\,$g$\,$cm$^{-3}$, respectively) and through the Earth. 
On the other hand, neutrino
conversion is also induced by the nonlinear self-interaction
associated with neutrino-neutrino forward scattering in the dense
neutrino background close to the neutron star (for an application
of these phenomena in the context of the neutrino signals from 
oxygen-neon-magnesium core supernovae, see Lunardini et al.\ 2008,
and from black hole formation in stellar collapse events, see
Nakazato et al. 2008; for a discussion of 
the flavor conversion physics, see the references to original
publications in the latter two papers). The flavor
composition of the neutrino burst in the detector and thus the 
detector response therefore do not only depend on the type and 
structure of the progenitor star but also on a number of degrees 
of freedom associated with unknown neutrino properties and the
location of the experiment, e.g.\ the
neutrino-mass hierarchy (normal or inverted), the still
undetermined mixing angle between the mass eigenstates 1 and 3,
and the nadir angle of the neutrino path through the Earth.

Because of the involved complexity we defer an investigation
of the flavor conversion effects and their consequences for the
neutrino measurements to a future paper. However, since the SASI
modulations affect the luminosities and mean energies of neutrinos
and antineutrinos of all flavors with similar strength, we can
ignore the flavor oscillation effects for a simple 
estimate to convince us that the SASI modulations of the radiated 
signal should be detectable in principle. Super-Kamiokande, for 
example, is expected to capture roughly 8000 electron antineutrinos
through their absorption by protons, if a Galactic supernova 
occurs at a distance of 10$\,$kpc and releases at total energy
of $5\times 10^{52}\,$erg in $\bar\nu_e$ with an average energy
of 15$\,$MeV. A luminosity of $4\times 10^{52}\,$erg$\,$s$^{-1}$,
which is a typical value for the emission maxima during the 
SASI phase, should account for an event rate of approximately 
7000 $\bar\nu_e$ captures per second, corresponding to about 70 
events in the $\sim$10$\,$ms intervals of peak emission.
If $L_\nu\langle\epsilon_\nu^2\rangle
\langle\epsilon_\nu\rangle^{-1}$ is about 50\%  
lower in luminosity minima, about one half of this event rate
and event number can be expected during the low-emission periods. 
Therefore the SASI modulations will lead to a variability
of the measured signal that can be larger, though not much, than
the statistical $\sqrt{N_{\nu,{\mathrm{det}}}}$ fluctuations, 
$N_{\nu,{\mathrm{det}}}$ being the number of 
detected neutrinos in a time interval of 10$\,$ms. 
For Hyper-Kamiokande, whose fiducial 
volume would be roughly 30 times bigger than that of 
Super-Kamiokande, the Poisson noise would be reduced to a few
percent of the event number within 10$\,$ms, so that the 
SASI modulations should create a clear
signature. Even more powerful for this purpose is the
IceCube experiment. It will be able to identify the supernova
neutrino burst by the Cherenkov glow caused 
by the neutrino energy deposition in the ice, which will show
up as time-correlated noise in all phototubes. For a supernova
at a distance of 10$\,$kpc, the $\bar\nu_e$ luminosity maxima
during the SASI phase are expected to cause a photon count
rate of more than $10^6$ per second in the 4800 optical 
modules (see Dighe et al.\ 2003). 
The $\sim$$1.2\times 10^4$ events within 10$\,$ms intervals and
the variations of several 1000 events between periods of 
luminosity peaks and minima (in the case of IceCube
these variations
scale with $L_\nu\langle\epsilon_\nu^3\rangle
\langle\epsilon_\nu\rangle^{-1}$, because the crucial quantity
is the energy deposition in the detector) 
dominate by far the Poisson fluctuations associated
with the background counting rate of 300$\,$Hz per optical
module; the Poisson variability of the background 
corresponds to $\sqrt{N_\gamma}\sim 120$ photon events in
all phototubes, which is $\sim$1\% of the supernova signal in
a 10$\,$ms interval (an impression of the detection capability
even for a supernova model with a lower fluctuation 
amplitude of the neutrino emission than in the models 
discussed here, can be obtained from
the upper lefthand panel of Fig.~2 in Dighe et al.\ 2003). 
 
The detectability of gravitational-wave signals very similar to
the ones presented in this work was investigated in much 
detail by M\"uller et al.\ (2004). Due to the restriction
to a 90-degree grid (one hemisphere) in the case of the 
15$\,M_\odot$ core-collapse model s15r considered in the 
latter paper, the contributions of 
dipolar SASI modes were not included in there. 
Moreover, since the collapse evolution was followed to later 
times in our most recent simulations (about 400$\,$ms instead 
of $\sim$270$\,$ms after bounce) and the quadrupole amplitudes
exhibit a trend in rise during the later post-bounce phases,
the signal-to-noise (S/N) ratios and maximum detection distances
given for LIGO~I and Advanced LIGO by M\"uller et al.\ (2004) 
are likely to be only lower limits. Since the S/N ratio scales
roughly linearly with the Fourier transform of the wave 
amplitude, the detection prospects grow with longer integration
periods of the signal and thus higher Fourier power. The
measurement of dominant gravitational-wave power at frequencies
of several hundred Hz with growing strength over a period of hundreds
of milliseconds would therefore be a strong indication of SASI and
convective activity around the nascent neutron star. M\"uller 
et al.\ (2004) estimated a S/N ratio of roughly 70 in Advanced 
LIGO for a 15$\,M_\odot$ supernova at 10$\,$kpc, corresponding 
to the possibility of capturing gravitational waves from such 
sources up to a distance of roughly 100$\,$kpc. A third-generation
underground interferometer facility (Einstein Telescope) would
lead to a gain in sensitivity of approximately a factor of 20
in the most relevant frequency range between 100$\,$Hz 
and 1000$\,$Hz (Spiering 2007). This would increase the 
expected detection rate of supernovae by up to a factor 
of 3 (Ando et al.\ 2005). A detailed analysis of the 
gravitational-wave signals and their observational implications
from a greater sample of core-collapse models will be presented 
in a forthcoming paper.

\begin{acknowledgements}
We thank C.D.~Ott and G.G.~Raffelt for useful discussions.
The project was supported by the Deutsche Forschungsgemeinschaft
through the Transregional Collaborative Research Centers SFB/TR~27
``Neutrinos and Beyond'' and SFB/TR~7 ``Gravitational Wave Astronomy'',
and the Cluster of Excellence EXC~153 ``Origin and Structure of the Universe''
({\tt http://www.universe-cluster.de}). 
Computer time grants at the John von Neumann Institute for
Computing (NIC) in J\"ulich, the H\"ochst\-leistungs\-re\-chen\-zentrum
of the Stuttgart University (HLRS) under grant number SuperN/12758, 
the Leib\-niz-Re\-chen\-zentrum
M\"unchen, and the RZG in Garching are acknowledged.
\end{acknowledgements}

{\small
\bibliographystyle{aa}

\begin{thebibliography}{38}
\expandafter\ifx\csname natexlab\endcsname\relax\def\natexlab#1{#1}\fi

\bibitem[{}()]{alex88}
Alexeyev, E.N., Alexeyeva, L.N., Krivosheina, I.V., \& Volchenko, V.I.
1988, Phys.~Lett., B205, 209

\bibitem[{}()]{ando05}
Ando, S., Beacom, J.F., \& Y\"uksel, H. 2005, \prl, 95, 171101

\bibitem[{}()]{bion87}
Bionta, R.M., et al. 1987, \prl, 58, 1494 (IMB collaboration)

\bibitem[{}()]{blanchet90}
Blanchet, L., Damour, T., \& Sch\"afer, G. 1990, \mnras, 242, 289

\bibitem[{}()]{blondin06}
Blondin, J.M. \& Mezzacappa, A. 2006, \apj, 642, 401

\bibitem[{}()]{blondin07}
Blondin, J.M. \& Mezzacappa, A. 2007, Nature, 445, 58

\bibitem[{}()]{blondin03}
Blondin, J.M., Mezzacappa, A., \& DeMarino, C. 2003, \apj, 584, 971 

\bibitem[{}()]{Bruenn.etal.2006}
Bruenn, S.W., Dirk, C.J., Mezzacappa, A., Hayes, J.C., Blondin, J.M.,
Hix, W.R., \& Messer, O.E.B. 2006,
in: SciDAC 2006, Scientific Discovery through Advanced Computing, Denver,
Colorado, USA, 25--29 June 2006, Eds.\ W.M. Tang, et al.,
Journ.\ Phys.\ Conf.\ Ser., 46, p. 393; arXiv0709.0537

\bibitem[{}()]{bura06a}
Buras, R., Rampp, M., Janka, H.-Th., \& Kifonidis, K., 2006a, \aap,
447, 1049

\bibitem[{}()]{bura06b}
Buras, R., Janka, H.-Th., Rampp, M., \& Kifonidis, K., 2006b, \aap,
457, 281

\bibitem[{}()]{burrows87}
Burrows, A. 1987, \apj, 318, L57

\bibitem[{}()]{burrows88}
Burrows, A. 1988, \apj, 334, 891

\bibitem[{}()]{burrows93}
Burrows, A. \& Fryxell, B.A. 1993, \apj, 418, L33

\bibitem[{}()]{burrows95}
Burrows, A., Hayes, J., \& Fryxell, B.A. 1995, \apj, 450, 830

\bibitem[{}()]{burr06}
Burrows, A., Livne, E., Dessart, L., Ott, C.D., \& Murphy, J. 2006,
\apj, 640, 878

\bibitem[{}()]{burr07a}
Burrows, A., Livne, E., Dessart, L., Ott, C.D., \& Murphy, J. 2007a,
\apj, 655, 416

\bibitem[{}()]{burr07b}
Burrows, A., Dessart, L., Livne, E., Ott, C.D., \& Murphy, J. 2007b,
\apj, 664, 416

\bibitem[{}()]{dess06}
Dessart, L., Burrows, A., Livne, E., \& Ott, C.D. 2006, \apj, 645, 534

\bibitem[{}()]{dighe03}
Dighe, A.S., Keil, M.Th., \& Raffelt, G.G. 2003, JCAP, 6, 005

\bibitem[{}()]{dimm08}
Dimmelmeier, H., Ott, C.D., Marek, A., \& Janka, H.-Th. 2008,
\prd, 78, 064056

\bibitem[{}()]{epstein78}
Epstein, R. 1978, \apj, 223, 1037

\bibitem[{}()]{epstein79}
Epstein, R. 1979, \mnras, 188, 305

\bibitem[{}()]{fisher08}
Fischer, T., G\"ogelein, P., Liebend\"orfer, M., Mezzacappa, A.,
\& Thielemann, F.-K. 2008, in {\em Origin of Matter and Evolution 
of Galaxies}, eds.\ T.~Suda, T.~Nozawa, A.~Ohnishi, K.~Kato, M.Y.~Fujimoto,
T.~Kajino, and S.~Kubono, American Institute of Physics Conf.\ Proc.,
Vol.~1016, p.~277

\bibitem[{}()]{Foglizzo.01}
Foglizzo, T., 2001, \aap, 368, 311

\bibitem[{}()]{Foglizzo.02}
Foglizzo, T., 2002, \aap, 392, 353

\bibitem[{}()]{foglizzo06}
Foglizzo, T., Scheck, L., \& Janka, H.-Th. 2006, \apj, 652, 1436

\bibitem[{}()]{foglizzo07}
Foglizzo, T., Galletti, P., Scheck, L., \& Janka, H.-Th.
2007, \apj, 654, 1006 

\bibitem[{}()]{FryWar02}
Fryer, C.L. \& Warren, M.S. 2002, \apjl, 574, L65

\bibitem[{}()]{FryWar04}
Fryer, C.L. \& Warren, M.S. 2004, \apj, 601, 391

\bibitem[{}()]{fryer02}
Fryer, C.L., Holz, D.E., \& Hughes, S.A. 2002, \apj, 565, 430

\bibitem[{}()]{fryer04}
Fryer, C.L., Holz, D.E., \& Hughes, S.A. 2004, \apj, 609, 288

\bibitem[{}()]{heger05}
Heger, A., Woosley, S.E., \& Spruit, H.C. 2005, \apj, 626, 350

\bibitem[{}()]{herant94}
Herant, M., Benz, W., Hix, W.R., Fryer, C.L., \& Colgate, S.A. 1994, \apj,
435, 339

\bibitem[{}()]{HiWo85}
Hillebrandt, W. \&  Wolff, R.G. 1985, in Nucleosynthesis:
Challenges and New Developments, ed.\ W.D.~Arnett \&
J.W.~Truran (Chicago: Univ. Chicago Press), 131

\bibitem[{}()]{Hillebrandt.etal:1984}
Hillebrandt, W., Nomoto, K., \& Wolff, R.G. 1984, \aap, 133, 175

\bibitem[{}()]{hira87}
Hirata, K., et al. 1987, \prl, 58, 1490 (Kamiokande II collaboration)

\bibitem[{}()]{Iwakami08}
Iwakami, W., Kotake, K., Ohnishi, N., Yamada, S., \& Sawada, K.
2008, \apj, 678, 1207

\bibitem[{}()]{janka96}
Janka, H.-Th. \& M\"uller, E. 1996, \aap, 306, 167

\bibitem[{}()]{janka05}
Janka, H.-Th., Buras, R., Kitaura, F.S., Marek, A., Rampp, M., 
\& Scheck, L. 2005, Nucl.\ Phys., A758, 19c

\bibitem[{}()]{janka07}
Janka, H.-Th., Langanke K., Marek A., Mart\'{\i}nez-Pinedo G.,
\& M\"uller B. 2007, Physics Reports, 442, 38

\bibitem[{}()]{keil95}
Keil, W. \& Janka, H.-Th. 1995, \aap, 296, 145

\bibitem[{}()]{keil96}
Keil, W., Janka, H.-Th., \& M\"uller, E. 1996, \apj, 473, L111

\bibitem[{}()]{keil03}
Keil, W., Raffelt, G., \& Janka, H.-Th. 2003, \apj, 590, 971

\bibitem[{}()]{kotake06}
Kotake, K., Sato, K., \& Takahashi, K. 2006, Rep.\ Prog.\ Phys.,
69, 971 

\bibitem[{}()]{kotake07}
Kotake, K., Ohnishi, N., \& Yamada, S. 2007, \apj, 655, 406

\bibitem[{}()]{Lattimer91}
Lattimer, J.M. \& Swesty, F.D. 1991, Nucl.\ Phys., A535, 331

\bibitem[{}()]{lieb03}
Liebend\"orfer, M., Mezzacappa, A., Messer, O.E.B., 
Mart\'{\i}nez-Pinedo, G., Hix., W.R., \& Thielemann, F.-K. 2003,
Nucl.\ Phys., A719, 144c

\bibitem[{}()]{luna08}
Lunardini, C., M\"uller, B., \& Janka, H.-Th. 2008, 
\prd, 78, 023016

\bibitem[{}()]{marek07}
Marek, A., 2007, PhD Thesis, Technische Universit\"at M\"unchen

\bibitem[{}()]{marek07b}
Marek, A. \& Janka, H.-Th. 2007, \apj, in press; arXiv:0708.3372

\bibitem[{}()]{marek06}
Marek, A., Dimmelmeier, H., Janka, H.-Th., M\"uller, E., \& Buras, R.
2006, \aap, 445, 273

\bibitem[{}()]{mueller97}
M\"uller, E. \& Janka, H.-Th. 1997, \aap, 317, 140

\bibitem[{}()]{mueller04}
M\"uller, E., Rampp, M., Buras, R., Janka, H.-Th.,
\& Shoemaker, D.H. 2004, \apj, 603, 221

\bibitem[{}()]{murphy08}
Murphy, J.W. \& Burrows, A. 2008, \apj, 688, 1159

\bibitem[{}()]{nakazato08}
Nakazato, K., Sumiyoshi, K., Suzuki, H., \& Yamada, S. 2008,
Phys.\ Rev.~D, 78, 083014 

\bibitem[{}()]{ohnishi06}
Ohnishi, N., Kotake, K., \& Yamada, S. 2006, \apj, 641, 1018

\bibitem[{}()]{ott08b}
Ott, C.D. 2008, Classical and Quantum Gravity, submitted; arXiv:0809.0695

\bibitem[{}()]{ott04}
Ott, C.D., Burrows, A., Livne, E., \& Walder, R. 2004, \apj, 600, 834

\bibitem[{}()]{ott06}
Ott, C.D., Burrows, A., Dessart, L., \& Livne, E. 2006, \prl, 96, 201102

\bibitem[{}()]{ott08}
Ott, C.D., Burrows, A., Dessart, L., \& Livne, E. 2008, \apj, 685, 1069

\bibitem[{}()]{raffelt01}
Raffelt, G.G. 2001, \apj, 561, 890

\bibitem[{}()]{rampp02}
Rampp, M., Janka, H.-Th. 2002, \aap, 396, 361

\bibitem[{}()]{sagert08}
Sagert, I., Hempel, M., Pagliara, G., Schaffner-Bielich, J., 
Fischer, T., Mezzacappa, A., Thielemann, F.-K., \& Liebend\"orfer, M.
2008, \prl, submitted

\bibitem[{}()]{scheck04}
Scheck, L., Plewa, T., Janka, H.-Th., Kifonidis, K., \& M\"uller, E.
2004, \prl, 92, 011103

\bibitem[{}()]{scheck06}
Scheck, L., Kifonidis, K., Janka, H.-Th., \& M\"uller, E. 2006,
\aap, 457, 963

\bibitem[{}()]{scheck08}
Scheck, L., Janka, H.-Th., Foglizzo, T., \& Kifonidis, K. 2008,
\aap, 477, 931

\bibitem[{}()]{shen98}
Shen, H., Toki, H., Oyamatsu, K., \& and Sumiyoshi K. 1998, Nucl. Phys.,
A637, 435

\bibitem[{}()]{spier07}
Spiering, C. 2007, The Messenger, 129, 33

\bibitem[{}()]{sumi06}
Sumiyoshi, K., Yamada, S., Suzuki, H., \& Chiba S. 2006, \prl,
97, 091101 

\bibitem[{}()]{sumi08}
Sumiyoshi, K., Yamada, S., \& Suzuki, H. 2007, \apj, 667, 382

\bibitem[{}()]{swesty05}
Swesty, F.D. \& Myra, E.S. 2005, Journal of Physics: Conference Series, 
16, 380

\bibitem[{}()]{thompson05}
Thompson, T.A., Quataert, E., \& Burrows, A. 2005, \apj, 620, 861

\bibitem[{}()]{yamasaki08}
Yamasaki, T. \& Foglizzo, T. 2008, \apj, 679, 607

\bibitem[{}()]{woosley06}
Woosley, S.E. \& Bloom, J.S. 2006, \araa, 44, 507

\bibitem[{}()]{woosley95}
Woosley, S.E. \& Weaver, T.A. 1995, \apjs, 101, 181

\end{thebibliography}


}

\end{document}